\documentclass[twocolumn]{aastex631}

\newcommand{\water}{H$_2$O}

\newcommand{\methanol}{CH$_3$OH}

\usepackage{hyperref}
\usepackage{float}
\usepackage{subfigure}
\usepackage{graphicx} 
\usepackage{scalerel}
\usepackage{natbib}
\usepackage{wrapfig}
\usepackage{tabularx}
\usepackage{placeins}

\begin{document}

\title{Laboratory Evidence that Methanol-Rich Ice Mantles Lower Methyl Formate Binding Energies}

\correspondingauthor{Rachel E. Gross}
\email{reg4ff@virginia.edu}

\author[0000-0002-0477-6047]{Rachel E. Gross}
\affiliation{University of Virginia, Department of Astronomy, Charlottesville, VA 22904, USA}

\author[0000-0002-3800-9639]{Jeroen Terwisscha van Scheltinga}
\affiliation{Photonics Research Group, The Hague University of Applied Sciences, 2628 AL Delft, The Netherlands}

\author[0000-0003-2076-8001]{L. Ilsedore Cleeves}
\affiliation{University of Virginia, Department of Astronomy, Charlottesville, VA 22904, USA}

\author[0000-0002-4549-7204]{Catherine A. Dukes}
\affiliation{University of Virginia, Laboratory for Astrophysics \& Surface Physics, Dept. Materials Science \& Engineering, Charlottesville, VA 22904, USA}

\author[0000-0001-7723-8955]{Robin T. Garrod}

\affiliation{University of Virginia, Department of Astronomy, Charlottesville, VA 22904, USA}

\author[0000-0002-5061-3054]{Alexia Simon}
\affiliation{University of Virginia, Department of Astronomy, Charlottesville, VA 22904, USA}

\author{Shane Sawyer}
\affiliation{University of Virginia, Department of Astronomy, Charlottesville, VA 22904, USA}

\author[0000-0002-6112-7502]{Adam Woodson}
\affiliation{University of Virginia, Laboratory for Astrophysics \& Surface Physics, Dept. Materials Science \& Engineering, Charlottesville, VA 22904, USA}




\begin{abstract}

Complex organic molecules are widely detected in star-forming regions and are important precursors to prebiotic species. Because these molecules are thought to form primarily within icy grain mantles, their desorption is assumed to be regulated by the sublimation of the surrounding H$_2$O ice matrix. However, observations increasingly reveal gas-phase abundances of methyl formate (MF) at temperatures well below the sublimation temperature of water ice ($T \sim 100$~K), challenging this view. To investigate how ice composition and phase influence MF desorption, we present laboratory temperature-programmed desorption measurements of MF from four astrophysically relevant substrates: amorphous and crystalline H$_2$O and CH$_3$OH. Desorption kinetics were analyzed using pre-exponential factors derived from transition-state theory with leading-edge analysis and non-negative least-squares inversion to recover binding-energy distributions. We find that MF binds more weakly to CH$_3$OH ice than to H$_2$O-rich substrates, with representative binding energies of $6247 \pm 43$~K for amorphous H$_2$O, $6213 \pm 38$ K for crystalline H$_2$O, and $5469 \pm 40$~K for amorphous CH$_3$OH. MF deposited on crystalline CH$_3$OH exhibits coverage-independent leading edges consistent with multilayer or island desorption, yielding an effective binding energy of $5333 \pm 17$~K. Enhanced surface mobility and reduced trapping efficiency on methanol ice, combined with weaker MF binding, provide a natural explanation for the early appearance of gas-phase MF observed in star-forming environments, with characteristic desorption temperatures of $\sim85$~K on CH$_3$OH and $\sim95$--100~K on H$_2$O. Together, these effects support a two-step thermal desorption pathway in which MF associated with CH$_3$OH-rich ice can be released prior to bulk H$_2$O sublimation.

\end{abstract}



\section{Introduction} \label{sec:intro}

Methyl formate (HCOOCH$_3$; hereafter MF) is one of the most abundant oxygen-bearing complex organic molecules (COMs) detected in star-forming regions and is observed in environments ranging from molecular clouds to protoplanetary disks \citep[e.g.,][]{jorgensen2016,McGuire2022,Busch2022, Booth2024}. COMs, defined as carbon-bearing species containing six or more atoms, are thought to form efficiently within the icy mantles of interstellar dust grains through grain-surface chemistry \citep{Fedoseev2015,Garrod2022}. As star-forming environments warm, molecules stored in these ice mantles sublimate into the gas phase through thermal desorption, where they become more readily observable at infrared and millimeter wavelengths \citep{herbst_vanDishoeck2009,Boogert2015}. Understanding the parameters that regulate when and how molecules desorb from interstellar ices is therefore essential for interpreting the chemical evolution of star and planet-forming environments.

High angular-resolution observations from the Atacama Large Millimeter/submillimeter Array (ALMA) and other radio facilities increasingly reveal that gas-phase COMs are present in environments far colder than traditional astrochemical models would predict. A clear example is MF in the prestellar core L1689B, where detections of several O-bearing COMs by \citet{Bacmann2012} have been reported at a kinetic temperature of only 10 K -well below the $\sim$100~K temperature typically associated with MF release in \water-rich ices \citep{Kirk2007}. In the high-mass hot-core Sagittarius B2(N1), ALMA observations show that certain COMs like MF maintain a detectable gas-phase abundance down to $\sim80$~K, while other O-bearing COMs such as methanol (\methanol) exhibit a strong abundance increase coincident with the \water-ice sublimation temperature ($\sim100$~K). This behavior suggests that methanol is closely coupled to \water\ ice co-desorption, whereas MF does not strictly follow this relationship and remains detectable at lower temperatures \citep{Busch2022}. Most recently, ALMA detections of MF in protoplanetary disks \citep{Booth2024} extend this “cold-COM puzzle” to the planet-formation stage. JWST observations have detected MF in the solid phase toward embedded protostars \citep[e.g.,][]{Yang2022,Rocha2024}, confirming that the molecule is present in interstellar ice mantles prior to the onset of thermal desorption. These observations suggest that MF may be released through desorption pathways not strictly tied to the sublimation of bulk \water\ ice. Possible explanations include non-thermal desorption mechanisms such as photodesorption and reactive desorption, or a two-step thermal desorption process; \citet{Busch2022} proposed, in the latter case, that the observed release of MF and other COMs at temperatures below 100 K could be due to their presence in the \water-poor and initially CO-rich outer layers of the dust-grain ice mantles. The initial, low-temperature build-up of CO and \methanol\ in those outer layers would also allow COMs to be formed, through non-diffusive reactions or via energetic processing. At higher temperatures, following the desorption of volatile CO at 20 – 30 K, the remaining CO- and \methanol-related COMs would experience lower binding energies than the same molecules in the deeper, \water-rich layers, allowing them to desorb at temperatures less than 100 K. Other COM material would remain trapped in the \water-rich layers beneath until higher temperatures were reached.

Interpretation of these observations requires a detailed physical understanding of how molecules desorb from interstellar ice mantles. Thermal desorption from grain surfaces is governed by a molecule’s surface binding energy ($E_{\rm des}$), defined as the energy required for an atom or molecule to escape the surface and desorb into gas. Molecules generally bind more strongly to polar substrates such as \water\ ice than to less polar surfaces such as CO-rich or organic ices \citep{Collings2004,Bisschop2007lab,Burke2015}. Laboratory measurements using temperature programmed desorption (TPD) experiments provide essential constraints on the temperatures at which molecules sublimate to the gas phase during star formation.

Interstellar ice mantles are structurally and compositionally heterogeneous, and these variations can influence molecular surface binding energies and desorption behavior \citep[][]{Boogert2015}. Pure \water\ ice may form as compact amorphous ice through energetic grain-surface reactions driven by heat or radiation, or as more porous amorphous structures when condensed from the gas phase onto very cold grains \citep{Kouchi1994,Jenniskens1995,Fraser2001,Accolla2013,Minissale2022}. During the star formation process these low-density amorphous ices compact and eventually crystallize, processes that affect molecular mobility, trapping efficiency, and desorption pathways \citep{Collings2004,Cazaux2003}. Mixed ice systems add further complexity. In particular, \methanol–\water\ mixtures can undergo phase segregation, in which an initially mixed ice separates into \methanol-rich and \water-rich domains. This restructuring leads to desorption behavior that differs from either pure ice constituent \citep{Oberg2009,Luna2018}. Consequently, both astrophysical ice composition and phase (amorphous versus crystalline) can influence molecular desorption behavior \citep{Garrod2013}.

The conventional model for COM desorption from icy grains is driven by \water\ ice kinetics. \water\ is the most abundant solid-phase constituent of interstellar ice mantles, and it is generally assumed to act as the primary trapping matrix for most COMs and simpler species \citep[e.g.,][]{Tielens1997, collings2003, Viti2004}. Laboratory investigations have demonstrated that water ice efficiently retains volatile molecules through strong hydrogen-bonding interactions or physical trapping, with release occurring either during bulk ice sublimation or through volcano desorption during the amorphous-to-crystalline phase transition \citep{Smith1997,Collings2004, Bisschop2007lab,Simon2023}. Under this view, COM abundances in the gas phase are not expected to be significant until dust temperatures reach the water sublimation threshold, typically $\sim100–160$~K, depending on the heating rate, vacuum pressure, and ice morphology \citep{Cazaux2003, Ceccarelli2004, Burke2015}.

The formation chemistry of MF provides important context for investigating its desorption pathways. \methanol, an important precursor for MF synthesis, is one of the most abundant molecules in interstellar ices, ranging from less than 1\% to over 25\% relative to \water\ in low-mass protostars and up to $\sim30$\% in massive young stellar objects \citep{Pontoppidan2004,Boogert2015,Luna2018}. \methanol\ forms efficiently on grain surfaces through sequential hydrogenation of CO at temperatures as low as $\sim10$ K \citep{Watanabe2002,Fuchs2009}. MF is subsequently synthesized within \methanol-bearing ices through several pathways,
including radical recombination during warm-up \citep{Garrod2006}, UV photolysis of solid methanol \citep{Oberg2009}, and cosmic-ray processing of CO:\methanol\ mixtures \citep{Modica2010}.

MF formation may be enhanced by the addition of \water\ in photolyzed \methanol-containing ice mixtures. 
\citet{Ishibashi2021} found that UV processing of \methanol\ on 
amorphous \water\ ice efficiently produced MF, extending the work to earlier \methanol-dominated experiments \citep{Oberg2009}. They attributed this enhancement to 
OH supplied by \water\ photolysis, which promotes CH$_3$O formation 
and facilitates an MF formation pathway through methoxymethanol 
(CH$_3$OCH$_2$OH). This work identifies \water\ as an active 
component in the grain-surface formation chemistry of MF.

Existing laboratory investigations of MF desorption have primarily focused on its interaction with water-rich substrates. \citet{Bertin2010_MF} studied MF adsorption and desorption on water ice using temperature-programmed desorption and FTIR spectroscopy, finding that the MF surface monolayer desorbs approximately 15 K below the onset of bulk water sublimation, suggesting partial independence from the surrounding water matrix. Similarly, \citet{Burke2015} showed that when MF is both layered and mixed with amorphous solid water (ASW), it exhibits both trapping within the \water\ ice and has a partially independent desorption component, indicating that a fraction of MF could escape prior to bulk water sublimation. However, the interaction of MF with a \water-poor ice matrix remains poorly constrained experimentally.

In this work, we present a systematic laboratory investigation of MF desorption from four astrophysically relevant ice substrates: amorphous and crystalline water ice, and amorphous and crystalline methanol ice. Using temperature-programmed desorption (TPD) experiments, we measure the desorption behavior of MF and determine its binding energies on each substrate. By directly comparing MF desorption from \methanol-dominated and \water-dominated ice surfaces across both amorphous and crystalline phases, we test the hypothesis that methanol-rich ice environments present a lower thermal barrier for MF sublimation, helping to explain the presence of cold MF detections. 

The structure of this paper is as follows. Section \ref{methods} describes the experimental apparatus and preparation, and the kinetic analysis framework used to derive MF binding energies. Section \ref{results} presents the temperature-programmed desorption spectra and compares MF desorption from \water\ and \methanol\ substrates. Section \ref{discussion} interprets these laboratory results in an astrophysical context and discusses their implications for the thermal release of MF in star-forming environments. Finally, Section \ref{summary} summarizes the main conclusions.

\section{Methods} \label{methods}
\subsection{Experimental Setup and Ice Preparation}

All experiments were conducted in the Integrated Chemical Experiments (ICE) UHV multi-technique cryogenic analytical chamber, part of the KEVION facility at the University of Virginia (see Fig.1). The chamber is a non-magnetic, stainless steel ultra-high vacuum (UHV) chamber with a base pressure of $\sim4\times10^{-10}$ Torr at room temperature \citep{Schaible2014}. At this base pressure, background deposition over the duration of the experiments is negligible compared with the intentionally deposited ice films. The chamber is equipped with a closed-cycle helium cryostat (Janis Model 205) with radiation shield coupled directly to the target that allows substrate temperatures between 10 and 300~K with a LakeShore 336 temperature controller. Temperature regulation is achieved using a resistive heater and monitored with two silicon diode sensors (precision $\pm0.1$~K; accuracy $\sim$1~K). Infrared spectra were collected with a N$_2$-coupled 670 Thermo Nicolet Fourier-transform infrared spectrometer covering the range 4000 to 900 wavenumbers with 2 cm$^{-1}$ resolution. Gas species were detected during TPD with a Hiden Analytical HAL 4 quadrupole mass spectrometer (QMS) equipped with an electrostatic energy analyzer (EQS), operated in multiple ion detection (MID) mode. 

\begin{figure*}[t]
\centering
\includegraphics[width=0.85\textwidth]{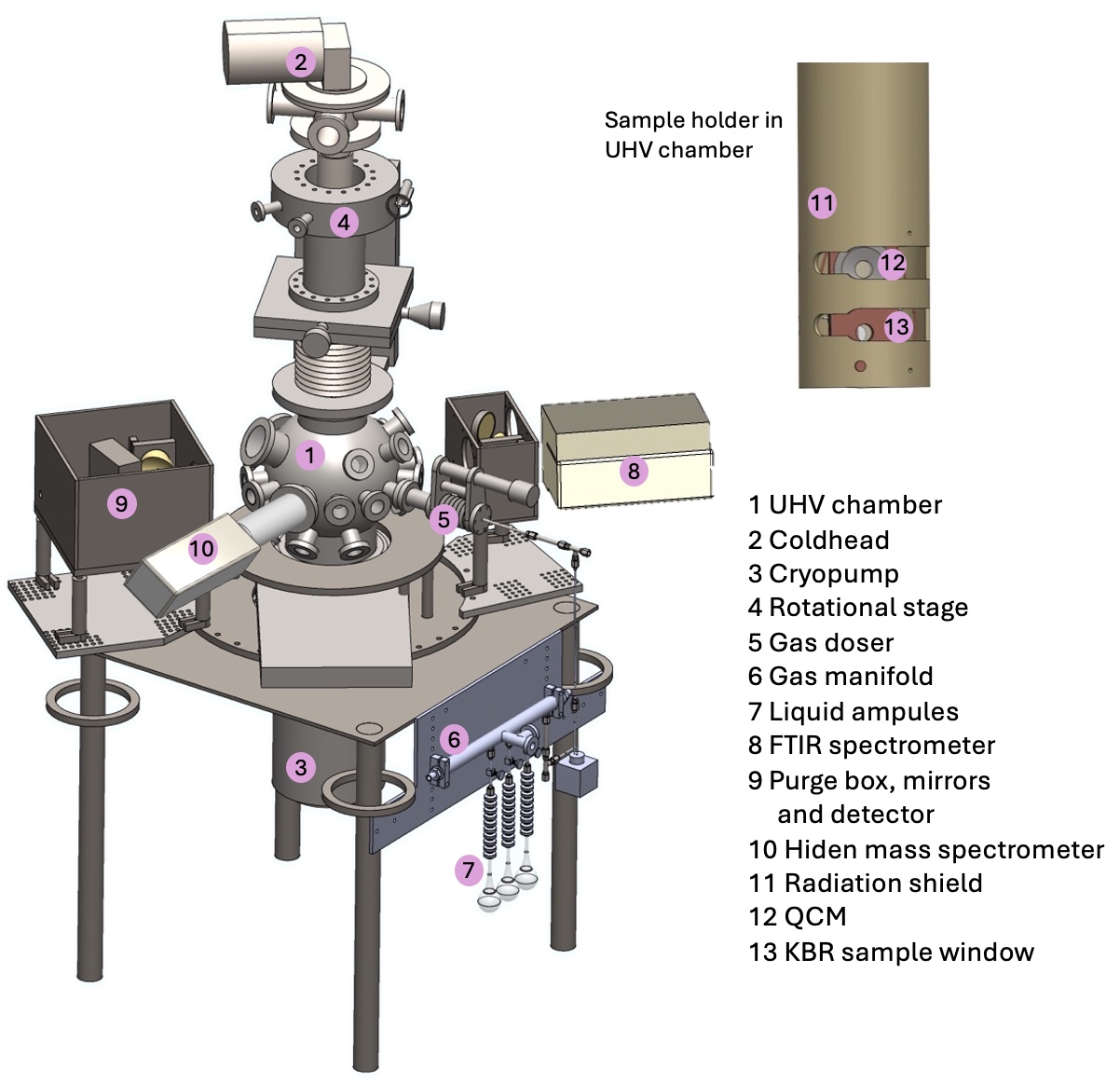}
\caption{Schematic of the experimental chamber setup.}
\label{fig:chamber_schematic}
\end{figure*}

Ice films were deposited primarily on a gold-coated 5MHz quartz crystal microbalance (QCM) mounted on the cryo-cooled target, allowing for measurement of the deposited mass and determination of the ice thickness via the Sauerbrey relation (see Section \ref{qcm}). In a subset of experiments, ices were deposited onto a potassium bromide (KBr) window on the same cryo-target to enable Fourier Transform Infrared (FTIR) spectroscopy in transmission mode, allowing independent verification of deposited ice phase and composition. Representative FTIR spectra confirming the structural phases of the ice substrates used in this work are presented in Appendix \ref{app:ftir}.

All chemicals used in the study were obtained at high purity: Deionized water, methanol (99.8\%, spectrophotometric grade, Alfa Aesar), and MF (Sigma-Aldrich, anhydrous, 99\%). Prior to deposition, each liquid was purified by three freeze pump thaw cycles in a liquid nitrogen bath to remove dissolved gases and volatile impurities.

Ice deposition was performed using a capillary-array gas doser connected to the gas manifold through a leak valve, positioned approximately 1.5 inches from the QCM or KBr window. Substrate temperature during deposition is maintained at a target value (described below) to control ice phase and morphology.

Amorphous solid water (ASW) and amorphous \methanol\ substrate ices were prepared by direct deposition of the respective vapor onto the QCM held at 35~K. This deposition temperature yields compact amorphous ice films \citep{Accolla2013}. These films are expected to contain a reduced level of open porosity compared to ice grown at even lower temperatures (e.g.,15 K), but may still contain buried microporous structures that become accessible upon heating. FTIR spectra confirm the amorphous character of these films (Appendix \ref{app:ftir}).

Crystalline ice substrates were prepared by direct deposition at temperatures where structural reordering occurs. \water\ vapor deposited at 130 K rapidly crystallizes into the cubic ice phase \citep{Jenniskens1994,Collings2004}, producing crystalline ice without the need for post-deposition annealing. \methanol\ deposited at 115 K readily forms the crystalline $\alpha$-methanol phase \citep{Boogert2015}.

After the substrate layer was grown, the sample was cooled to 35 K before depositing MF. Cooling to the same temperature for all substrates guarantees that the subsequent MF adsorption occurs under identical surface‑temperature conditions, eliminating any systematic bias in the initial coverage and sticking.

To confirm that direct deposition and annealing routes produce equivalent crystalline structures, we performed a control series in which amorphous ice was first deposited at 35 K, then annealed to the appropriate crystallization temperature (130 K for H$_2$O, 115 K for CH$_3$OH), after which the temperature was lowered back to 35 K. The resulting TPD profiles and FTIR spectra were indistinguishable from those obtained by direct‑deposition, within the experimental noise. Consequently, all crystalline substrates reported in the manuscript were prepared by the direct‑deposition method.

Mixed MF:H$_2$O and MF:CH$_3$OH ices were prepared by introducing the gases simultaneously into the dosing manifold. Mixing ratios were determined using the ideal gas law, and the relative deposition rates of MF and the co-deposited species were verified through sequential layering experiments prior to preparing mixed ices. The full list of experiments performed for this manuscript can be found in Table \ref{tab:experiments}.


\begin{deluxetable*}{l c c c c}
\tablecaption{Summary of experiments performed in this work.\label{tab:experiments}}
\tablewidth{0pt}
\tablehead{
\colhead{Experiment} &
\colhead{Substrate ML} &
\colhead{MF coverage (ML)} &
\colhead{MF:X} &
\colhead{$\beta$ (K min$^{-1}$)}
}
\startdata
\multicolumn{5}{c}{\textit{TPD / QCM desorption experiments}} \\
MF on ASW           & $\sim$50 & 0.20, 0.30, 0.55, 1.00, 1.50, 2.50 & \dots & 5 \\
MF on c-H$_2$O      & $\sim$50 & 0.20, 0.30, 0.40, 0.60, 1.00, 1.50 & \dots & 5 \\
MF on a-CH$_3$OH    & $\sim$50 & 0.25, 0.45, 0.60, 1.00, 2.00 & \dots & 5 \\
MF on c-CH$_3$OH    & $\sim$50 & 0.10, 0.25, 0.36, 0.55, 0.80, 1.30, 2.60 & \dots & 5 \\
MF multilayer       & \dots & 31, 41, 53 & \dots & 5 \\
MF:H$_2$O mixture   & $\sim$50 & \dots & 1:50 & 5 \\
MF:CH$_3$OH mixture & $\sim$50 & \dots & 1:50 & 5 \\
\hline
\multicolumn{5}{c}{\textit{Heating-rate verification TPD experiments}} \\
MF on ASW           & $\sim$50 & 1.0 & \dots & 2, 10 \\
MF on c-H$_2$O      & $\sim$50 & 1.0 & \dots & 2, 10 \\
MF on a-CH$_3$OH    & $\sim$50 & 1.0 & \dots & 2, 10 \\
MF on c-CH$_3$OH    & $\sim$50 & 1.0 & \dots & 2, 10 \\
\hline
\multicolumn{5}{c}{\textit{FTIR KBr experiments}} \\
Pure H$_2$O ice      & $\sim$50 & \dots & \dots & 5 \\
Pure CH$_3$OH ice    & $\sim$50 & \dots & \dots & 5 \\
Pure MF ice          & $\sim$50 & \dots & \dots & 5 \\
MF on a-CH$_3$OH     & $\sim$50 & layered & \dots & 5 \\
MF:CH$_3$OH mixture  & $\sim$50 & \dots & 1:50 & 5 \\
MF on c-CH$_3$OH     & $\sim$50 & layered & \dots & 5 \\
\enddata
\tablenotetext{}{\textit{Note.} ASW denotes compact amorphous solid water. TPD experiments were primarily performed with a heating rate of 5 K min$^{-1}$. For each substrate, one representative MF coverage of $\sim$1 ML was additionally measured at heating rates of 2 and 10 K min$^{-1}$ to verify the TST prefactor and test heating-rate invariance of the derived kinetic parameters. FTIR spectra were collected during paused warm-up increments during heating at 5 K min$^{-1}$.}
\end{deluxetable*}

\subsection{Quartz Crystal Microbalance}

\label{qcm}

\begin{figure*}
  \centering
  \includegraphics[scale=0.5]{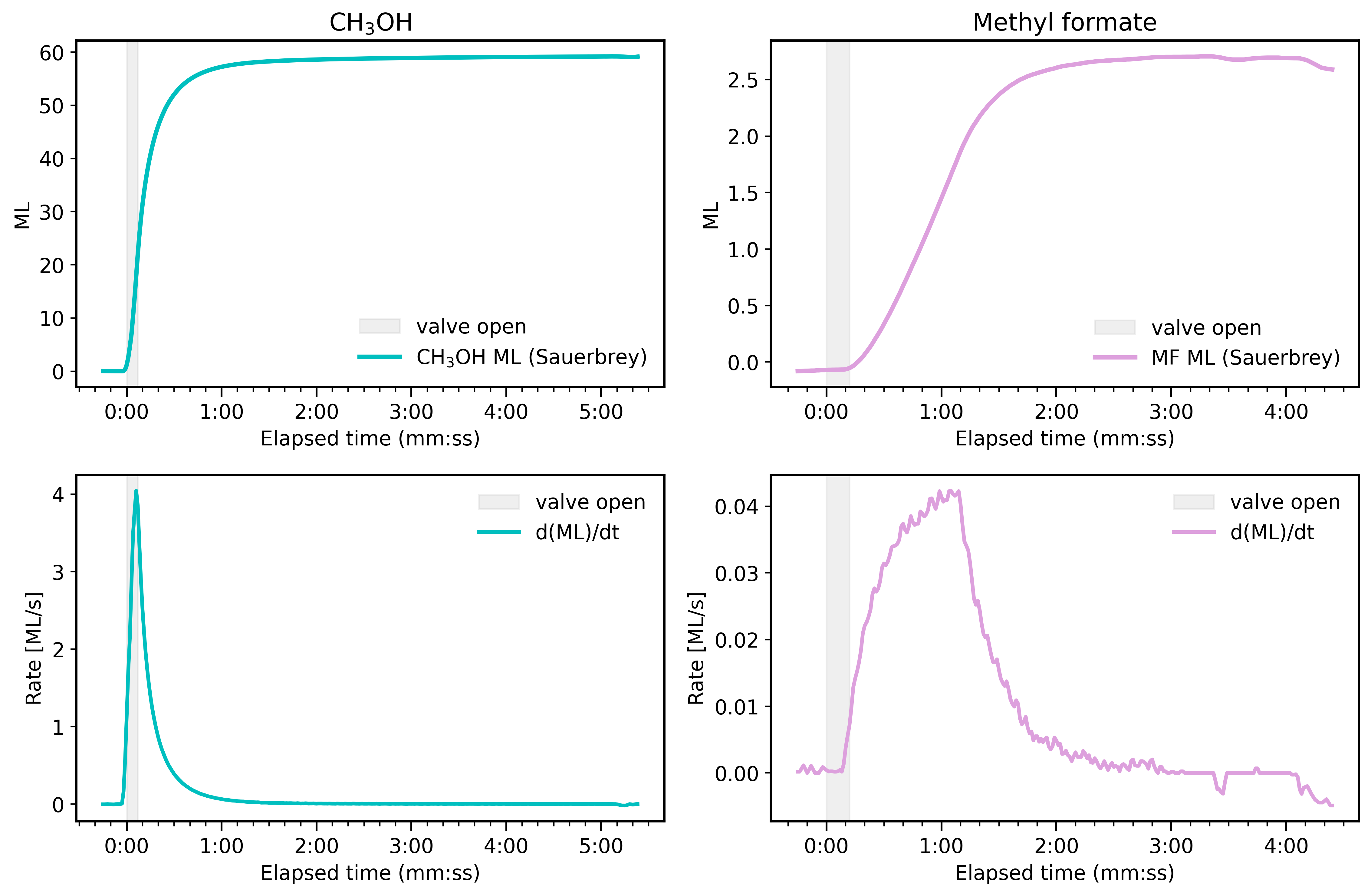}
  \caption{
Conversion of QCM frequency shifts into surface coverage in monolayers using the Sauerbrey relation.
Top panels show cumulative coverage during deposition for methanol (left) and MF (right), plotted on a common time axis for direct comparison, with the valve-open deposition window indicated. Bottom panels show the corresponding deposition rates 
derived from the time derivative of the smoothed coverage. These measurements establish the absolute 
coverage scale used for all subsequent TPD experiments.
}
  \label{qcm_fig}
\end{figure*}

Ice deposition was quantified with a QCM by converting the measured resonance frequency shift into an areal mass loading via the Sauerbrey relation \citep{Sauerbrey1959}. Under this approximation, the deposited film is treated as a rigid, laterally uniform mass load fully coupled to the oscillating crystal. For a rigid, thin film coupled to the crystal, the change in areal mass density is

\begin{equation}
\frac{\Delta m}{A} = - C \, \frac{\Delta f}{n},
\end{equation}

where $\Delta m/A$ is the mass per unit area, $\Delta f$ is the frequency shift, $n$ is the harmonic number (here the fundamental, $n=1$), and $C$ is the sensitivity constant of the crystal. For a 5~MHz AT-cut quartz crystal in vacuum/air, $C = 17.7~\mathrm{ng~cm^{-2}~Hz^{-1}}$ \citep{Huang2017qcm}. Because $C \propto f_0^{-2}$, the sensitivity constant was adjusted using the measured baseline frequency $f_0$ as

\begin{equation}
C(f_0) = 17.7 \left(\frac{5~\mathrm{MHz}}{f_0}\right)^2 \mathrm{ng~cm^{-2}~Hz^{-1}}.
\end{equation}

Frequency baselines were determined from manually selected stable time intervals immediately prior to deposition, typically spanning $\sim$30--120~s. To account for slow instrumental drift, $f_0$ was estimated either as the mean frequency within this baseline interval or from a linear fit evaluated at the segment start time.

The deposited areal mass density was converted to molecular surface coverage using the molar mass and Avogadro's number. One monolayer (ML) is defined as $1\times10^{15}$ molecules cm$^{-2}$, consistent with the surface site density commonly adopted in laboratory astrochemistry and TPD studies \citep{Collings2004,Bisschop2007lab,Burke2015,Ligterink2023,Kruczkiewicz2024desorption}. For a species with molar mass $M$, the mass corresponding to one monolayer is $\left(\frac{m}{A}\right)_{\mathrm{1~ML}} = \frac{M}{N_A}\left(1\times10^{15}\right)$, yielding the surface coverage

\begin{equation}
\theta(\mathrm{ML}) =
\frac{\Delta m/A}{(m/A)_{\mathrm{1~ML}}}.
\end{equation}

\noindent This approach expresses ice deposition in monolayers without requiring an assumed bulk ice density, as would be needed for conversion to a physical thickness.

Uncertainties in coverage were estimated from the standard deviation of the baseline frequency and propagated through the Sauerbrey conversion. Although the Sauerbrey relation strictly applies to rigid films, the approximation is valid at low-temperature, vacuum conditions and for submicron ice thicknesses used here \citep{Huang2017qcm,Johannsmann2021qcm,Burke2015}.
The resulting QCM-derived coverages establish the boundary between sub-monolayer and multilayer regimes used throughout the TPD analysis (Figure \ref{qcm_fig}).

\subsection{TPD Peak Processing}
\label{peak_processing}

TPD experiments were performed by heating the substrate at constant rates of 2, 5, or 10 K min$^{-1}$ following ice deposition. Desorbing neutral species were monitored with the Hiden QMS using 70 eV electron impact ionization. Unfragmented MF molecules were tracked at $m/z = 60$ with time and temperature, while complete methanol (\methanol) molecules were monitored at $m/z = 31$ (with fragments 15, 29, and 31 used for tracking of both species). \water\ was tracked with $m/z = 18$.

Fragmentation patterns were determined from pure MF and pure \methanol\ reference experiments conducted under identical ionization and QMS tuning conditions. These reference spectra were used to confirm fragment assignments and to deconvolve overlapping fragment contributions. Details are provided in Appendix \ref{sec:cracking}, Table \ref{tab:cracking_ratios}.

To remove instrumental background in the TPD data, a first-order polynomial was fitted to signal regions where no desorption was detected (typically 80–95 K and 190–195 K) \citep{Collings2004,Burke2015,Santos2025}. The fitted baseline was subtracted from the QMS signal. The measured signal is proportional to the density of desorbing molecules entering the mass spectrometer; however, the desorption flux depends on the mean molecular velocity, which scales as $\sqrt{T}$. Therefore, the desorption rate is proportional to the measured signal multiplied by $\sqrt{T}$, and this correction was applied prior to further analysis \citep{Nekrylova1993}.

The desorption features in the rate vs. temperature plots were noted and identified within specific temperature regions: MF–MF multilayer desorption (105–125 K), MF–substrate desorption (125–150 K), and the molecular volcano feature (145–175 K), concurrent with \water\ sublimation. Our analysis focused on the MF–substrate peak, which was defined as the dominant desorption feature occurring after multilayer sublimation but before reaching temperatures that induce substrate restructuring (i.e., crystallization, etc.). The peak desorption rate temperature $T_{\rm peak}$ was identified as the local maximum within this feature.

\subsection{Kinetic Analysis Framework}

Binding energies ($E_{\rm des}$) for the MF to an icy substrate were derived using three methods depending on kinetic behavior: the Redhead method (for characteristic energy estimates), non-negative least-squares (NNLS) inversion (to recover binding-energy distributions at low coverage), and leading-edge analysis (for multilayer regimes). To avoid covariance between the pre-exponential factor $\nu$ and $E_{\rm des}$, the prefactor was calculated and fixed using transition-state theory (TST).

\subsubsection{Transition-State-Theory Prefactor}
\label{sec:TST}

Thermal desorption from surfaces is described by the Polanyi–Wigner equation,

\begin{equation}
r_{\text{des}} = -\frac{d\theta}{dt} = \nu \theta^n \exp\left(-\frac{E_{\text{des}}}{RT}\right),
\label{polanyi}
\end{equation}

where $r_{\rm des}$ is the desorption rate, $\theta$ is the surface coverage, $n$ is the desorption order, $\nu$ is the pre-exponential factor (hereafter called the prefactor, $\nu$), and $E_{\rm des}$ is the molecular surface binding energy. Binding energies are reported in Kelvin units ($E_{\rm des}/k_B$) and electron-volts (eV). Rather than assume an arbitrary value for $\nu$, we computed it with transition‑state theory (TST) exactly as described by \citet{Ligterink2023}:

\begin{equation}
\nu(T) = \frac{k_B T}{h} q_{\rm tr,2D}^{\ddagger} q_{\rm rot,3D}^{\ddagger},
\end{equation}

with the two‑dimensional translational and three‑dimensional rotational partition functions and necessary molecular constants (mass, symmetry number, principal moments of inertia $I_{x},I_{y},I_{z}$) taken directly from their Table A.1; they obtained the moments of inertia from the ChemSpider database. Molecular parameters used for MF in this work are listed in Table \ref{tab:TST_params}. All parameters were converted to SI units prior to evaluating the partition functions.

In the full transition-state-theory formalism, the prefactor is given by $\nu = (k_B T/h)(q^{\ddagger}/q_{\rm ads})$, where $q^{\ddagger}$ and $q_{\rm ads}$ are the partition functions of the transition state and adsorbed (reactant) state, respectively. Following \citet{Ligterink2023}, the adsorbed state is assumed to be localized at a binding site, such that $q_{\rm ads} \approx 1$. Under this assumption, the prefactor is dominated by the translational and rotational degrees of freedom gained upon desorption, as captured by the transition-state partition functions in the equation above. The implications of this assumption for the absolute binding-energy scale are discussed in Appendix~\ref{app:qads}.

TST assumes that the ensemble of molecules in the transition state follows a Boltzmann distribution. Recent laboratory studies have shown that this assumption begins to break down for molecules containing more than 20 atoms \citep{Ligterink2023,Minissale2022}.  MF has only eight atoms, well below this empirical limit, so the TST‑derived prefactor, $\nu(T)$, is expected to be valid for the present work.

The temperature-dependent prefactor $\nu(T)$ was evaluated at the experimentally measured peak desorption rate temperature $T_{\rm peak}$ for each heating-rate experiment. The experiment-specific $\nu_{\rm TST}(T_{\rm peak})$ was used in each Redhead calculation, whereas the mean $\langle\nu_{\rm TST}\rangle$ across heating rates was used for the NNLS inversion (Table \ref{tab:nu_E_all_leadingedge}).

\begin{table}[ht!] \centering \caption{Molecular constants used to compute the transition-state-theory prefactor $\nu_{\rm TST}$ for MF.}
\label{tab:TST_params} 
\begin{tabular}{lcc}
\hline 
Symbol & Value & Units \\ 
\hline $m$ & $60.0$ & amu \\ 
$\sigma$ & $1$ & \dots \\ 
$I_x$ & $107.99$ & amu \AA$^2$ \\ 
$I_y$ & $102.32$ & amu \AA$^2$ \\ 
$I_z$ & $8.84$ & amu \AA$^2$ \\ 
$N_{\rm site}$ & $1.0\times10^{15}$ & sites cm$^{-2}$ \\ \hline \end{tabular}
\vspace{2pt} \parbox{0.7\textwidth}{\footnotesize \textbf{Note.} Molecular constants adopted from \citet{Ligterink2023}.}
\end{table}

\subsubsection{Redhead Analysis}

A representative binding energy was estimated from the peak
desorption-rate temperature using the Redhead method
\citep{Redhead1962}. The general desorption rate is described by
Polanyi--Wigner in equation \ref{polanyi}. For first-order desorption with a constant binding energy under a linear temperature ramp, the condition that the desorption rate reaches a maximum at $T_{\rm peak}$ gives the first-order Redhead peak equation,

\begin{equation}
\frac{E_{\rm des}}{T_{\rm peak}^{2}}
=
\frac{\nu}{\beta}
\exp\left(-\frac{E_{\rm des}}{T_{\rm peak}}\right),
\end{equation}

\noindent where $T_{\rm peak}$ is the peak desorption temperature, $\beta$ is the heating rate, and $\nu$ is the pre-exponential factor. This equation is equivalent to Equation~(5a) of \citet{Redhead1962}, with the binding energy expressed in Kelvin as $E_{\rm des}=E/R$. The heating rate $\beta$ is expressed in K~s$^{-1}$. This relation allows the binding energy to be estimated from the measured peak temperature without fitting the full TPD profile.

For each experiment, the pre-exponential factor was calculated from
transition-state theory at the measured $T_{\rm peak}$, as described in Section~\ref{sec:TST}. The experiment-specific value of
$\nu_{\rm TST}$ was then used to solve the equation numerically for
$E_{\rm des}$.

The resulting Redhead-derived binding energies are reported in
Table~\ref{tab:nu_E_all_leadingedge}. These values provide an independent estimate of the characteristic binding-energy scale and serve as a consistency check for the median energies recovered by the NNLS inversion (Section~\ref{sec:dist_inversion}). The Redhead estimates are not used directly in the inversion but inform the choice of the binding-energy grid explored by the NNLS model.

\subsubsection{Binding energy distribution from NNLS inversion}
\label{sec:dist_inversion}

Sub-monolayer desorption of COMs from heterogeneous surfaces cannot generally be described by a single binding energy. Instead, the TPD profile reflects a distribution of adsorption sites that can be represented as a superposition of first-order desorption components spanning a range of binding energies \citep{Noble2012,Doronin2015,Collings2015,Fayolle2016}.
We adopt a non-parametric inversion approach similar to that of \citet{Behmard2019}, in which the TPD curve is modeled as a linear combination of Polanyi–Wigner desorption components evaluated over a grid of binding energies. The binding-energy distribution is recovered using a non-negative least-squares (NNLS) inversion.

For each experiment, the isolated MF-substrate desorption peak (Section \ref{peak_processing}) was extracted from the baseline-subtracted QMS signal and normalized to unit area,

\begin{equation}
R_{\rm exp}(T)
=
\frac{y_{\rm sub}(T)}
{\int y_{\rm sub}(T)\, dT},
\end{equation}

so that the inversion is constrained by the shape of the desorption feature rather than the absolute signal magnitude.

The normalized profile was modeled as a linear combination of basis functions,

\begin{equation}
R_{\rm model}(T_i)
=
\sum_j f_j\,G(T_i,E_j),
\end{equation}

where $G(T_i,E_j)$ is the predicted desorption-rate curve for a trial binding energy $E_j$. Each basis function was generated by numerically solving the first-order Polanyi--Wigner desorption equation using the experimental heating rate and a fixed pre-exponential factor $\nu$, and was normalized to unit area. The coefficients $f_j$ represent the fractional population of adsorption sites with binding energies near $E_j$.

The coefficients were obtained with a non-negative least-squares (NNLS) fit, enforcing $f_j \ge 0$ and therefore guaranteeing physically meaningful site populations. The inversion was performed on a uniform binding-energy grid spanning $E_{\rm min}=4500$~K to $E_{\rm max}=7500$~K with a bin width of $\Delta E=90$~K. This bin size was chosen as a compromise between resolving structure in the distribution and avoiding degeneracies in the NNLS inversion; tests with 50 and 100~K yielded qualitatively similar results \citep[][]{Santos2025}. The grid bounds were selected to bracket the Redhead estimate and avoid boundary artifacts.

From the recovered distribution $f(E)$ we computed several summary statistics describing the characteristic binding energy and spread of adsorption sites. For each individual TPD experiment, we adopt the median of the recovered binding-energy distribution as the representative binding energy for that run. For substrates where multiple low-coverage experiments were analyzed, the representative binding energy reported in Table \ref{tab:binding_energy_summary} corresponds to the mean of the individual run medians.

The uncertainty on the binding-energy distribution was estimated by propagating three independent sources of uncertainty: experimental noise, temperature calibration, and coverage-to-coverage scatter. Experimental noise was quantified using a Monte Carlo (MC) approach. For each QMS run, the noise level, $s_{\rm QMS}$, was estimated from the median absolute deviation (MAD) of the baseline-subtracted signal in signal-free regions surrounding the desorption feature. We then generated 1000 synthetic TPD curves by adding Gaussian noise with standard deviation $s_{\rm QMS}$ to the baseline-subtracted desorption signal. The synthetic curves were area-normalized and inverted using the same NNLS procedure as the experimental data. The standard deviation of the resulting median binding energies was adopted as the MC noise uncertainty.

A temperature-scale uncertainty was evaluated by repeating the inversion with the temperature axis shifted by $+\Delta T$ and $-\Delta T$, where $\Delta T=0.8$~K represents the estimated uncertainty in the experimental
temperature scale. The temperature contribution was taken as half the absolute difference between the median binding energies recovered from the two shifted temperature scales. For each run, the MC and temperature contributions were combined in quadrature to obtain the total per-run
uncertainty. Individual run results and uncertainty components are listed in Appendix \ref{app:NNLS_validation_table}, Table \ref{tab:nnls_runs}.

For substrates where multiple low-coverage experiments were analyzed, the standard deviation of the median binding energies among the independent runs was included as an additional run-to-run uncertainty. The uncertainty reported for the representative binding energy in Table \ref{tab:binding_energy_summary} was obtained by combining the mean per-run uncertainty and the run-to-run scatter in quadrature.

\subsubsection{Leading-edge analysis}
\label{sec:leading_edge}

For multilayer (thick-film) desorption, the kinetics follow the zeroth-order form of the Polanyi-Wigner equation, in which the desorption rate is independent of surface coverage. The desorption rate remains approximately constant until the multilayer reservoir is depleted, and the peak temperature of the TPD profile is therefore governed primarily by depletion of the ice film rather than by the intrinsic desorption energies \citep{King1975,Tait2005_I}. As a result, max temperature peak-based analyses are not applicable for determining binding energies in the multilayer limit.

Instead, the binding energy is extracted from the leading edge of the TPD curve, where depletion effects are negligible and the desorption rate reflects the intrinsic temperature dependence of the desorption process. Over this restricted interval, the zeroth-order Polanyi--Wigner equation reduces to an Arrhenius form:

\begin{equation}
\ln r_{\rm des}(T) = \ln \nu - \frac{E_{\rm des}}{T},
\end{equation}

where $r_{\rm des}$ is the calibrated desorption rate, $\nu$ is the pre-exponential factor, and $E_{\rm des}$ is the binding energy expressed in Kelvin units. In this approach, the leading edge is treated as an exponential temperature-dependent desorption rate over the restricted interval selected for fitting \citep{King1975,Redhead1962}.

The leading-edge fitting interval was selected within the rising portion of the desorption peak as the largest contiguous region that (i) exceeds five times the standard deviation of the baseline residuals, (ii) lies between fixed lower and upper fractions of the peak intensity to avoid baseline-dominated and near-peak points, and (iii) maintains a positive slope. A representative temperature, $T_{\rm rep}$, was defined as the maximum temperature within this selected interval.

For each run, the QMS signal was first truncated to the temperature range containing the MF desorption peak. A linear baseline was then fit to the low-temperature portion of the TPD profile ($T \leq 105$ K) and subtracted from the signal. The resulting desorption profile was scaled so that its integral over the desorption window matched the independently measured deposited coverage $\Theta$, yielding a calibrated desorption rate in units of ML K$^{-1}$. This rate was then converted to ML s$^{-1}$ using the experimental heating rate $\beta$ before constructing the Arrhenius plots.

To reduce covariance between $\nu$ and $E_{\rm des}$, the prefactor was fixed to the explicit TST value evaluated at $T_{\rm rep}$, $\nu=\nu_{\rm TST}(T_{\rm rep})$ (Section \ref{sec:TST}), rather than being treated as a free fit parameter \citep{Doronin2015,Luna2018}. With $\nu$ fixed, $E_{\rm des}$ was obtained by linear least-squares fitting of $\ln r_{\rm des}$ versus $1/T$ over the selected leading-edge interval. 

Uncertainties in $E_{\rm des}$ include both the statistical uncertainty from the fixed-$\nu$ Arrhenius fit and the systematic contribution from the absolute temperature accuracy ($\pm 0.8$ K). The temperature contribution was estimated by repeating the fit after shifting the selected temperatures by $\pm 0.8$ K, recomputing both $T_{\rm rep}$ and $\nu_{\rm TST}(T_{\rm rep})$ for the shifted intervals. The statistical and temperature-derived uncertainties were then combined in quadrature.

\begin{table*}[ht!]
\centering
\caption{Transition-state-theory prefactors and corresponding binding energies derived from Redhead and leading-edge analyses.}
\label{tab:nu_E_all_leadingedge}
\setlength{\tabcolsep}{4pt}
\begin{tabular*}{\textwidth}{@{\extracolsep{\fill}}lccccc@{}}
\hline
 & $\beta$ & $T_{\rm peak}$  & $\nu_{\rm TST}$ & $E_{\rm des}$ & $E_{\rm des}$ \\
 & (K min$^{-1}$) & (K) & (s$^{-1}$) & (K) & (eV) \\

\hline

\multicolumn{6}{l}{\textbf{Redhead analysis (first-order)}} \\
\hline

\multicolumn{6}{l}{\textbf{Amorphous H$_2$O}} \\
 & 2.0  & 127.3 & $4.43\times10^{18}$ & 6024 & 0.519 \\
 & 5.0  & 132.5 & $5.10\times10^{18}$ & 6176 & 0.532 \\
 & 10.0 & 131.8 & $5.00\times10^{18}$ & 6051 & 0.521 \\
Mean & \dots & \dots & $(4.84\pm0.36)\times10^{18}$ & $6084\pm81$ & $0.524\pm0.007$ \\
\hline

\multicolumn{6}{l}{\textbf{Crystalline H$_2$O}} \\
 & 2.0  & 128.3 & $4.55\times10^{18}$ & 6078 & 0.524 \\
 & 5.0  & 132.4 & $5.09\times10^{18}$ & 6172 & 0.532 \\
 & 10.0 & 134.5 & $5.36\times10^{18}$ & 6184 & 0.533 \\
Mean & \dots & \dots & $(5.00\pm0.41)\times10^{18}$ & $6145\pm59$ & $0.530\pm0.005$ \\
\hline

\multicolumn{6}{l}{\textbf{Amorphous CH$_3$OH}} \\
 & 2.0  & 115.3 & $3.14\times10^{18}$ & 5409 & 0.466 \\
 & 5.0  & 117.8 & $3.38\times10^{18}$ & 5431 & 0.468 \\
 & 10.0 & 120.5 & $3.66\times10^{18}$ & 5485 & 0.473 \\
Mean & \dots & \dots & $(3.39\pm0.26)\times10^{18}$ & $5441\pm39$ & $0.469\pm0.003$ \\
\hline

\multicolumn{6}{l}{\textbf{Leading-edge analysis
($\nu$ constrained by TST for each run)}} \\
\hline
 & Coverage (ML) & $T_{\rm rep}$ & $\nu_{\rm TST}$ & $E_{\rm des}$ & $E_{\rm des}$ \\
          &               & (K)           & (s$^{-1}$)      & (K)           & (eV) \\
\hline

\multicolumn{6}{l}{\textbf{Pure MF multilayer}} \\
 & 53  & 126.53 & $4.34\times10^{18}$ & 5520 & 0.476 \\
 & 41  & 125.87 & $4.26\times10^{18}$ & 5508 & 0.475 \\
 & 31  & 125.21 & $4.18\times10^{18}$ & 5500 & 0.474 \\
Mean & \dots & \dots & $(4.26\pm0.08)\times10^{18}$ & $5509\pm10$ & $0.475\pm0.001$ \\
\hline

\multicolumn{6}{l}{\textbf{MF on crystalline CH$_3$OH}} \\
 & 0.25 & 109.33 & $2.60\times10^{18}$ & 5335 & 0.460 \\
 & 0.36 & 109.80 & $2.64\times10^{18}$ & 5324 & 0.459 \\
 & 0.55 & 111.09 & $2.75\times10^{18}$ & 5331 & 0.459 \\
 & 0.80 & 111.39 & $2.77\times10^{18}$ & 5316 & 0.458 \\
 & 1.30 & 112.62 & $2.88\times10^{18}$ & 5331 & 0.459 \\
 & 2.60 & 114.55 & $3.06\times10^{18}$ & 5362 & 0.462 \\
Mean & \dots & \dots & $(2.78\pm0.17)\times10^{18}$ & $5333\pm17$ & $0.459\pm0.001$ \\
\hline
\end{tabular*}

\vspace{2pt}
\parbox{0.95\textwidth}{\footnotesize
\textbf{Note.}
For amorphous and crystalline H$_2$O and amorphous CH$_3$OH, binding energies were derived using Redhead analysis assuming first-order desorption. For pure MF multilayers and MF on crystalline CH$_3$OH, binding energies were obtained from fixed-$\nu$ leading-edge Arrhenius fits. The prefactor $\nu_{\rm TST}$ was evaluated at the relevant temperature ($T_{\rm peak}$ for Redhead analysis and $T_{\rm rep}$ for leading-edge fits). The uncertainties in the mean rows represent the standard deviation across experiments.}
\end{table*}

For pure MF multilayers and MF on crystalline CH$_3$OH,
binding energies were obtained from leading-edge Arrhenius
fits in which $\nu_{\rm TST}$ was calculated at
$T_{\rm rep}$ and held fixed within each individual fit.
Because $T_{\rm rep}$ differs among experiments, the
corresponding $\nu_{\rm TST}$ values also vary between runs.

\begin{table*}[t]
\centering
\caption{Representative binding energy distributions of MF.}
\label{tab:binding_energy_summary}
\begin{tabular}{lcccc}
\hline
Substrate & Coverage used (ML) & Method & $E_{\rm des}$ (K) & $E_{\rm des}$ (eV) \\
\hline

Amorphous \methanol\
& 0.25, 0.45, 0.60, 1.00
& Distribution inversion
& $5469 \pm 40$
& $0.471 \pm 0.003$ \\

Amorphous \water\ 
& 0.20, 0.30, 0.55
& Distribution inversion
& $6247 \pm 43$
& $0.538 \pm 0.004$ \\

Crystalline \water\
& 0.20
& Distribution inversion
& $6213 \pm 38$
& $0.535 \pm 0.003$ \\

\hline
\end{tabular}

\vspace{0.2cm}

\parbox{0.95\textwidth}{\footnotesize
\textbf{Note.}
Energies are reported in Kelvin (K) and electron-volts (eV) using $E_{\rm des}(\mathrm{eV}) = k_B E_{\rm des}(\mathrm{K})$, with $k_B = 8.61733\times10^{-5}\ \mathrm{eV\,K^{-1}}$. For distribution inversion, $E_{\rm des}$ corresponds to the median of the recovered binding-energy distribution. 
For multi-run substrates, the representative value is the mean of the selected run medians, and the quoted uncertainty is the quadrature sum of the mean per-run uncertainty and the run-to-run scatter. 
Per-run uncertainties include Monte Carlo noise propagation and the $\pm0.8$~K temperature systematic. 
For crystalline \water, only the single-run uncertainty is quoted.}
\end{table*}

\section{Results}
\label{results}

\subsection{Binding Energies and Prefactors from Redhead, Distribution inversion and Leading edge Analysis}

Table \ref{tab:nu_E_all_leadingedge} summarizes the heating-rate-dependent MF peak temperature measurements, the calculated TST pre-exponential factors $\nu_{\rm TST}(T_{\rm peak})$, and the Redhead-derived binding energies for substrates displaying first-order desorption behavior; these substrates include: amorphous and crystalline H$_2$O, and amorphous CH$_3$OH. MF deposited on crystalline CH$_3$OH exhibits coverage-independent leading edges characteristic of island formation (Section \ref{sec:crysMeOH}), thus binding energies were derived via leading-edge analysis rather than the Redhead/NNLS approach.

Binding energy distributions obtained via NNLS inversion for the first-order experiments are displayed in Figure \ref{fig:binding_energy_distributions}. Table \ref{tab:binding_energy_summary} lists the median values of these distributions, alongside the single-energy values derived from the Redhead method \citep{Redhead1962}. For MF on amorphous CH$_3$OH, the Redhead single-energy estimates are consistent with the median values from the NNLS inversion. For H$_2$O substrates, however, the inversion analysis reveals broader, asymmetric distributions with extended low- and high-energy tails, such that the median is offset from the single-energy Redhead estimate.

The recovered binding-energy distributions for the water substrates evolve systematically with increasing MF coverage (Figure \ref{fig:binding_energy_distributions}). As coverage increases, these distributions broaden, indicating a transition in the dominant binding environment from exclusively MF–substrate interactions at sub-monolayer coverage to a regime where MF–MF surface intermolecular interactions begin to contribute significantly to the total desorption profile.

\subsection{Multilayer Pure MF Desorption}

TPD spectra for thick MF films without a water or methanol substrate are shown in Figure \ref{fig:leading_edge_comparison} (top panel). 
Arrhenius plots constructed from the objectively selected leading-edge intervals display linear behavior for pure MF multilayers across coverages of 31-53 ML (Appendix \ref{appendix:arrhenius}). Using the prefactor fixed to $\nu_{\rm TST}(T_{\rm rep})$, the derived binding energies are consistent across all three films, indicating that the characteristic multilayer binding energy is independent of thickness, as expected.

Because these three experiments yield statistically indistinguishable results, we adopt their weighted mean value of $E_{\rm des}=5509 \pm 10$ K as the representative MF–MF binding energy used in subsequent comparisons and reported in Table \ref{tab:nu_E_all_leadingedge}.

\subsection{Low Coverage MF Desorption}

\begin{figure*}
\centering

\includegraphics[width=0.48\textwidth]{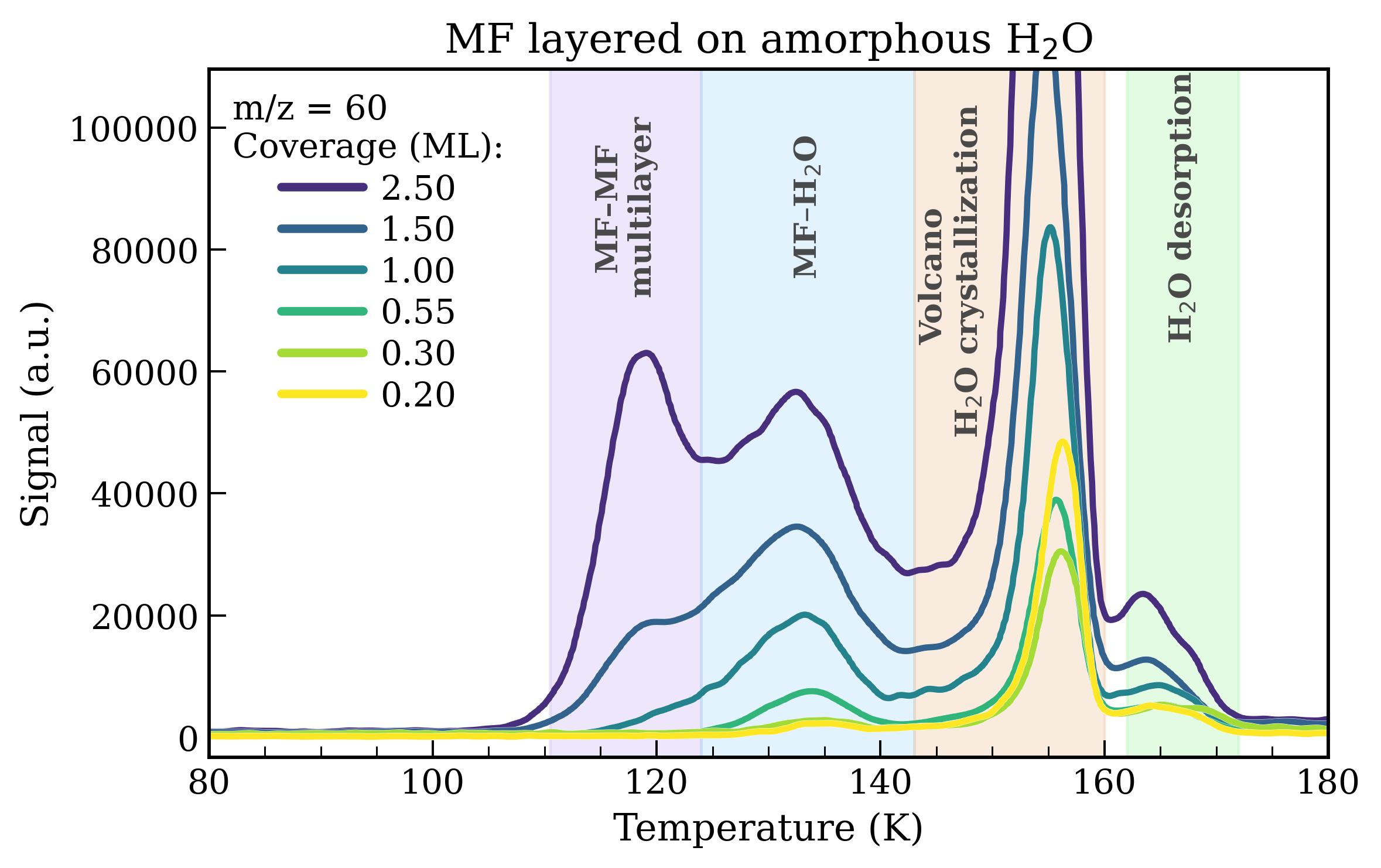}
\hfill
\includegraphics[width=0.48\textwidth]{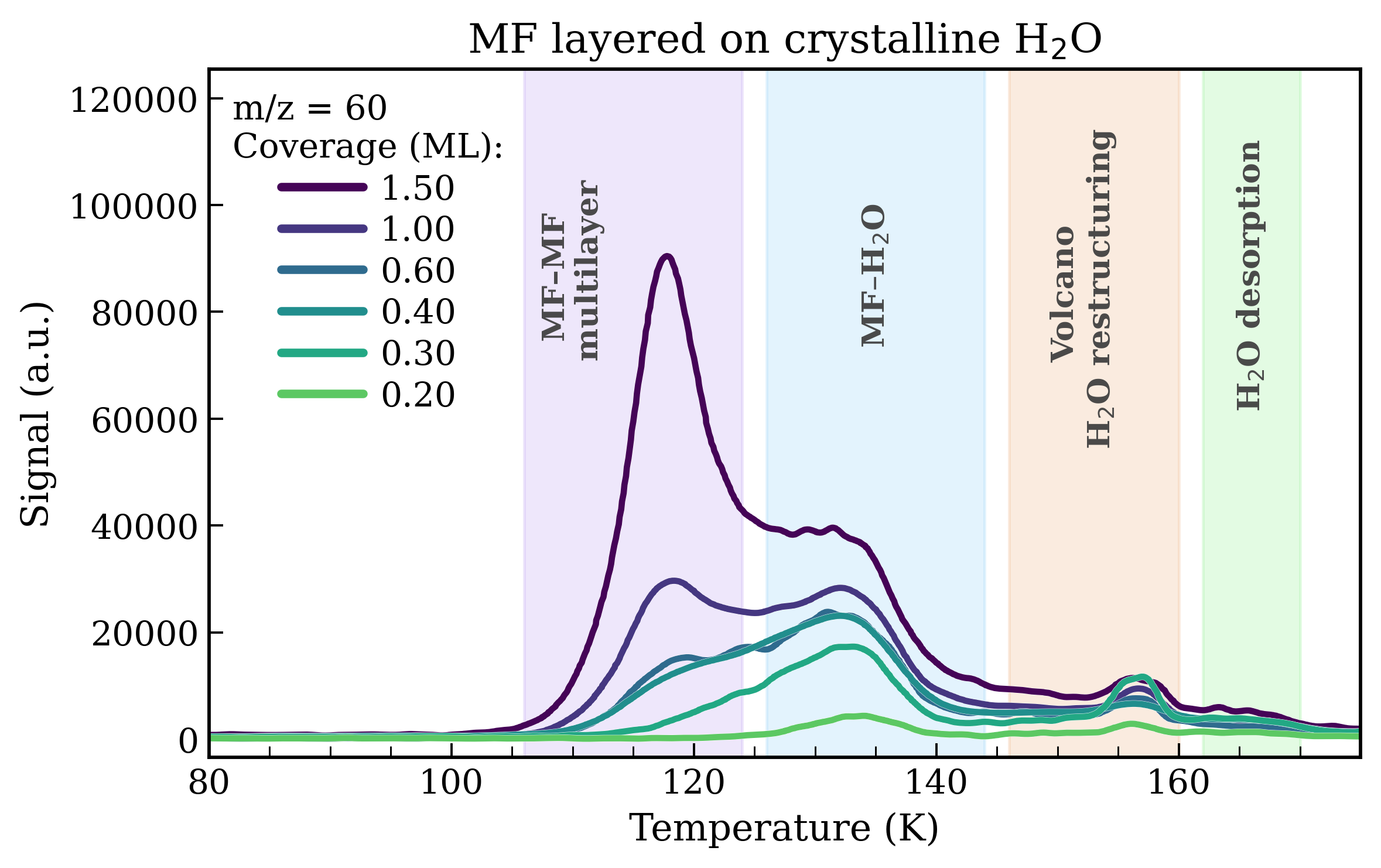}

\vspace{0.4cm}

\includegraphics[width=0.48\textwidth]{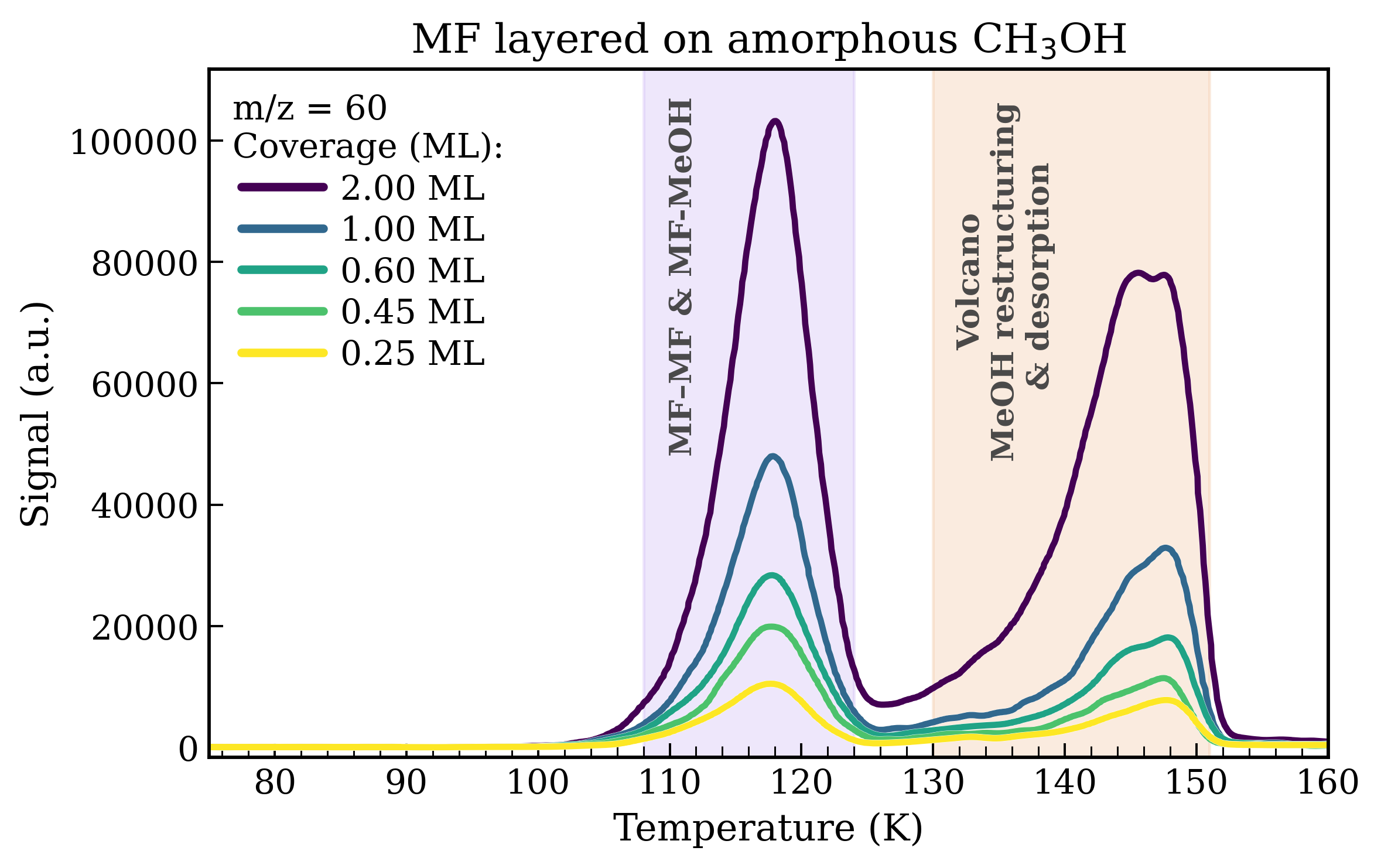}
\hfill
\includegraphics[width=0.48\textwidth]{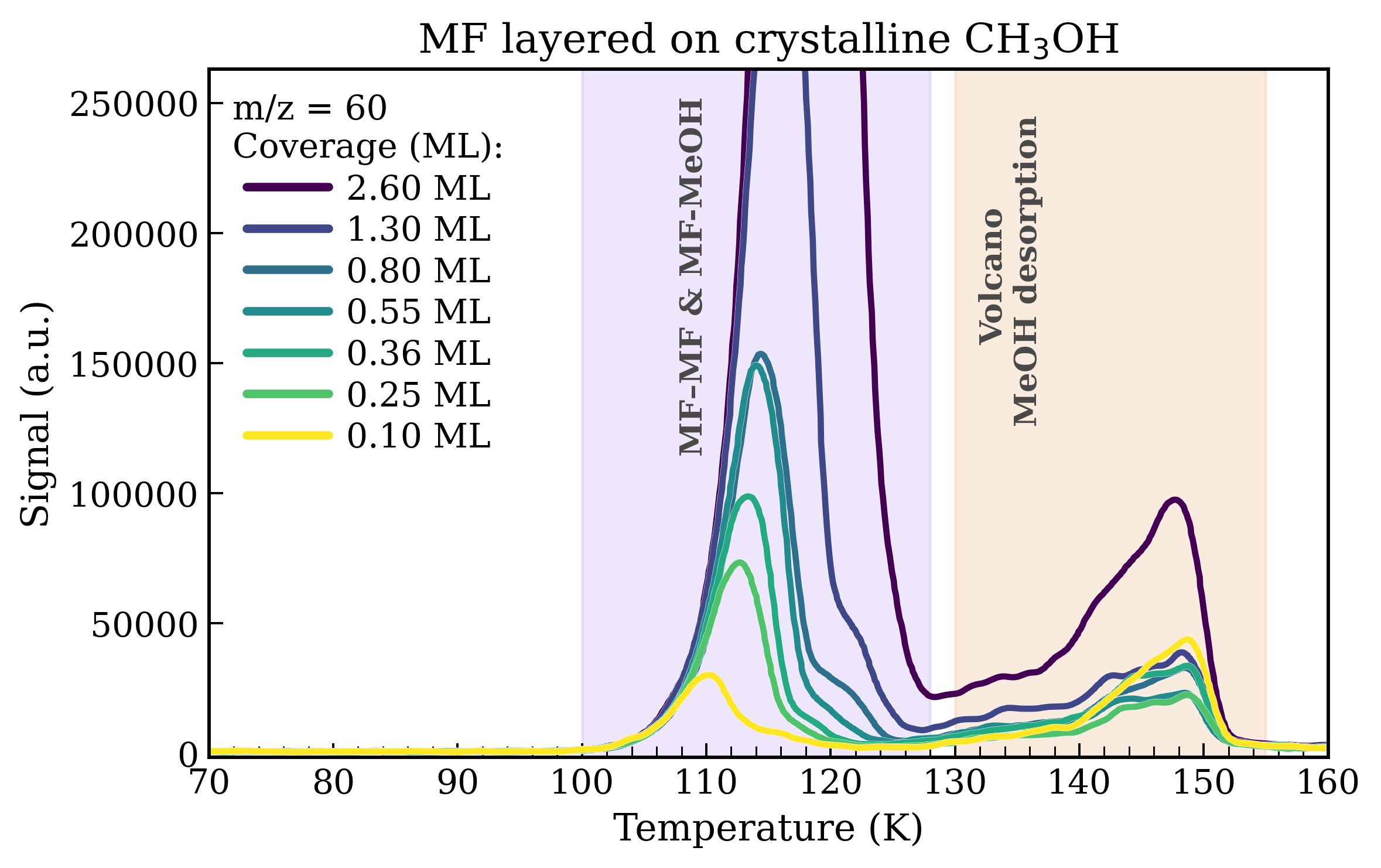}

\caption{
Temperature-programmed desorption (TPD) spectra of methyl formate (MF; $m/z = 60$) for varying initial coverages on amorphous and crystalline H$_2$O (top panels) and CH$_3$OH (bottom panels) substrates. Colored curves indicate different MF coverages in monolayers (ML), as labeled in each panel. Shaded regions mark distinct desorption regimes, including MF multilayer desorption, MF–ice interactions, and higher-temperature desorption associated with substrate restructuring and co-desorption (e.g., volcano desorption and bulk ice release). Differences in peak structure and temperature reflect variations in binding sites, trapping efficiency, and ice morphology between substrates.
}
\label{fig:MF_all_substrates}

\end{figure*}

\subsubsection{MF Desorption from Amorphous \water}

TPD spectra of MF deposited on compact ASW at coverages between 0.30 and 2.50 ML are shown in Figure \ref{fig:MF_all_substrates} (top left). At higher coverages ($\ge 1.0$ ML), a low-temperature desorption peak appears between 115 and 125~K corresponding to multilayer MF–MF desorption \citep{Collings2004,Burke2015}.

At sub-monolayer coverages (0.20–0.55 ML), multilayer desorption is absent and a higher-temperature peak emerges between 130 and 145~K, increasing with coverage. This component corresponds to first-order desorption of MF from ASW. The peak position remains approximately constant with decreasing coverage, confirming desorption directly from ASW substrate binding sites rather than from MF–MF interactions associated with multilayer or clustered MF. This behavior is consistent with previous studies of MF thermal desorption from amorphous water ice \citep{Burke2015,Ligterink2023}.

A second desorption peak appears between 145 and 165~K. This feature is caused by emission via the "volcano effect" that occurs when amorphous water ice crystallizes during warm-up, releasing MF molecules that were trapped within the ASW matrix and would create discontinuities within the \water\ crystal structure \citep{Smith1997,Fraser2001,Collings2004,Burke2015}. The presence of this peak even at low MF coverages indicates that a fraction of the surface molecules diffuse into the ice, becoming entrapped within the porous ASW structure.

To determine the MF–H$_2$O interaction energy, we restrict the binding-energy analysis to coverages where first-order MF–substrate desorption dominates. At higher coverages the TPD profiles develop shoulders and broadened features indicative of mixed MF–MF and MF–H$_2$O contributions. The analysis therefore focuses on the sub-monolayer coverages of 0.20, 0.30, and 0.55 ML. NNLS inversions for higher coverages are shown for completeness in Appendix \ref{app:distribution_inversion} but are not used to define the MF–H$_2$O binding energy.

Distribution-inversion analysis of these three coverages yields a representative MF–H$_2$O binding energy of $E_{\rm des}=6247$~K (0.538 eV; Table \ref{tab:binding_energy_summary}). The reported value corresponds to the arithmetic mean of the median binding energies recovered from the individual NNLS distributions.

\begin{figure*}[t]
\centering
\includegraphics[width=0.85\textwidth]{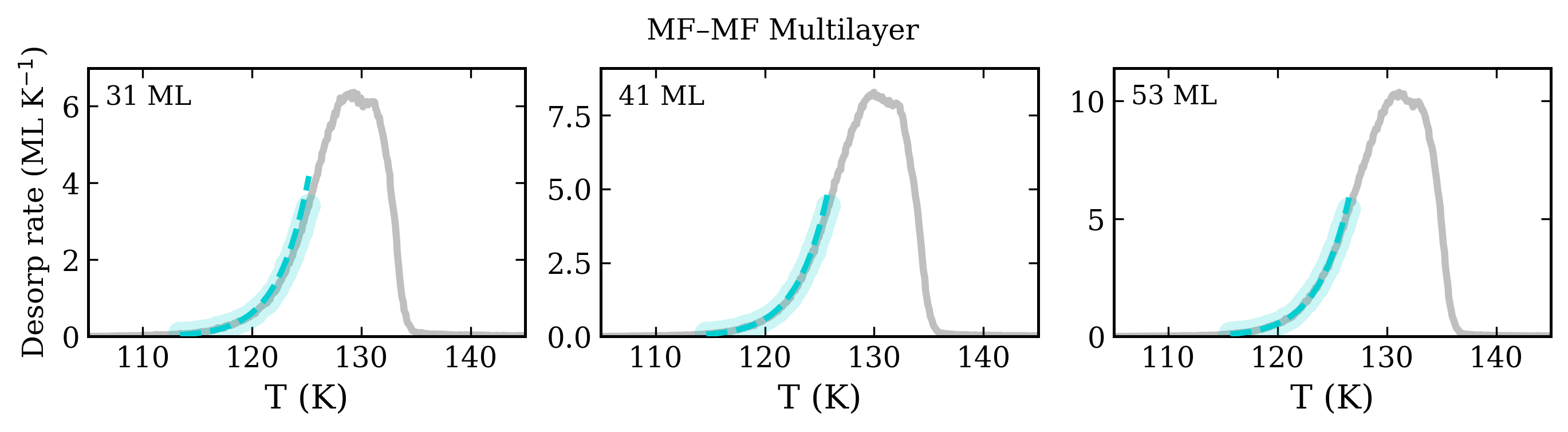}
\vspace{0.5cm}
\includegraphics[width=0.85\textwidth]{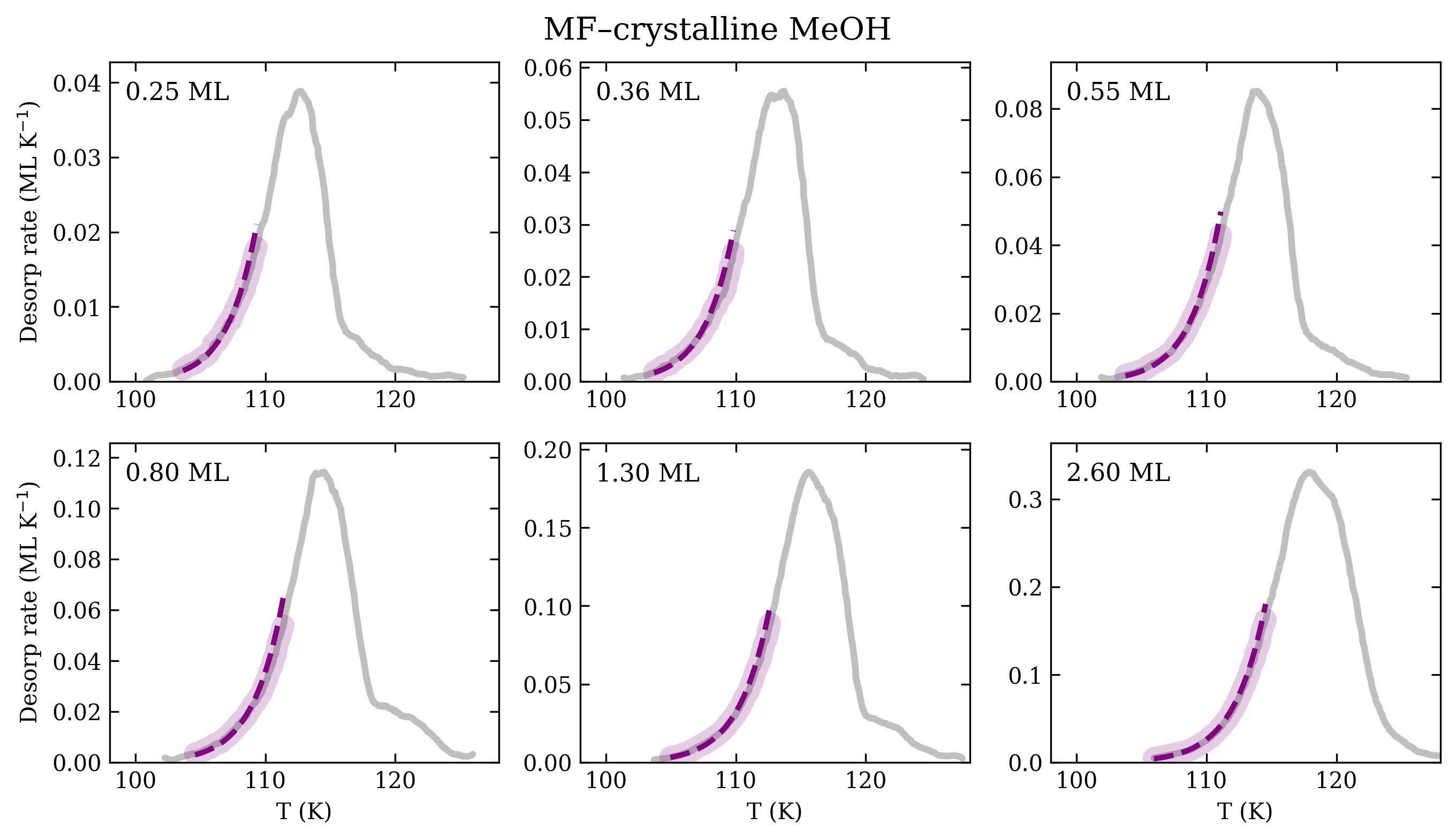}
\caption{
Leading-edge analysis of MF desorption.
\textit{Top:} Pure MF multilayer desorption.
\textit{Bottom:} MF deposited on crystalline \methanol\ (labeled MeOH in the panels).
Gray curves show the calibrated desorption rate, the highlighted regions indicate the leading-edge intervals used for Arrhenius fitting, and the dashed curves show the Arrhenius model using the derived binding energy with the prefactor fixed to $\nu(T_{rep})$. The coverage-independent leading edge observed for MF on crystalline methanol indicates multilayer-like desorption behavior.}
\label{fig:leading_edge_comparison}
\end{figure*}

\begin{figure}[t]
\centering

\includegraphics[width=\columnwidth]{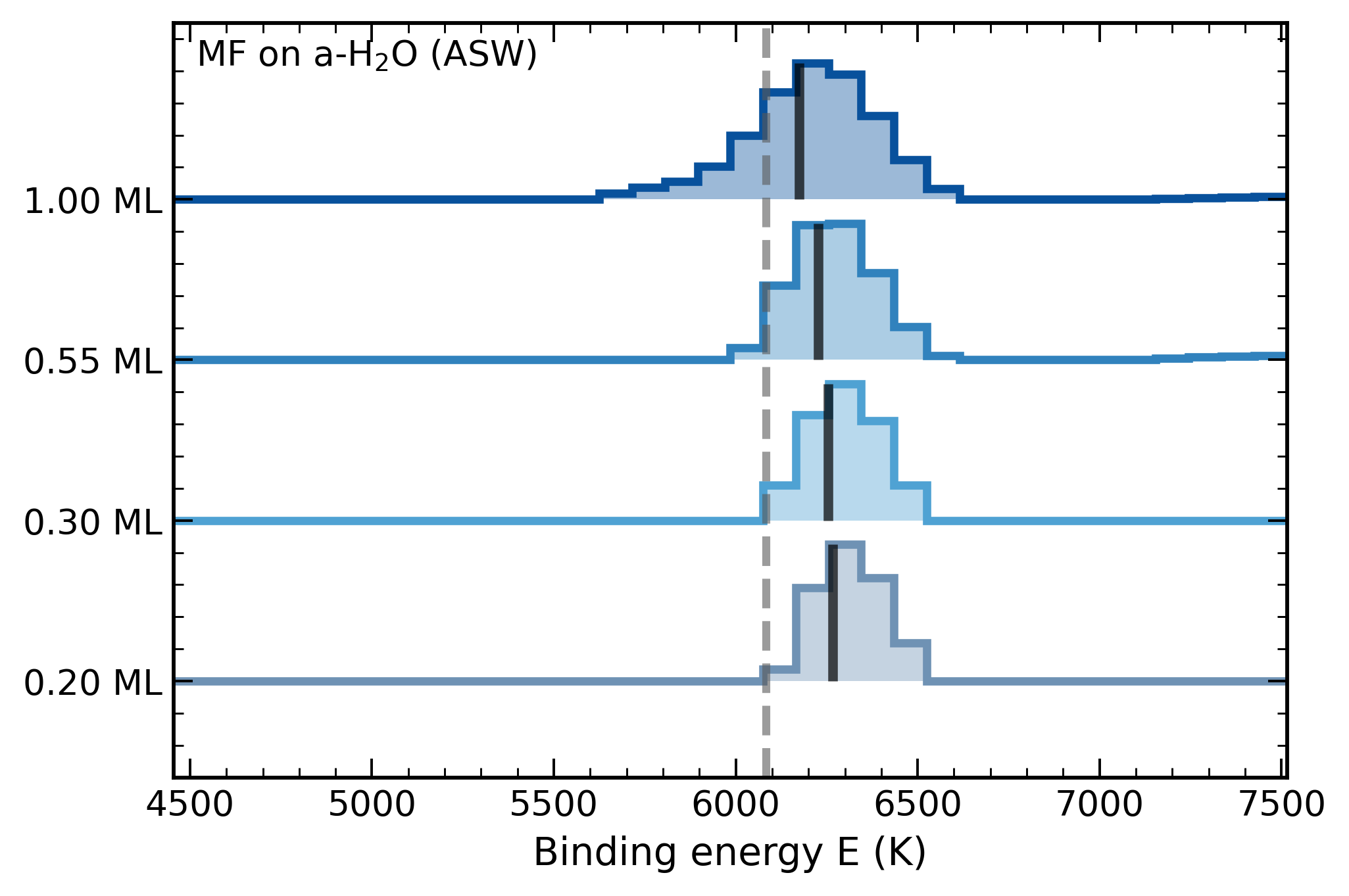}

\vspace{0.25cm}

\includegraphics[width=\columnwidth]{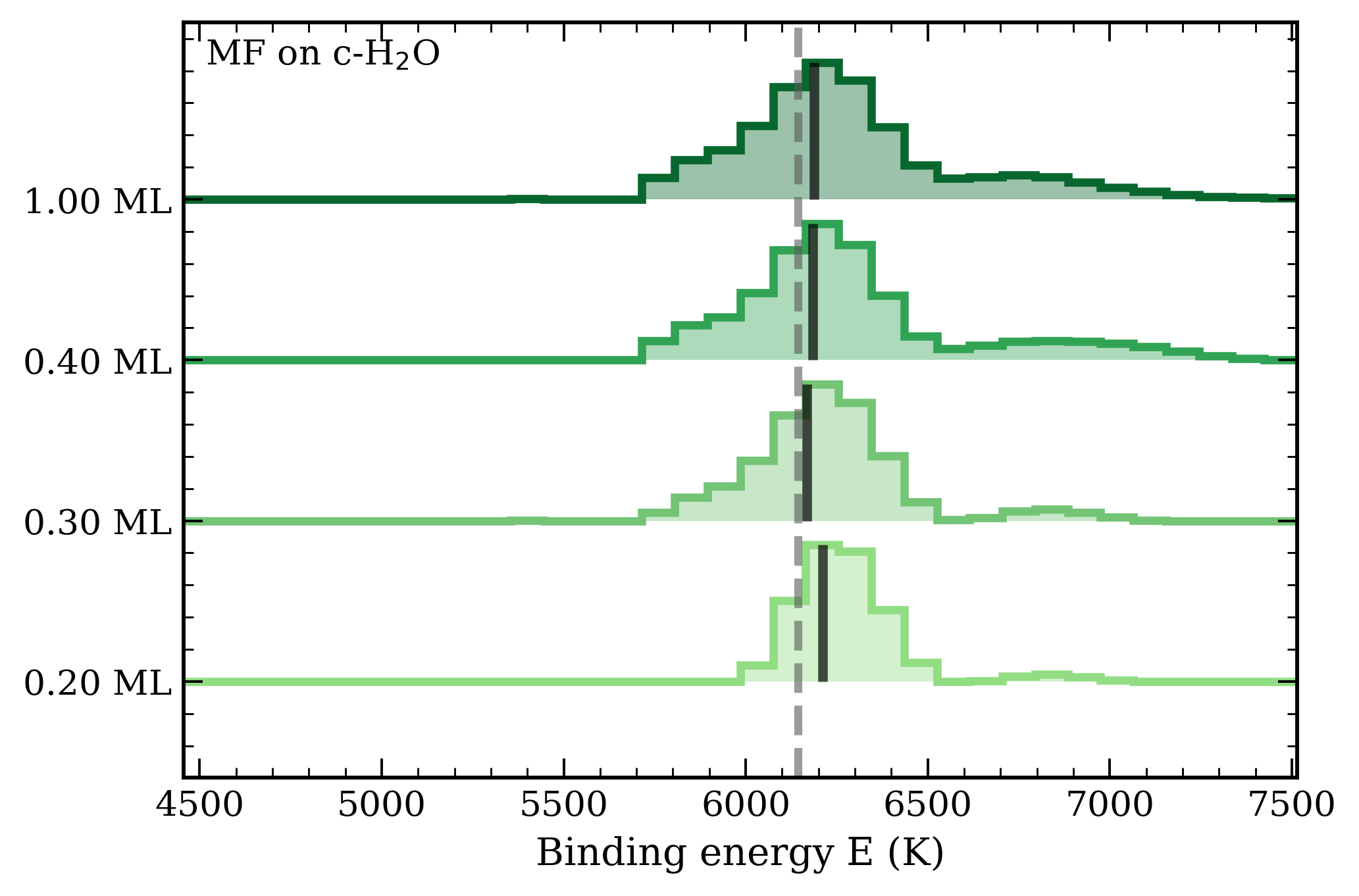}

\vspace{0.25cm}

\includegraphics[width=\columnwidth]{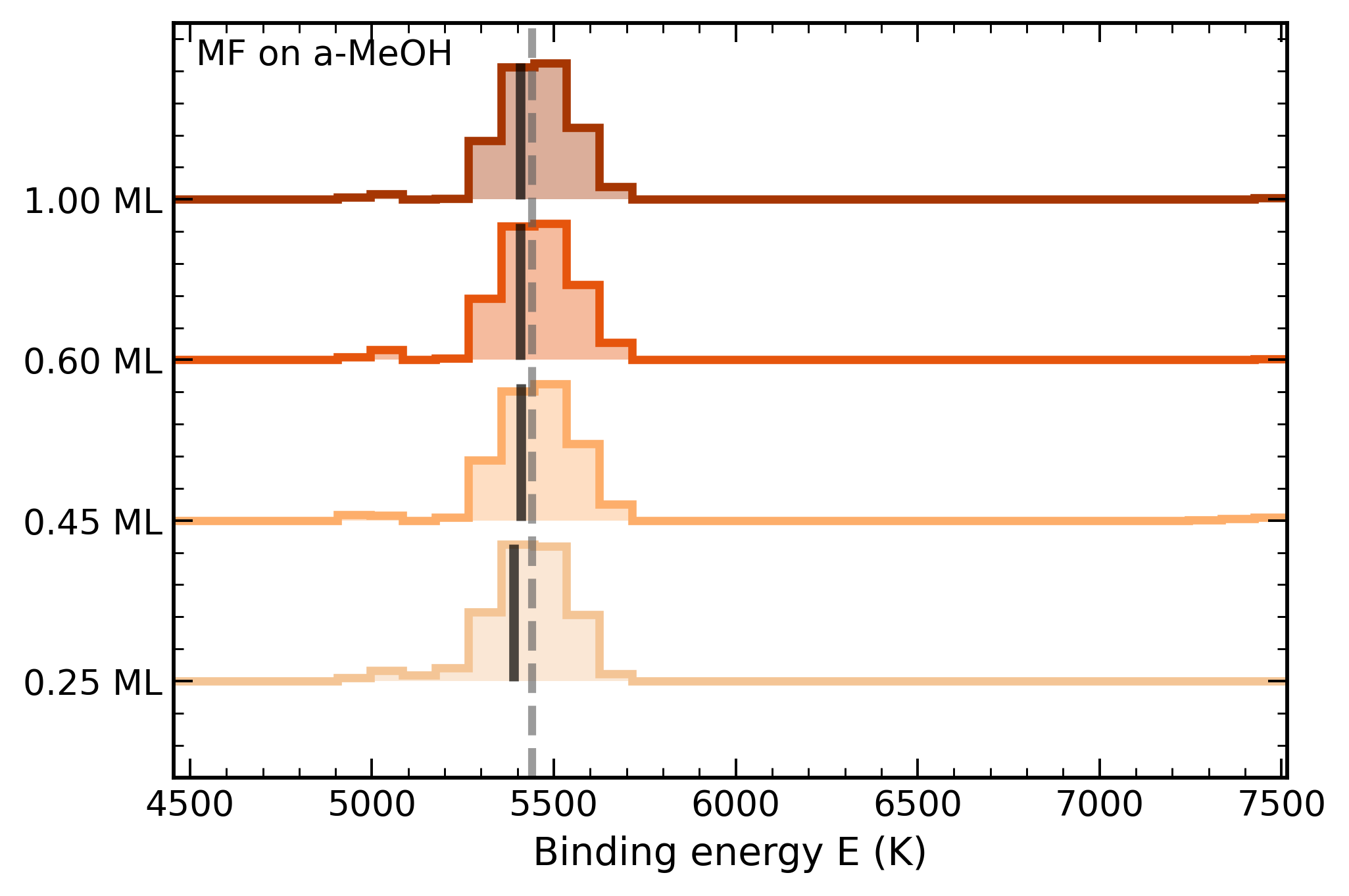}

\caption{
Binding-energy distributions of MF derived from NNLS inversion of the TPD spectra. Top: MF on amorphous H$_2$O (ASW). Middle: MF on crystalline H$_2$O. Bottom: MF on amorphous CH$_3$OH. Colored stepped profiles show the recovered NNLS weights as a function of binding energy for the MF coverages labeled at left. Black vertical lines mark the median binding energy of each distribution (defined as the energy at which 50\% of the cumulative desorption contribution is reached), while dashed gray vertical lines show the corresponding Redhead binding-energy estimate for MF on each substrate.
}

\label{fig:binding_energy_distributions}

\end{figure}

The binding-energy distributions for the three sub-monolayer coverages are tightly clustered and exhibit similar median energies, indicating that the MF–H$_2$O interaction energy is largely independent of coverage in this regime. In contrast, the distribution recovered for the 1.00~ML spectrum becomes noticeably broader (Figure \ref{fig:binding_energy_distributions}), reflecting the increasing contribution of MF–MF interactions as the coverage approaches a full monolayer.

\subsubsection{MF Desorption from Crystalline \water}

TPD spectra of MF deposited on crystalline \water\ ice are shown in Figure \ref{fig:MF_all_substrates}. In contrast to amorphous water ice, the volcano desorption component is significantly reduced, reflecting the lower porosity, diminished in-diffusion and reduced trapping efficiency of crystalline ice \citep{Collings2004,Burke2015}. As a result, MF desorption from crystalline water occurs primarily through direct desorption from surface binding sites.

However, surface intermolecular MF–MF interactions appear at relatively low coverages. As shown in the top right panel of Figure \ref{fig:MF_all_substrates}, the 0.30 ML spectrum already develops a shoulder near $\sim120$~K corresponding to multilayer MF–MF desorption. This feature indicates that even modest increases in coverage introduce mixed MF–MF and MF–H$_2$O contributions to the desorption profiles.

Because of this early onset of MF-MF intermolecular interactions, isolating the MF–H$_2$O desorption component becomes difficult for coverages above $\sim0.2$ ML. The NNLS distribution inversions reflect this behavior: while the 0.20 ML spectrum produces a relatively well-defined binding-energy peak, the recovered distributions become progressively broader for higher coverages, indicating mixed MF–MF and MF–H$_2$O adsorption environments. This broadening is clearly visible in the binding-energy distributions shown in Figure \ref{fig:binding_energy_distributions}, where the distributions for coverages $\ge$0.20 ML are noticeably wider and less sharply peaked. Representative NNLS reconstructions for all crystalline-water coverages are shown in Appendix \ref{app:distribution_inversion}.

For this reason, the MF–H$_2$O binding energy on crystalline water is determined using only the 0.20 ML spectrum. The desorption peak between $\sim125$ and 140~K was used in the inversion analysis, yielding a binding energy of $E_{\rm des}=6213\pm38$~K (Table \ref{tab:binding_energy_summary}).

\subsubsection{MF Desorption from Amorphous \methanol}

TPD spectra of MF deposited on amorphous \methanol\ ice are shown in Figure \ref{fig:MF_all_substrates} (bottom left) as a function of coverage. Across all investigated coverages (0.25–2.0 ML), the desorption profiles are dominated by a single peak between $\sim105$ and 125~K. Unlike the \water\ substrates discussed above, the TPD spectra do not exhibit a clear transition between MF–MF multilayer desorption and MF–substrate desorption regimes.

Because MF–MF and MF–\methanol\ desorption occur within the same temperature range, their individual contributions cannot be uniquely separated in the TPD profiles. We therefore interpret this peak as representing the MF desorption from amorphous \methanol\ ice.

A second feature appears near $\sim146$~K and coincides with crystallization and desorption of the underlying \methanol\ ice. This peak corresponds to a volcano-type release of molecules that diffused and were trapped within the \methanol\ matrix during restructuring of the ice.

Binding-energy distributions derived from NNLS inversion of the TPD spectra are shown in Figure \ref{fig:binding_energy_distributions}. The recovered distributions are similar across all investigated coverages, indicating that the effective interaction energy is largely independent of coverage in this regime. The representative binding energy derived from these distributions is $E_{\rm des}=5469\pm40$~K (Table \ref{tab:binding_energy_summary}); the corresponding NNLS reconstructions are shown in Appendix \ref{app:distribution_inversion}.

The binding energy derived for MF on amorphous \methanol\ ($E_{\rm des}=5469\pm40$~K) is very similar to the MF multilayer binding energy ($E_{\rm des}\sim5509$~K), reflective of the analogous molecular structures of the two species. This similarity indicates that MF–MF and MF–\methanol\ interactions have comparable strengths, which explains why a distinct transition between multilayer and monolayer desorption is not observed in the TPD profiles. To our knowledge, this work provides the first experimental measurement of MF desorption from \methanol\ ice substrates.

\subsubsection{MF Desorption from Crystalline \methanol} \label{sec:crysMeOH}

TPD spectra of MF deposited on crystalline CH$_3$OH are shown as a function of coverage in Figure \ref{fig:MF_all_substrates}. Compared to the amorphous substrates, the MF TPD spectra exhibit reduced evidence for trapping and a visibly weaker volcano desorption feature, consistent with previous studies showing that crystalline ice structures provide fewer trapping sites than amorphous matrices \citep{Burke2015,MartinDomenech2014}.

Across the investigated coverages (0.10–2.6 ML), the first desorption peak displays a common leading edge near $\sim105$~K (Figure \ref{fig:MF_all_substrates}, bottom right). The overlap of the leading edges with increasing peak energy indicates coverage-independent desorption behavior characteristic of multilayer, or zeroth-order, kinetics \citep{Collings2015}. This behavior suggests that MF–MF intermolecular interactions dominate the desorption process even at sub-monolayer coverages, and that the desorption rate is controlled by the same kinetic process across the entire coverage range.

Because the desorption profiles do not exhibit a clear transition between multilayer and monolayer regimes, the MF–substrate interaction cannot be isolated using the distribution-inversion approach applied to the water substrates. Instead, binding energies were derived using leading-edge analysis of the desorption peak between 100 and 128~K (Figure \ref{fig:leading_edge_comparison}, bottom panels).

The resulting leading-edge analysis yields an effective binding energy of 
$E_{\rm des}=5333\pm17$~K (Table \ref{tab:nu_E_all_leadingedge}). 
The corresponding Polanyi-Wigner coverage-independent Arrhenius plots are shown in Appendix \ref{appendix:arrhenius}. 
The coverage-independent leading edge and the similarity of this value to the MF multilayer binding energy indicate that desorption from crystalline \methanol\ is dominated by MF–MF interactions, consistent with MF aggregate formation on the crystalline \methanol\ surface.

\subsection{Mixed MF Ices}

\begin{figure*}[t]
\centering
\begin{minipage}{0.48\textwidth}
\centering
\includegraphics[width=\textwidth]{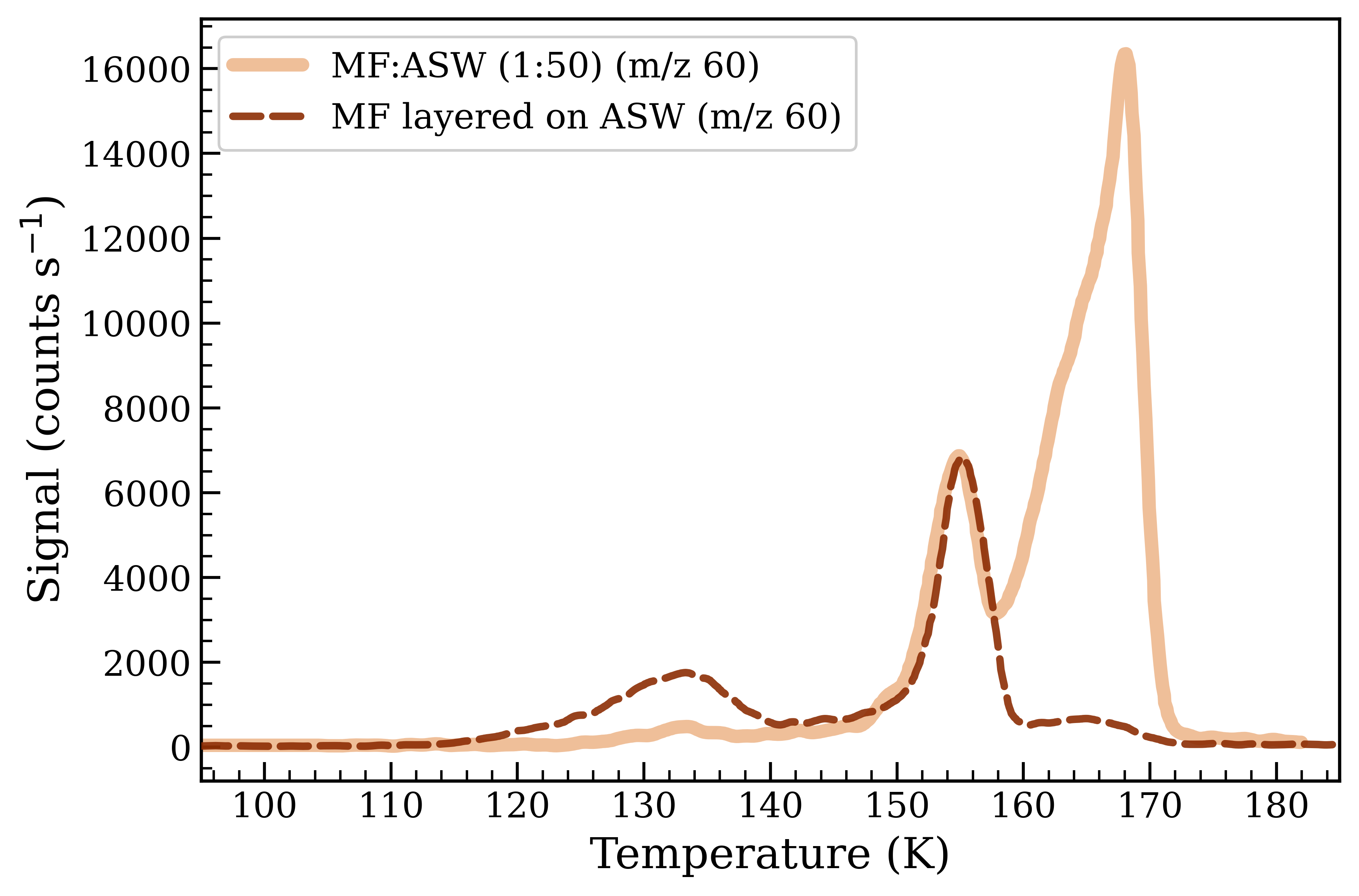}
\vspace{0.2cm}
\textbf{(a) MF:H$_2$O}
\end{minipage}
\hfill
\begin{minipage}{0.48\textwidth}
\centering
\includegraphics[width=\textwidth]{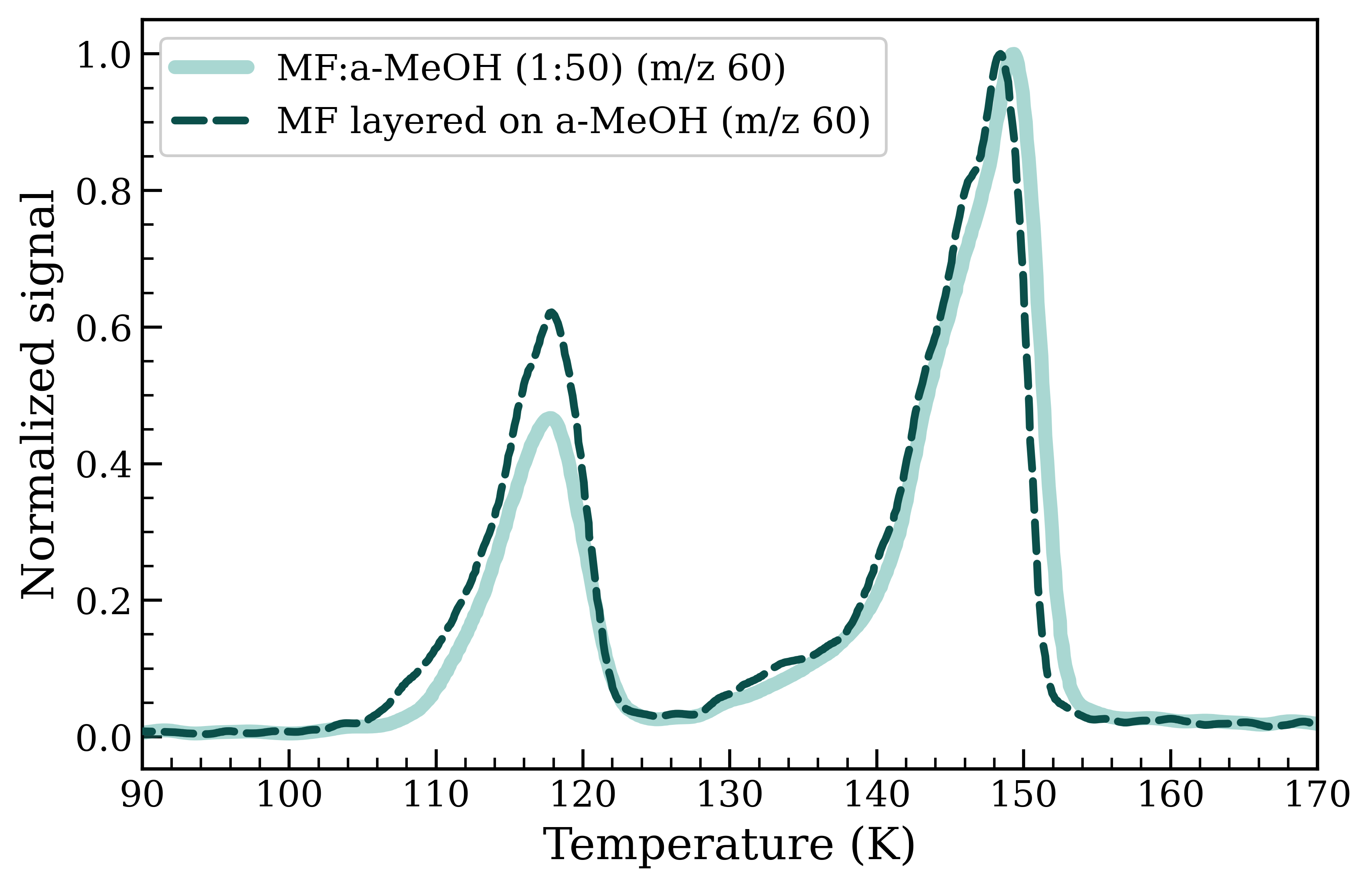}
\vspace{0.2cm}
\textbf{(b) MF:CH$_3$OH}
\end{minipage}
\caption{
Normalized TPD profiles of MF in mixed and layered ice systems.
(a) MF co-deposited with \water\ compared with 1.0 ML MF deposited as a layer on top of 50 ML of ASW.
(b) MF co-deposited with amorphous CH$_3$OH compared with 1.0 ML of MF deposited as a layer on top of 50 ML of amorphous CH$_3$OH.
The H$_2$O-containing ices show a strong dependence on deposition geometry, with enhanced retention of MF to the \water-desorption temperature in the mixed ice.
In contrast, the mixed and layered CH$_3$OH profiles are nearly identical and exhibit both a low-temperature, multilayer-like MF feature and a higher-temperature feature associated with \methanol\ crystallization.
}
\label{fig:mixing_combined}
\end{figure*}

Figure \ref{fig:mixing_combined}a compares the TPD profiles of MF in mixed and layered ASW ice systems. When MF is co-deposited with \water, the low-temperature MF desorption feature is strongly suppressed, and the dominant release occurs at $170\pm5$~K, coincident with bulk water sublimation. In the layered geometry, where MF is deposited on top of a pre-formed ASW ice, a pronounced desorption feature is instead observed near $155$~K, with only a smaller fraction of the MF retained until water sublimation. This intermediate-temperature feature is associated with water-ice restructuring and the release of trapped MF. Overall, a substantially larger fraction of MF is retained to high temperatures in the mixed \water\ ice than in either of the \methanol-containing systems (see Figure \ref{fig:mf_trapping}).

Figure \ref{fig:mixing_combined}b compares MF deposited with amorphous CH$_3$OH in two scenarios: (i) a single monolayer of MF deposited on top of a pre-deposited \methanol\ ice and (ii) MF co-deposited with \methanol. The two TPD profiles are nearly identical, indicating that the deposition geometry has little effect on the resulting MF desorption behavior. Both exhibit a low-temperature feature centered at $115\pm5$~K and a more intense feature centered at $148\pm3$~K. The low‑temperature peak lies at the same temperature as MF multilayer desorption, whereas the high‑temperature peak coincides with the crystallization of methanol (see \citet{Collings2004}). For comparison, normalized TPD profiles of pure MF, CH$_3$OH, and H$_2$O ices are provided in Appendix \ref{app:reference_tpd} (Figure \ref{reference_tpd}).

\section{Discussion}
\label{discussion}

\subsection{Binding Energies of MF and Comparison with Literature}
\label{binding_energy_compare}

Our TPD experiments establish a clear hierarchy of MF binding strengths across astrophysically relevant ice analogues. From the NNLS inversion analysis we derive median binding energies of $6247 \pm 43$~K for MF on amorphous \water\ ice (ASW) and $6213 \pm 38$~K on crystalline \water. For pure MF multilayers the analysis yields a lower binding energy of $5509 \pm 10$~K.

These values are higher than several earlier experimental and theoretical estimates. For example, \citet{Bertin2010_MF} reported binding energies of 4500–4620~K for MF on \water\ ice, while \citet{Burke2015} derived 4210~K for pure MF. Similarly, \citet{Lattelais2011} reported adsorption energies of 9.2~kcal~mol$^{-1}$ from theoretical calculations and $8.9 \pm 1.0$~kcal~mol$^{-1}$ from TPD experiments, corresponding to $4630$~K and $4480 \pm 500$~K, respectively.

The discrepancy arises from the treatment of the pre-exponential factor $\nu$. Because $\nu$ and $E_{\rm des}$ are strongly covariant in the Polanyi–Wigner equation, adopting fixed, lower canonical pre-factors of $\nu \sim (10^{12}$–$10^{13}$~s$^{-1}$) systematically biases the derived binding energy toward lower values \citep{Ligterink2023}. In this work we adopt the transition state theory (TST) framework to explicitly calculate a physically motivated prefactor of $\nu$ on the order of 10$^{18}$~s$^{-1}$. Using this prefactor naturally yields larger binding energies for the same desorption temperature. Consistent with this interpretation, census reviews by \citet{Minissale2022} find higher binding energies for COMs on \water\ ice when using TST-based prefactors. Modeling by \citet{Ligterink2023} predicts a binding energy of $E_{\rm des}\approx6.3\times10^{3}$~K for MF on \water\ ice, in close agreement with our higher $E_{\rm des}$ measurements. Similar first-principle parametric studies for benzene desorbing from graphene were measured by \citet{smith_kay2018}. This team invoked a series of partial and full “inversion” methodologies where solutions to the Polanyi-Wigner equation were used to calculate synthetic TPD spectra over multiple coverages. These computationally-derived data sets were subsequently compared with measured desorption data to infer the optimal prefactor, a  $\nu$ of 10$^{17}$ \citep{smith_kay2018}. Due to the enhanced number of molecular degrees of freedom in large organic molecules (e.g., benzene), the entropy of the desorbing (or adsorbing) species is significantly greater, a result reflected in a larger prefactor \citep{Fichthorn2002,Tait2005_II}.

The benefit of the NNLS inversion over traditional single-energy models for heterogeneous ices is the recovery of a distribution of binding-energies for MF that are reflective of the breadth of available surface sites (Figure \ref{fig:binding_energy_distributions}). In each panel the dashed gray line shows the Redhead estimate, which is obtained from the temperature of the TPD peak maximum ($T_{\rm peak}$; Table \ref{tab:nu_E_all_leadingedge}), while the colored distributions represent the NNLS-derived weights as a function of binding energy, where the height reflects the contribution of each energy component to the overall desorption profile. The solid black line marks the median of the NNLS-derived distribution, defined as the binding energy at which 50\% of the cumulative desorption contribution is reached, and therefore does not necessarily coincide with the peak of the distribution. For MF on amorphous \methanol\ the NNLS median is slightly higher than the Redhead value. The Redhead approach uses only the peak temperature, whereas the NNLS inversion reproduces the full TPD curve.

For the water substrates the Redhead method also fails to capture the high‑energy tails that appear in the NNLS $E_{\rm des}$ distribution; these tails reflect the intrinsic heterogeneity of adsorption sites on the ice, where both shallow and deep binding sites coexist \citep[e.g.,][]{Behmard2019}. The forward‑model reconstructions of the TPD curves, which are plotted over the experimental data in Appendix \ref{app:distribution_inversion} Figure \ref{fig:nnls_validation_all}, demonstrate that the NNLS‑derived distributions faithfully reproduce the observed desorption temperature profiles. This validation confirms that the median values shown in Figure \ref{fig:binding_energy_distributions} are not artifacts of the inversion but genuine descriptors of the underlying energy landscape.

This physical heterogeneity is also evidenced by the systematic broadening of the distributions with coverage. At low sub-monolayer coverages, MF molecules preferentially occupy the most energetically favorable substrate sites. As coverage increases toward a full monolayer, the distribution widens as MF–MF surface interactions begin to compete with MF–substrate binding \citep[e.g.,][]{Fayolle2016}. This transition from isolated molecules to an interacting surface film naturally broadens the energetic landscape, a feature that single-value models like the Redhead method cannot characterize \citep{Santos2025}. The resulting high-energy tail in the ASW distribution is particularly significant for astrochemistry, as it represents a population of MF that remains trapped until the structural reorganization of the bulk water mantle.

\subsection{The Role of Ice Phase and Composition in Molecular Trapping}

The physical phase of \water\ ice strongly influences whether a volatile species remains on the surface or diffuses readily, becoming trapped within the bulk of the ice mantle. When \water\ is deposited at very low temperature ($\sim20$K or below) it forms porous amorphous solid water, a highly disordered network that contains micropores capable of sequestering guest molecules \citep{Fraser2001,Collings2004,Burke2015}. Molecules incorporated into these pores remain immobilised until the ice undergoes structural re‑organisation during heating, at which point they may be released in a characteristic “volcano’’ desorption peak, or  substrate sublimation occurs \citep{Smith1997,MartinDomenech2014}.

In the present study, the ASW \water\ ice was grown at 35K, a temperature that yields relatively compact ASW. The FTIR spectrum of the pure \water\ layer (Figure \ref{fig:ftir_all}, Appendix \ref{app:ftir}) does not display the dangling–OH band that is diagnostic of highly porous ASW \citep{Kouchi1994, Raut2015}.  Nevertheless, the absence of this feature does not imply a completely non‑porous ice, only that the concentration is below the sensitivity of the analytical method. Compact ASW can retain a population of “hidden’’ micropores that collapse during the amorphous‑to‑crystalline transition, releasing trapped species in the volcano desorption feature \citep[e.g.,][]{Kouchi1994,Isokoski2014,Bossa2014}.  This mechanism accounts for the $\sim145$–160K desorption component observed in our TPD experiments (Figure \ref{fig:MF_all_substrates}) and is widely reported for molecules trapped in water ice \citep{Smith1997,Viti2004,Collings2004,Burke2015}.

\begin{figure}[t]
\centering
\includegraphics[width=\columnwidth]{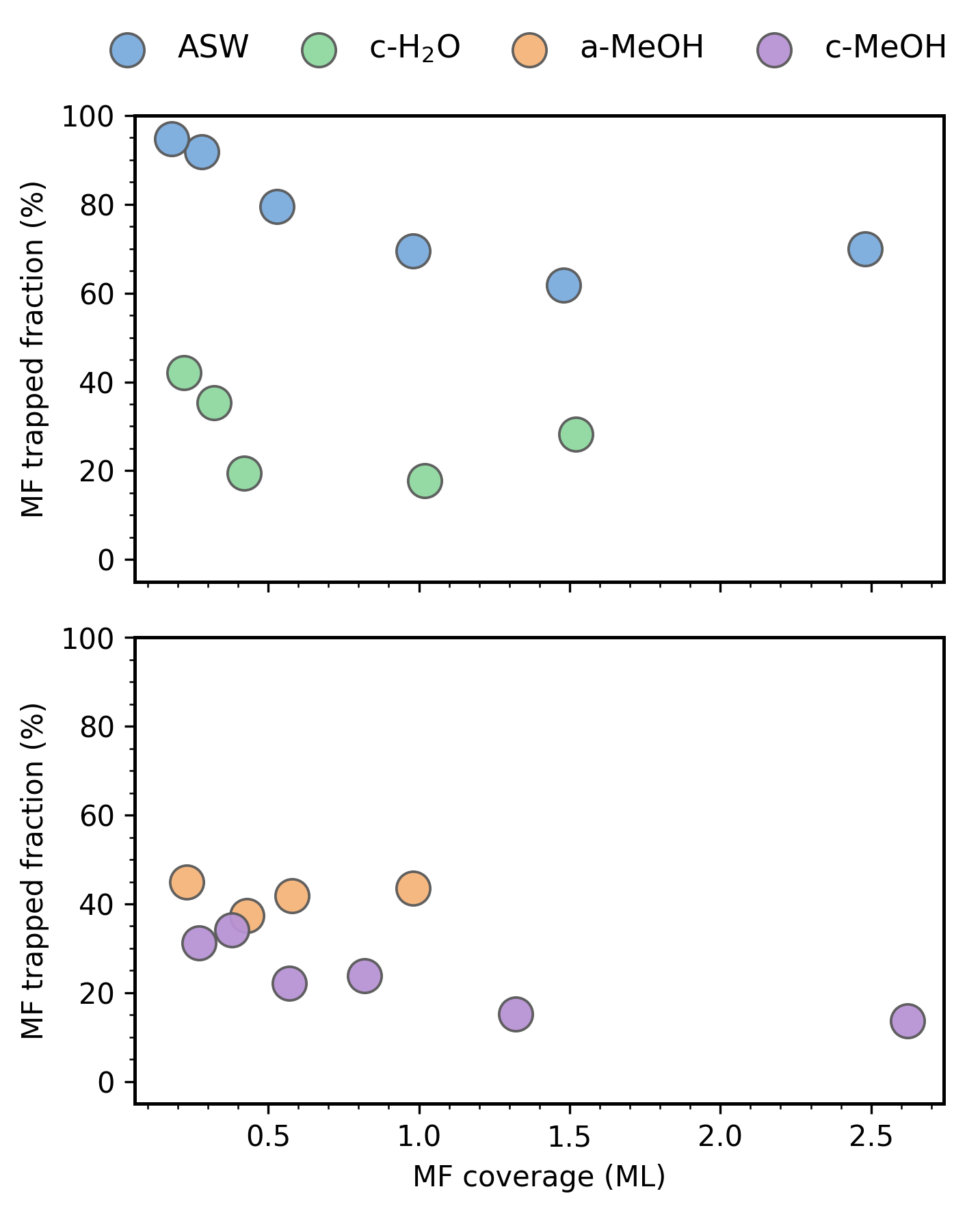}
\caption{
Fraction of MF remaining trapped in the substrate as a function of MF coverage.
Top: \water\ substrates (ASW and crystalline).
Bottom: \methanol\ substrates (amorphous and crystalline CH$_3$OH).
}
\label{fig:mf_trapping}
\end{figure}

Consistent with the above behavior, the MF TPD curves reveal substantial trapping within the ASW matrix (Figure  \ref{fig:mf_trapping}).  Quantitative analysis shows that at the lowest coverages $\sim90$–$95\%$ of the deposited MF remains trapped until the crystallization event, and even at the highest coverages a significant $60$–$80\%$ is retained.  
In stark contrast, MF deposited on pre‑formed crystalline \water\ ice exhibits much weaker trapping: only $\sim20$–$40\%$ of the MF diffuses into the structure and survives to the volcano stage. Crystalline water, grown at elevated temperature, possesses an ordered lattice with far fewer structural cavities, so the volcano desorption peak is strongly suppressed.

Methanol substrates display trapping efficiencies where both amorphous and crystalline phases are comparable to crystalline \water.  For MF on amorphous \methanol\ the trapped fraction lies in the range $35$–$45\%$, while on crystalline \methanol\ it drops further to $10$–$35\%$ depending on initial coverage (Figure \ref{fig:mf_trapping}). The high‑temperature MF desorption feature appearing near the methanol crystallization and desorption temperatures (Figure \ref{fig:MF_all_substrates}), indicates that a fraction of MF remains associated with the methanol matrix and is released only during thermal restructuring or sublimation of the \methanol\ ice \citep[e.g.,][]{Brown_Bolina2007,Burke2015}.  However, this feature is considerably weaker than the pronounced volcano release observed for ASW, confirming that trapping in \methanol\ is far less efficient than in amorphous \water\ ice.

\subsection{Surface Mobility and Island Formation on Methanol Ice}

\begin{figure}
\centering

\begin{minipage}{0.48\textwidth}
\centering
\includegraphics[width=\linewidth]{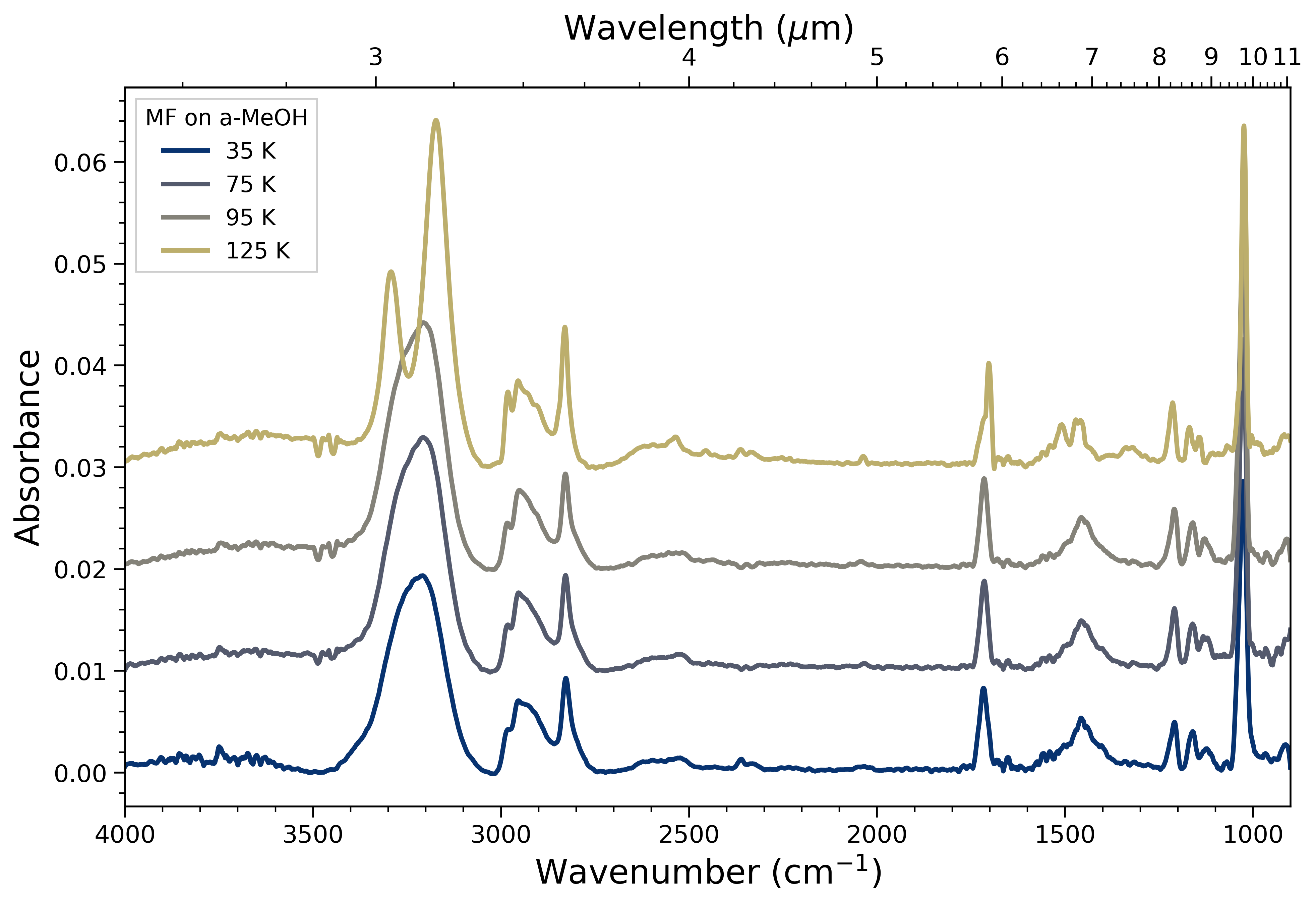}
\small (a) MF layered on amorphous \methanol
\end{minipage}

\vspace{0.5cm}

\begin{minipage}{0.48\textwidth}
\centering
\includegraphics[width=\linewidth]{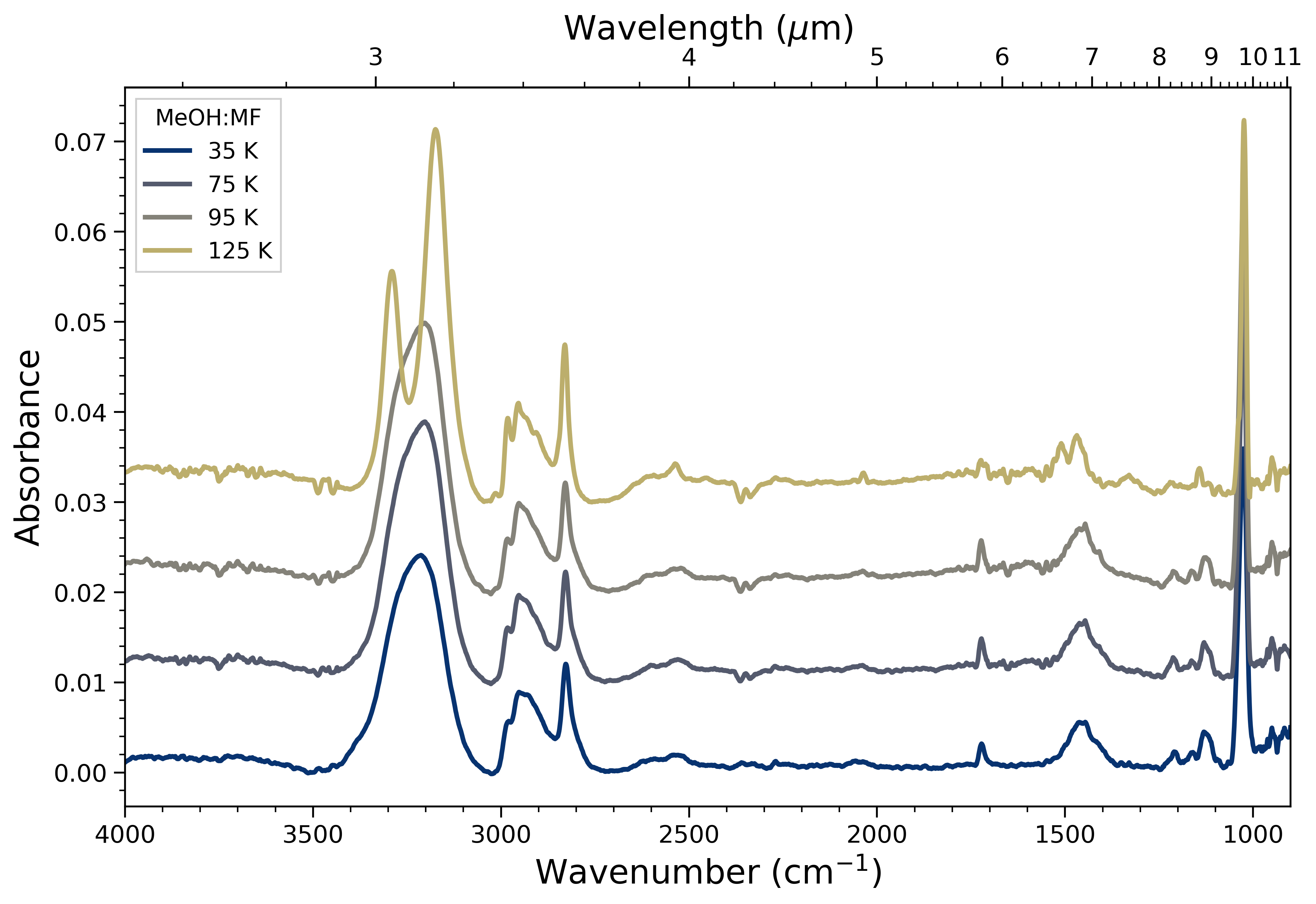}
\small (b) Co-deposited MF:\methanol
\end{minipage}

\begin{minipage}{0.48\textwidth}
\centering
\includegraphics[width=\linewidth]{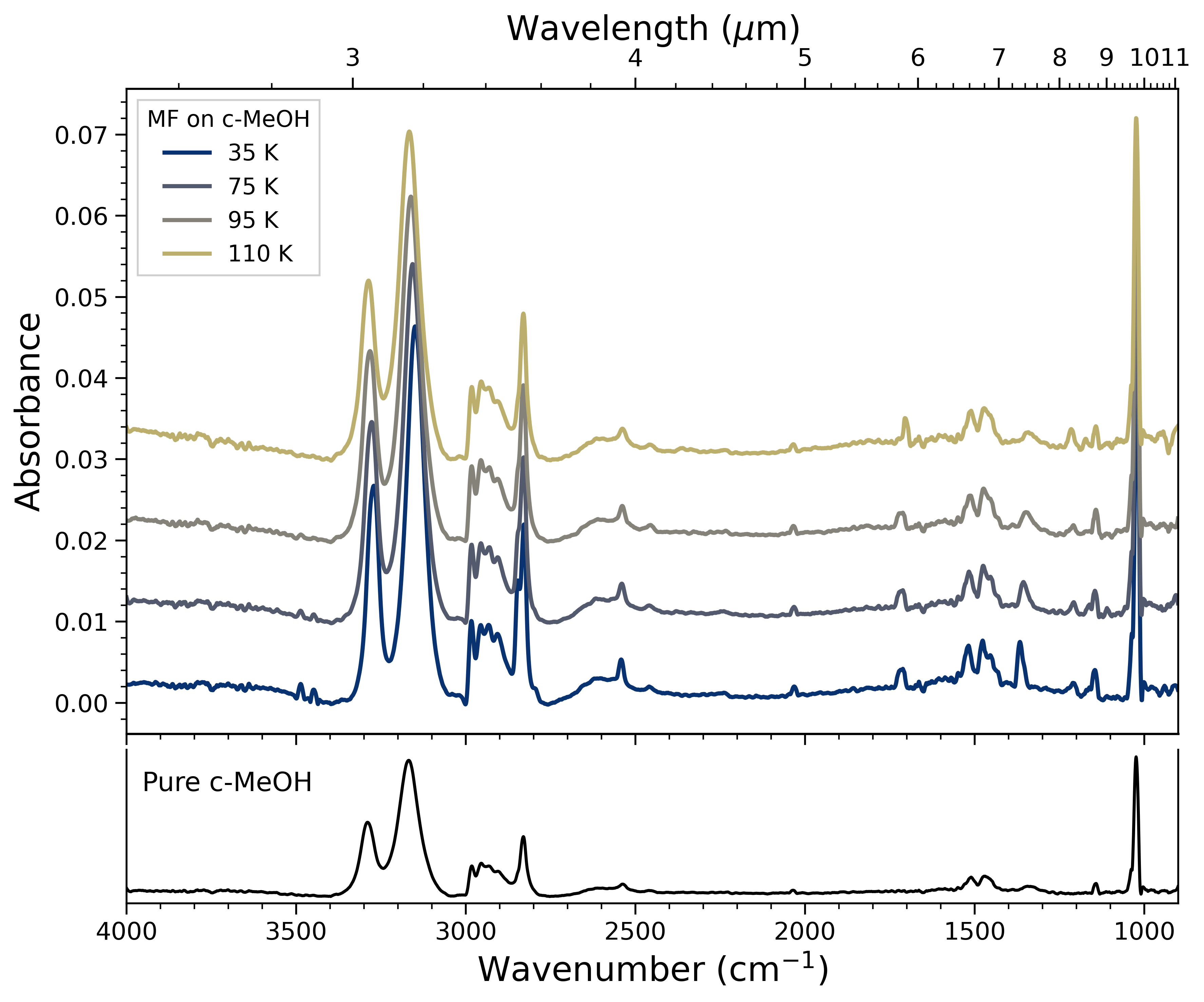}
\small (c) MF layered on crystalline \methanol
\end{minipage}

\vspace{0.5cm}

\caption{
FTIR spectra of MF interacting with \methanol\ ice.
(a) MF deposited on amorphous \methanol.
(b) Co-deposited MF:\methanol\ ice.
(c) MF deposited on crystalline \methanol. The pure crystalline \methanol\ spectra is directly under for comparison with the MF spectra.}
\label{fig:ftir_mf_meoh_comparison}
\end{figure}

A distinct kinetic behavior is observed for MF deposited on crystalline \methanol. Even at sub-monolayer coverages, the TPD profiles exhibit coverage-independent leading edges characteristic of zeroth-order desorption kinetics \citep{King1975,Collings2004}. This phenomenon, known as islanding, suggests that the desorption rate is governed primarily by intermolecular MF–MF interactions rather than adsorption to isolated substrate sites.

The chemical structure of the substrate is the primary driver of this behavior. Crystalline \methanol\ (specifically the $\alpha$-phase stable at these temperatures) consists of long, hydrogen-bonded molecular chains in which the hydroxyl groups are largely involved in internal lattice bonding, resulting in a surface with few available hydrogen-bonding sites for adsorbates \citep{Baber2011,Brown_Bolina2007}. In our experiments, the effective desorption energy for MF on crystalline \methanol\ ($5333\pm17$~K) is lower than on ASW ($6247\pm43$~K) and approaches the pure MF–MF multilayer interaction energy ($5509\pm10$~K), indicating that MF–substrate interactions are relatively weak. Consequently, MF molecules exhibit high surface mobility and diffuse across the ordered surface until they encounter other MF molecules, leading to aggregation into three-dimensional islands dominated by MF–MF interactions, similar to clustering behavior observed for other weakly interacting species on hydrogen-bonded ice surfaces \citep[e.g.,][]{TvS2022}.

In contrast, MF deposited on amorphous \methanol\ exhibits first-order desorption kinetics. The disordered amorphous matrix provides a heterogeneous distribution of binding geometries, including structural defects and crevices that can stabilize isolated MF molecules through available physical or hydrogen-bond interactions. These sites promote adsorption of MF directly onto the substrate, suppressing the mobility required for island formation and allowing the MF to sample a broader binding-energy distribution ($5469\pm40$~K). Under these conditions, adsorption occurs at a multitude of available sites and desorption proceeds from molecules bound directly to the methanol substrate rather than from segregated MF domains.

The mobility of MF within amorphous \methanol\ ice is revealed by the nearly indistinguishable TPD curves obtained for a layered sample (MF deposited on an amorphous \methanol\ film) and for a co-deposited MF:\methanol\ mixture (Figure \ref{fig:mixing_combined}) \citep{Collings2004,Minissale2022}. The near overlap of the leading edge, the intermediate volcano feature, and the high-temperature desorption peak indicates that MF molecules redistribute throughout the methanol bulk prior to significant desorption. This behavior contrasts with \water-rich ices, where the tetrahedral hydrogen-bond network of ASW forms a rigid three-dimensional cage that traps species until the amorphous-to-crystalline transition, producing the characteristic volcano desorption signal \citep{Smith1997,Fraser2001,Minissale2022}. The oxygen-bearing functional groups of MF can strongly influence its interaction with the hydrogen-bonded water surface and therefore its subsequent desorption behavior \citep{Burke2015,Kruczkiewicz2024desorption}. \methanol\ ice, by contrast, forms a lattice of relatively linear hydrogen-bonded chains that is structurally more flexible and contains fewer persistent trapping sites \citep{Brown_Bolina2007,Burke2015}. Modest heating can therefore reorganize the \methanol\ framework and open transient pathways that allow MF to percolate through the structure and equilibrate with the surrounding matrix \citep{Viti2004,Garrod2008}.

FTIR spectra recorded during the warm-up of the layered and co-deposited ices show that MF remains largely mixed with the amorphous \methanol\ substrate (Figure \ref{fig:ftir_mf_meoh_comparison}). In both configurations the spectra are dominated by the broad \methanol\ O–H stretching band near $\sim3300$~cm$^{-1}$ and the strong C–O stretch near $\sim1020$~cm$^{-1}$. The MF carbonyl (C=O) vibration remains detectable around $\sim1740$~cm$^{-1}$ but does not display the pronounced peak splitting or strong narrowing expected for segregated crystalline MF domains \citep{TvS2021methylformate}. A weak shoulder appears near $\sim1700$~cm$^{-1}$ only at the highest temperatures in the layered and mixed experiments. Pure MF crystals exhibit a characteristic redshift of the C=O mode from $\sim1723$~cm$^{-1}$ in amorphous MF to $\sim1698$~cm$^{-1}$ in crystalline MF, so the emergence of this shoulder marks the onset of MF ordering within the ice. However, the crystallization feature is far weaker than in the pure-MF spectra (Appendix \ref{app:ftir}), indicating that only a small fraction of MF forms ordered domains.

The layered and co-deposited MF on amorphous \methanol\ spectra evolve nearly identically during warm-up, indicating that MF deposited as a surface layer rapidly redistributes within the \methanol\ matrix. In particular, the MF carbonyl band remains clearly detectable at temperatures up to $\sim125$~K in the amorphous-\methanol\ experiment, well above the temperature at which MF desorbs from the surface in the TPD experiments (Figure \ref{fig:MF_all_substrates}). This persistence indicates that a substantial fraction of MF becomes incorporated within the methanol ice during the early stages of heating. When MF is deposited on crystalline \methanol, however, the $\sim1700$~cm$^{-1}$ crystallization component disappears by $\sim130$~K, consistent with more limited intermixing within the ordered methanol lattice \citep{Bertin2010_MF}. Together with the TPD results, these observations indicate that MF mixes efficiently within amorphous methanol ice but that the methanol matrix does not trap MF as effectively as ASW.

\subsection{Astrophysical Implications}

Thermal desorption temperatures derived from laboratory TPD experiments cannot be directly applied to interstellar environments because desorption is a kinetic process that depends on the heating rate. While the binding energy $E_{\rm des}$ and prefactor $\nu$ describe the intrinsic desorption properties of a molecule--surface system, the temperature at which desorption occurs decreases substantially under the much slower heating conditions relevant to interstellar ices \citep{Brown_Bolina2007,Viti2004,Ligterink2023}.

To compare the laboratory binding energies with astrophysical warm-up conditions, we estimate the surface residence time implied by the same first-order desorption kinetics used in the laboratory analysis:

\begin{equation}
\tau_{\rm des}(T)=\nu^{-1}\exp\left(\frac{E_{\rm des}}{T}\right),
\end{equation}

where $E_{\rm des}$ is expressed in Kelvin units, $\nu$ is the desorption prefactor in s$^{-1}$, and $\tau_{\rm des}$ is the characteristic time a molecule remains bound to the surface at temperature $T$. Because $E_{\rm des}$ and $T$ are both in Kelvin, the ratio $E_{\rm des}/T$ is dimensionless. This expression is the first-order Polanyi--Wigner desorption rate rewritten as a timescale; the same activated form has been used to estimate thermal evaporation timescales for species on interstellar grains \citep[e.g.,][]{hasegawa1992}.

For an adopted astrophysical heating rate of $\sim1$~K~century$^{-1}$, the ice spends approximately one century within each 1 K temperature interval. We therefore estimate the characteristic desorption temperature by solving the residence-time equation for the temperature at which $\tau_{\rm des}\sim100$ yr, or $3.16\times10^9$ s. Using the binding energies and transition-state-theory prefactors derived in this work, this gives $T_{\rm des}\approx85$ K for MF on amorphous \methanol\ using $E_{\rm des}=5469$ K and $\nu=3.39\times10^{18}$ s$^{-1}$. For MF on amorphous and crystalline \water, using $E_{\rm des}=6247$ and 6213 K with $\nu\simeq(4.8$--$5.0)\times10^{18}$ s$^{-1}$ gives $T_{\rm des}\approx96$ K. We therefore report characteristic astrophysical desorption temperatures of $\sim85$ K for MF associated with amorphous \methanol\ and $\sim95$--100 K for MF associated with \water-rich substrates.

However, the much larger temperature (energy) separation seen in the laboratory between \methanol-associated MF desorption and the high-temperature MF release from amorphous \water\ ice arises because MF trapped in the \water\ matrix is retained until volcano desorption and eventual co-desorption with the restructuring water mantle. This distinction is important for interpreting MF release from interstellar ice mantles. MF associated with \methanol-rich regions may begin to desorb at relatively low temperatures, comparable to those inferred for the low-temperature MF component observed in Sagittarius B2(N1), whereas MF trapped in deeper \water-rich layers can remain bound until substantially higher temperatures (see \citet{Busch2022}). This scenario would be consistent with the formation of CO-/CH$_3$OH-rich outer ice layers in which MF and other COMs are produced, allowing them to desorb at lower temperatures than COMs trapped in the deeper \water-ice layers.

The lower binding energy of MF on \methanol\ may account for the early MF abundance increase seen below $\sim100$~K in Sag B2 observations by \citet{Busch2022}, while the stronger adsorption and trapping in \water-rich amorphous ice provides a natural explanation for higher-temperature release of MF during water-matrix evolution. At the same time, detections of MF in very cold environments ($T\sim10$~K) still requires non-thermal desorption pathways such as photodesorption or reactive desorption \citep{Bacmann2012}.

Many astrochemical models historically adopted a single binding energy for each species, implicitly assuming desorption from a uniform substrate or a single adsorption site on a homogeneous ice. More recent three-phase models \citep{Garrod2013,Garrod2022, Kalvans2024} treat the surface and bulk ice separately and allow molecules to become trapped within the water-ice matrix. Our results show that both the binding energy and the trapping efficiency of MF depend strongly on the composition and structure of the surrounding ice. Accounting for these matrix-dependent binding energies and trapping mechanisms are therefore important for astrochemical models that aim to reproduce the observed abundance structure of MF in star-forming regions \citep[e.g.,][]{Viti2004,Collings2004,Minissale2022}.

Finally, laboratory ice experiments necessarily simplify the processes by which astrophysical ices form. In the present work, molecular layers are deposited directly onto substrates, whereas interstellar and protostellar ices are thought to grow through grain-surface chemistry, such as the hydrogenation of O and CO to form \water\ and \methanol, respectively. This difference in formation pathway may lead to variations in ice morphology, including porosity, molecular mixing or segregation, and the degree of crystallinity. In particular, chemically formed ices are expected to be more compact and less porous than laboratory-deposited ices \citep[e.g.,][]{Garrod2013}. These structural differences could influence trapping efficiencies and diffusion within the ice, and therefore affect desorption behavior. While our experiments isolate the role of ice composition in regulating MF binding and release, the extent to which these results apply to astrophysical environments will depend on the detailed structure and formation history of the ices.

\section{Summary and Conclusions}
\label{summary}

We performed a systematic laboratory investigation of the thermal desorption behavior of MF from astrophysically relevant ice substrates, \water\ and \methanol, using TPD experiments under UHV conditions. By combining leading-edge kinetic analysis with NNLS inversion of the desorption profiles, we derived representative desorption energies and binding-energy distributions for MF interacting with amorphous and crystalline \water\ and \methanol\ ices. Our main findings are summarized as follows:

\begin{enumerate}

\item Analysis of the TPD profiles for MF interacting with amorphous \methanol\ yields a representative desorption energy of $E_{\rm des}=5469\pm40$~K. This value is $\sim800$~K lower than that derived for MF on \water\ substrates, indicating that MF is less strongly retained on amorphous \methanol\ under the experimental conditions of this work.

\item Leading-edge analysis of MF desorption from crystalline \methanol\ yields an effective desorption energy of $E_{\rm des}=5333\pm17$~K. The coverage-independent leading edges indicate multilayer-like, zeroth-order desorption kinetics, consistent with MF–MF interactions dominating over MF–substrate interactions and promoting island formation on the surface.

\item Using transition-state theory, we derive a pre-exponential factor of $\nu_{\rm TST}\sim10^{18}$~s$^{-1}$ for MF desorption from both \water\ and \methanol\ substrates. This value is about five to six orders of magnitude larger than the commonly assumed prefactor of $10^{12}$--$10^{13}$~s$^{-1}$ often used in binding-energy estimates, emphasizing the importance of using physically motivated prefactors when deriving desorption energies for COMs and larger molecules. Because the prefactor enters the desorption analysis only logarithmically, however, this very large increase in $\nu$ produces a more modest upward shift in the derived binding energies, bringing them into the $\sim 5.3\times10^{3}$--$6.3\times10^{3}$ K range rather than the lower $\sim 4.2\times10^{3}$--$4.6\times10^{3}$ K values often reported when canonical prefactors are assumed.

\item The TPD measurements show that MF interacts with \water\ ice over a range of binding environments rather than a single characteristic energy. Median desorption energies of $E_{\rm des}=6247\pm43$~K and $6213\pm38$~K are derived for amorphous and crystalline \water\ substrates, respectively. The similar spread in energies indicates that MF experiences comparable surface heterogeneity in both cases.

\item The desorption behavior of MF also depends on the phase and structure of the substrate. MF associated with amorphous \water\ becomes partially trapped within the ice matrix and is released during structural evolution of the ice through restructuring (crystallization), volcano desorption, and co-desorption with the \water\ mantle. In contrast, trapping is weaker on crystalline \water\ and on \methanol\ substrates, while MF on crystalline \methanol\ forms islands even at sub-monolayer coverages.

\item The matrix-dependent binding energies and trapping efficiencies measured here on icy-grain analogues naturally produce a staged thermal desorption sequence as interstellar ice mantles are heated. MF associated with \methanol-rich regions of the mantle can desorb at characteristic temperatures of $\sim$85~K, while MF incorporated into \water-dominated ice layers remains trapped until higher temperatures of at least $\sim$95--100~K , where structural evolution of the \water\ mantle enables its release. This behavior moves toward resolving the cold COM problem by providing a physically motivated pathway for MF desorption below the canonical \water\ sublimation temperature for the early appearance of gas-phase MF in hot-core sources such as Sagittarius~B2(N1) \citep{Busch2022}; however, it does not fully account for detections at even lower temperatures ($\lesssim$50~K), indicating that additional non-thermal or chemical desorption mechanisms could still be at play.

\end{enumerate}


\section{Acknowledgments}

The authors thank the anonymous reviewer for constructive comments that improved the manuscript. R.E.G. and A.S. acknowledge support from the UVA Interconnected Universe Initiation Fellowship program. L.I.C. acknowledges support from the Research Corporation for Science Advancement Cottrell Scholarship Award 28249, the David and Lucille Packard Foundation, and NSF AAG Award 2205698. C.A.D., J.S.T., L.I.C., \& A.K.W. thank NASA's PSEF program (Award number 80NSSC23K0200) for development and support of the KEVION Ion Irradiation Facility for Space Science, located in the University of Virginia’s Laboratory for Astrophysics and Surface Physics (LASP), where our team made use of the associated ICE Chamber. R.T.G. thanks the National Science Foundation for funding through the Astronomy \& Astrophysics program (grant numbers 2206516 and 2510129).

%

\vspace{5mm}
\facilities{KEVION ICE chamber at UVA's Laboratory for Astrophysics and Surface Physics (LASP).}



\begin{thebibliography}{}
\expandafter\ifx\csname natexlab\endcsname\relax\def\natexlab#1{#1}\fi
\providecommand{\url}[1]{\href{#1}{#1}}
\providecommand{\dodoi}[1]{doi:~\href{http://doi.org/#1}{\nolinkurl{#1}}}
\providecommand{\doeprint}[1]{\href{http://ascl.net/#1}{\nolinkurl{http://ascl.net/#1}}}
\providecommand{\doarXiv}[1]{\href{https://arxiv.org/abs/#1}{\nolinkurl{https://arxiv.org/abs/#1}}}

\bibitem[{{Accolla} {et~al.}(2013){Accolla}, {Congiu}, {Manic{\`o}}, {Dulieu}, {Chaabouni}, {Lemaire}, \& {Pirronello}}]{Accolla2013}
{Accolla}, M., {Congiu}, E., {Manic{\`o}}, G., {et~al.} 2013, \mnras, 429, 3200, \dodoi{10.1093/mnras/sts578}

\bibitem[{{Baber} {et~al.}(2011){Baber}, {Lawton}, \& {Sykes}}]{Baber2011}
{Baber}, A.~E., {Lawton}, T.~J., \& {Sykes}, E. C.~H. 2011, The Journal of Physical Chemistry C, 115, 9157, \dodoi{10.1021/jp2006075}

\bibitem[{{Bacmann} {et~al.}(2012){Bacmann}, {Taquet}, {Faure}, {Kahane}, \& {Ceccarelli}}]{Bacmann2012}
{Bacmann}, A., {Taquet}, V., {Faure}, A., {Kahane}, C., \& {Ceccarelli}, C. 2012, \aap, 541, L12, \dodoi{10.1051/0004-6361/201219207}

\bibitem[{{Behmard} {et~al.}(2019){Behmard}, {Fayolle}, {Graninger}, {Bergner}, {Mart{\'\i}n-Dom{\'e}nech}, {Maksyutenko}, {Rajappan}, \& {{\"O}berg}}]{Behmard2019}
{Behmard}, A., {Fayolle}, E.~C., {Graninger}, D.~M., {et~al.} 2019, \apj, 875, 73, \dodoi{10.3847/1538-4357/ab0e7b}

\bibitem[{Bertin {et~al.}(2011)Bertin, Romanzin, Michaut, Jeseck, \& Fillion}]{Bertin2010_MF}
Bertin, M., Romanzin, C., Michaut, X., Jeseck, P., \& Fillion, J.-H. 2011, The Journal of Physical Chemistry C, 115, 12920, \dodoi{10.1021/jp201487u}

\bibitem[{{Bisschop} {et~al.}(2007){Bisschop}, {Fuchs}, {Boogert}, {van Dishoeck}, \& {Linnartz}}]{Bisschop2007lab}
{Bisschop}, S.~E., {Fuchs}, G.~W., {Boogert}, A.~C.~A., {van Dishoeck}, E.~F., \& {Linnartz}, H. 2007, \aap, 470, 749, \dodoi{10.1051/0004-6361:20077464}

\bibitem[{{Boogert} {et~al.}(2015){Boogert}, {Gerakines}, \& {Whittet}}]{Boogert2015}
{Boogert}, A.~C.~A., {Gerakines}, P.~A., \& {Whittet}, D. C.~B. 2015, \araa, 53, 541, \dodoi{10.1146/annurev-astro-082214-122348}

\bibitem[{{Booth} {et~al.}(2024){Booth}, {Leemker}, {van Dishoeck}, {Evans}, {Ilee}, {Kama}, {Keyte}, {Law}, {van der Marel}, {Nomura}, {Notsu}, {{\"O}berg}, {Temmink}, \& {Walsh}}]{Booth2024}
{Booth}, A.~S., {Leemker}, M., {van Dishoeck}, E.~F., {et~al.} 2024, \aj, 167, 164, \dodoi{10.3847/1538-3881/ad2700}

\bibitem[{{Bossa} {et~al.}(2014){Bossa}, {Isokoski}, {Paardekooper}, {Bonnin}, {van der Linden}, {Triemstra}, {Cazaux}, {Tielens}, \& {Linnartz}}]{Bossa2014}
{Bossa}, J.-B., {Isokoski}, K., {Paardekooper}, D.~M., {et~al.} 2014, \aap, 561, A136, \dodoi{10.1051/0004-6361/201322549}

\bibitem[{{Brown} \& {Bolina}(2007)}]{Brown_Bolina2007}
{Brown}, W.~A., \& {Bolina}, A.~S. 2007, \mnras, 374, 1006, \dodoi{10.1111/j.1365-2966.2006.11216.x}

\bibitem[{{Burke} {et~al.}(2015){Burke}, {Puletti}, {Woods}, {Viti}, {Slater}, \& {Brown}}]{Burke2015}
{Burke}, D.~J., {Puletti}, F., {Woods}, P.~M., {et~al.} 2015, \jcp, 143, 164704, \dodoi{10.1063/1.4934264}

\bibitem[{{Busch} {et~al.}(2022){Busch}, {Belloche}, {Garrod}, {M{\"u}ller}, \& {Menten}}]{Busch2022}
{Busch}, L.~A., {Belloche}, A., {Garrod}, R.~T., {M{\"u}ller}, H. S.~P., \& {Menten}, K.~M. 2022, \aap, 665, A96, \dodoi{10.1051/0004-6361/202243383}

\bibitem[{{Cazaux} {et~al.}(2003){Cazaux}, {Tielens}, {Ceccarelli}, {Castets}, {Wakelam}, {Caux}, {Parise}, \& {Teyssier}}]{Cazaux2003}
{Cazaux}, S., {Tielens}, A.~G.~G.~M., {Ceccarelli}, C., {et~al.} 2003, \apjl, 593, L51, \dodoi{10.1086/378038}

\bibitem[{{Ceccarelli}(2004)}]{Ceccarelli2004}
{Ceccarelli}, C. 2004, in Astronomical Society of the Pacific Conference Series, Vol. 323, Star Formation in the Interstellar Medium: In Honor of David Hollenbach, ed. D.~{Johnstone}, F.~C. {Adams}, D.~N.~C. {Lin}, D.~A. {Neufeeld}, \& E.~C. {Ostriker}, 195

\bibitem[{{Collings} {et~al.}(2004){Collings}, {Anderson}, {Chen}, {Dever}, {Viti}, {Williams}, \& {McCoustra}}]{Collings2004}
{Collings}, M.~P., {Anderson}, M.~A., {Chen}, R., {et~al.} 2004, \mnras, 354, 1133, \dodoi{10.1111/j.1365-2966.2004.08272.x}

\bibitem[{{Collings} {et~al.}(2003){Collings}, {Dever}, {Fraser}, {McCoustra}, \& {Williams}}]{collings2003}
{Collings}, M.~P., {Dever}, J.~W., {Fraser}, H.~J., {McCoustra}, M.~R.~S., \& {Williams}, D.~A. 2003, \apj, 583, 1058, \dodoi{10.1086/345389}

\bibitem[{{Collings} {et~al.}(2015){Collings}, {Frankland}, {Lasne}, {Marchione}, {Rosu-Finsen}, \& {McCoustra}}]{Collings2015}
{Collings}, M.~P., {Frankland}, V.~L., {Lasne}, J., {et~al.} 2015, \mnras, 449, 1826, \dodoi{10.1093/mnras/stv425}

\bibitem[{{Doronin} {et~al.}(2015){Doronin}, {Bertin}, {Michaut}, {Philippe}, \& {Fillion}}]{Doronin2015}
{Doronin}, M., {Bertin}, M., {Michaut}, X., {Philippe}, L., \& {Fillion}, J.-H. 2015, \jcp, 143, 084703, \dodoi{10.1063/1.4929376}

\bibitem[{{Fayolle} {et~al.}(2016){Fayolle}, {Balfe}, {Loomis}, {Bergner}, {Graninger}, {Rajappan}, \& {{\"O}berg}}]{Fayolle2016}
{Fayolle}, E.~C., {Balfe}, J., {Loomis}, R., {et~al.} 2016, \apjl, 816, L28, \dodoi{10.3847/2041-8205/816/2/L28}

\bibitem[{{Fedoseev} {et~al.}(2015){Fedoseev}, {Cuppen}, {Ioppolo}, {Lamberts}, \& {Linnartz}}]{Fedoseev2015}
{Fedoseev}, G., {Cuppen}, H.~M., {Ioppolo}, S., {Lamberts}, T., \& {Linnartz}, H. 2015, \mnras, 448, 1288, \dodoi{10.1093/mnras/stu2603}

\bibitem[{{Fichthorn} \& {Miron}(2002)}]{Fichthorn2002}
{Fichthorn}, K.~A., \& {Miron}, R.~A. 2002, \prl, 89, 196103, \dodoi{10.1103/PhysRevLett.89.196103}

\bibitem[{{Fraser} {et~al.}(2001){Fraser}, {Collings}, {McCoustra}, \& {Williams}}]{Fraser2001}
{Fraser}, H.~J., {Collings}, M.~P., {McCoustra}, M. R.~S., \& {Williams}, D.~A. 2001, \mnras, 327, 1165, \dodoi{10.1046/j.1365-8711.2001.04835.x}

\bibitem[{{Fuchs} {et~al.}(2009){Fuchs}, {Cuppen}, {Ioppolo}, {Romanzin}, {Bisschop}, {Andersson}, {van Dishoeck}, \& {Linnartz}}]{Fuchs2009}
{Fuchs}, G.~W., {Cuppen}, H.~M., {Ioppolo}, S., {et~al.} 2009, \aap, 505, 629, \dodoi{10.1051/0004-6361/200810784}

\bibitem[{{Garrod}(2013)}]{Garrod2013}
{Garrod}, R.~T. 2013, \apj, 778, 158, \dodoi{10.1088/0004-637X/778/2/158}

\bibitem[{{Garrod} \& {Herbst}(2006)}]{Garrod2006}
{Garrod}, R.~T., \& {Herbst}, E. 2006, \aap, 457, 927, \dodoi{10.1051/0004-6361:20065560}

\bibitem[{{Garrod} {et~al.}(2022){Garrod}, {Jin}, {Matis}, {Jones}, {Willis}, \& {Herbst}}]{Garrod2022}
{Garrod}, R.~T., {Jin}, M., {Matis}, K.~A., {et~al.} 2022, \apjs, 259, 1, \dodoi{10.3847/1538-4365/ac3131}

\bibitem[{{Garrod} {et~al.}(2008){Garrod}, {Weaver}, \& {Herbst}}]{Garrod2008}
{Garrod}, R.~T., {Weaver}, S.~L.~W., \& {Herbst}, E. 2008, \apj, 682, 283, \dodoi{10.1086/588035}

\bibitem[{{Hasegawa} {et~al.}(1992){Hasegawa}, {Herbst}, \& {Leung}}]{hasegawa1992}
{Hasegawa}, T.~I., {Herbst}, E., \& {Leung}, C.~M. 1992, \apjs, 82, 167, \dodoi{10.1086/191713}

\bibitem[{{Herbst} \& {van Dishoeck}(2009)}]{herbst_vanDishoeck2009}
{Herbst}, E., \& {van Dishoeck}, E.~F. 2009, \araa, 47, 427, \dodoi{10.1146/annurev-astro-082708-101654}

\bibitem[{{Huang} {et~al.}(2017){Huang}, {Bai}, {Hu}, \& {Hou}}]{Huang2017qcm}
{Huang}, X., {Bai}, Q., {Hu}, J., \& {Hou}, D. 2017, Sensors, 17, 1785, \dodoi{10.3390/s17081785}

\bibitem[{{Ishibashi} {et~al.}(2021){Ishibashi}, {Hidaka}, {Oba}, {Kouchi}, \& {Watanabe}}]{Ishibashi2021}
{Ishibashi}, A., {Hidaka}, H., {Oba}, Y., {Kouchi}, A., \& {Watanabe}, N. 2021, \apjl, 921, L13, \dodoi{10.3847/2041-8213/ac3005}

\bibitem[{{Isokoski} {et~al.}(2014){Isokoski}, {Bossa}, {Triemstra}, \& {Linnartz}}]{Isokoski2014}
{Isokoski}, K., {Bossa}, J.-B., {Triemstra}, T., \& {Linnartz}, H. 2014, Physical Chemistry Chemical Physics (Incorporating Faraday Transactions), 16, 3456, \dodoi{10.1039/C3CP54481H}

\bibitem[{{Jenniskens} \& {Blake}(1994)}]{Jenniskens1994}
{Jenniskens}, P., \& {Blake}, D.~F. 1994, Science, 265, 753, \dodoi{10.1126/science.11539186}

\bibitem[{{Jenniskens} {et~al.}(1995){Jenniskens}, {Blake}, {Wilson}, \& {Pohorille}}]{Jenniskens1995}
{Jenniskens}, P., {Blake}, D.~F., {Wilson}, M.~A., \& {Pohorille}, A. 1995, \apj, 455, 389, \dodoi{10.1086/176585}

\bibitem[{{Johannsmann} {et~al.}(2021){Johannsmann}, {Langhoff}, \& {Leppin}}]{Johannsmann2021qcm}
{Johannsmann}, D., {Langhoff}, A., \& {Leppin}, C. 2021, Sensors, 21, 3490, \dodoi{10.3390/s21103490}

\bibitem[{{J{\o}rgensen} {et~al.}(2016){J{\o}rgensen}, {van der Wiel}, {Coutens}, {Lykke}, {M{\"u}ller}, {van Dishoeck}, {Calcutt}, {Bjerkeli}, {Bourke}, {Drozdovskaya}, {Favre}, {Fayolle}, {Garrod}, {Jacobsen}, {{\"O}berg}, {Persson}, \& {Wampfler}}]{jorgensen2016}
{J{\o}rgensen}, J.~K., {van der Wiel}, M.~H.~D., {Coutens}, A., {et~al.} 2016, \aap, 595, A117, \dodoi{10.1051/0004-6361/201628648}

\bibitem[{{Kalv{\={a}}ns} {et~al.}(2024){Kalv{\={a}}ns}, {Kalni{\c{n}}a}, \& {Veitners}}]{Kalvans2024}
{Kalv{\={a}}ns}, J., {Kalni{\c{n}}a}, A., \& {Veitners}, K. 2024, \aap, 687, A296, \dodoi{10.1051/0004-6361/202450015}

\bibitem[{{King}(1975)}]{King1975}
{King}, D.~A. 1975, Surface Science, 47, 384, \dodoi{10.1016/0039-6028(75)90302-7}

\bibitem[{{Kirk} {et~al.}(2007){Kirk}, {Ward-Thompson}, \& {Andr{\'e}}}]{Kirk2007}
{Kirk}, J.~M., {Ward-Thompson}, D., \& {Andr{\'e}}, P. 2007, \mnras, 375, 843, \dodoi{10.1111/j.1365-2966.2006.11250.x}

\bibitem[{Knopf \& Ammann(2021)}]{KnopfAmmann2021}
Knopf, D.~A., \& Ammann, M. 2021, Atmospheric Chemistry and Physics, 21, 15725, \dodoi{10.5194/acp-21-15725-2021}

\bibitem[{{Kouchi} {et~al.}(1994){Kouchi}, {Yamamoto}, {Kozasa}, {Kuroda}, \& {Greenberg}}]{Kouchi1994}
{Kouchi}, A., {Yamamoto}, T., {Kozasa}, T., {Kuroda}, T., \& {Greenberg}, J.~M. 1994, \aap, 290, 1009

\bibitem[{{Kruczkiewicz} {et~al.}(2024){Kruczkiewicz}, {Dulieu}, {Ivlev}, {Caselli}, {Giuliano}, {Ceccarelli}, \& {Theul{\'e}}}]{Kruczkiewicz2024desorption}
{Kruczkiewicz}, F., {Dulieu}, F., {Ivlev}, A.~V., {et~al.} 2024, \aap, 686, A236, \dodoi{10.1051/0004-6361/202346948}

\bibitem[{{Lattelais} {et~al.}(2011){Lattelais}, {Bertin}, {Mokrane}, {Romanzin}, {Michaut}, {Jeseck}, {Fillion}, {Chaabouni}, {Congiu}, {Dulieu}, {Baouche}, {Lemaire}, {Pauzat}, {Pilm{\'e}}, {Minot}, \& {Ellinger}}]{Lattelais2011}
{Lattelais}, M., {Bertin}, M., {Mokrane}, H., {et~al.} 2011, \aap, 532, A12, \dodoi{10.1051/0004-6361/201016184}

\bibitem[{{Ligterink} \& {Minissale}(2023)}]{Ligterink2023}
{Ligterink}, N.~F.~W., \& {Minissale}, M. 2023, \aap, 676, A80, \dodoi{10.1051/0004-6361/202346436}

\bibitem[{{Luna} {et~al.}(2018){Luna}, {Satorre}, {Domingo}, {Mill{\'a}n}, {Luna-Ferr{\'a}ndiz}, {Gisbert}, \& {Santonja}}]{Luna2018}
{Luna}, R., {Satorre}, M.~{\'A}., {Domingo}, M., {et~al.} 2018, \mnras, 473, 1967, \dodoi{10.1093/mnras/stx2473}

\bibitem[{{Mart{\'\i}n-Dom{\'e}nech} {et~al.}(2014){Mart{\'\i}n-Dom{\'e}nech}, {Mu{\~n}oz Caro}, {Bueno}, \& {Goesmann}}]{MartinDomenech2014}
{Mart{\'\i}n-Dom{\'e}nech}, R., {Mu{\~n}oz Caro}, G.~M., {Bueno}, J., \& {Goesmann}, F. 2014, \aap, 564, A8, \dodoi{10.1051/0004-6361/201322824}

\bibitem[{{McGuire}(2022)}]{McGuire2022}
{McGuire}, B.~A. 2022, \apjs, 259, 30, \dodoi{10.3847/1538-4365/ac2a48}

\bibitem[{{Minissale} {et~al.}(2022){Minissale}, {Aikawa}, {Bergin}, {Bertin}, {Brown}, {Cazaux}, {Charnley}, {Coutens}, {Cuppen}, {Guzman}, {Linnartz}, {McCoustra}, {Rimola}, {Schrauwen}, {Toubin}, {Ugliengo}, {Watanabe}, {Wakelam}, \& {Dulieu}}]{Minissale2022}
{Minissale}, M., {Aikawa}, Y., {Bergin}, E., {et~al.} 2022, ACS Earth and Space Chemistry, 6, 597, \dodoi{10.1021/acsearthspacechem.1c00357}

\bibitem[{{Modica} \& {Palumbo}(2010)}]{Modica2010}
{Modica}, P., \& {Palumbo}, M.~E. 2010, \aap, 519, A22, \dodoi{10.1051/0004-6361/201014101}

\bibitem[{Nekrylova {et~al.}(1993)Nekrylova, French, Artsyukhovich, Ukraintsev, \& Harrison}]{Nekrylova1993}
Nekrylova, J.~V., French, C., Artsyukhovich, A.~N., Ukraintsev, V.~A., \& Harrison, I. 1993, Surface Science Letters, 295, L987, \dodoi{10.1016/0039-6028(93)90897-U}

\bibitem[{{Noble} {et~al.}(2012){Noble}, {Congiu}, {Dulieu}, \& {Fraser}}]{Noble2012}
{Noble}, J.~A., {Congiu}, E., {Dulieu}, F., \& {Fraser}, H.~J. 2012, \mnras, 421, 768, \dodoi{10.1111/j.1365-2966.2011.20351.x}

\bibitem[{{Pontoppidan} {et~al.}(2004){Pontoppidan}, {van Dishoeck}, \& {Dartois}}]{Pontoppidan2004}
{Pontoppidan}, K.~M., {van Dishoeck}, E.~F., \& {Dartois}, E. 2004, \aap, 426, 925, \dodoi{10.1051/0004-6361:20041276}

\bibitem[{{Raut} {et~al.}(2007){Raut}, {Fam{\'a}}, {Teolis}, \& {Baragiola}}]{Raut2007}
{Raut}, U., {Fam{\'a}}, M., {Teolis}, B.~D., \& {Baragiola}, R.~A. 2007, \jcp, 127, 204713, \dodoi{10.1063/1.2796166}

\bibitem[{{Raut} {et~al.}(2015){Raut}, {Mitchell}, \& {Baragiola}}]{Raut2015}
{Raut}, U., {Mitchell}, E.~H., \& {Baragiola}, R.~A. 2015, \apj, 811, 120, \dodoi{10.1088/0004-637X/811/2/120}

\bibitem[{{Redhead}(1962)}]{Redhead1962}
{Redhead}, P.~A. 1962, Vacuum, 12, 203, \dodoi{10.1016/0042-207X(62)90978-8}

\bibitem[{{Rocha} {et~al.}(2024){Rocha}, {van Dishoeck}, {Ressler}, {van Gelder}, {Slavicinska}, {Brunken}, {Linnartz}, {Ray}, {Beuther}, {Caratti o Garatti}, {Geers}, {Kavanagh}, {Klaassen}, {Justtanont}, {Chen}, {Francis}, {Gieser}, {Perotti}, {Tychoniec}, {Barsony}, {Majumdar}, {le Gouellec}, {Chu}, {Lew}, {Henning}, \& {Wright}}]{Rocha2024}
{Rocha}, W.~R.~M., {van Dishoeck}, E.~F., {Ressler}, M.~E., {et~al.} 2024, \aap, 683, A124, \dodoi{10.1051/0004-6361/202348427}

\bibitem[{{Rowland} {et~al.}(1991){Rowland}, {Fisher}, \& {Devlin}}]{Rowland1991}
{Rowland}, B., {Fisher}, M., \& {Devlin}, J.~P. 1991, \jcp, 95, 1378, \dodoi{10.1063/1.461119}

\bibitem[{{Santos} {et~al.}(2025){Santos}, {Piacentino}, {Bergner}, {Rajappan}, \& {{\"O}berg}}]{Santos2025}
{Santos}, J.~C., {Piacentino}, E.~L., {Bergner}, J.~B., {Rajappan}, M., \& {{\"O}berg}, K.~I. 2025, \aap, 698, A254, \dodoi{10.1051/0004-6361/202554068}

\bibitem[{{Sauerbrey}(1959)}]{Sauerbrey1959}
{Sauerbrey}, G. 1959, Zeitschrift fur Physik, 155, 206, \dodoi{10.1007/BF01337937}

\bibitem[{{Schaible} \& {Baragiola}(2014)}]{Schaible2014}
{Schaible}, M.~J., \& {Baragiola}, R.~A. 2014, Journal of Geophysical Research (Planets), 119, 2017, \dodoi{10.1002/2014JE004650}

\bibitem[{{Simon} {et~al.}(2023){Simon}, {Rajappan}, \& {{\"O}berg}}]{Simon2023}
{Simon}, A., {Rajappan}, M., \& {{\"O}berg}, K.~I. 2023, \apj, 955, 5, \dodoi{10.3847/1538-4357/aceaf8}

\bibitem[{{Smith} {et~al.}(1997){Smith}, {Huang}, {Wong}, \& {Kay}}]{Smith1997}
{Smith}, R.~S., {Huang}, C., {Wong}, E.~K.~L., \& {Kay}, B.~D. 1997, \prl, 79, 909, \dodoi{10.1103/PhysRevLett.79.909}

\bibitem[{Smith \& Kay(2018)}]{smith_kay2018}
Smith, R.~S., \& Kay, B.~D. 2018, The Journal of Physical Chemistry B, 122, 587, \dodoi{10.1021/acs.jpcb.7b05102}

\bibitem[{{Stevenson} {et~al.}(1999){Stevenson}, {Kimmel}, {Dohnalek}, {Smith}, \& {Kay}}]{Stevenson1999}
{Stevenson}, K.~P., {Kimmel}, G.~A., {Dohnalek}, Z., {Smith}, R.~S., \& {Kay}, B.~D. 1999, Science, 283, 1505, \dodoi{10.1126/science.283.5407.1505}

\bibitem[{{Tait} {et~al.}(2005{\natexlab{a}}){Tait}, {Dohn{\'a}lek}, {Campbell}, \& {Kay}}]{Tait2005_I}
{Tait}, S.~L., {Dohn{\'a}lek}, Z., {Campbell}, C.~T., \& {Kay}, B.~D. 2005{\natexlab{a}}, \jcp, 122, 164707, \dodoi{10.1063/1.1883629}

\bibitem[{{Tait} {et~al.}(2005{\natexlab{b}}){Tait}, {Dohn{\'a}lek}, {Campbell}, \& {Kay}}]{Tait2005_II}
---. 2005{\natexlab{b}}, \jcp, 122, 164708, \dodoi{10.1063/1.1883630}

\bibitem[{{Terwisscha van Scheltinga} {et~al.}(2022){Terwisscha van Scheltinga}, {Ligterink}, {Bosman}, {Hogerheijde}, \& {Linnartz}}]{TvS2022}
{Terwisscha van Scheltinga}, J., {Ligterink}, N.~F.~W., {Bosman}, A.~D., {Hogerheijde}, M.~R., \& {Linnartz}, H. 2022, \aap, 666, A35, \dodoi{10.1051/0004-6361/202142181}

\bibitem[{{Terwisscha van Scheltinga} {et~al.}(2021){Terwisscha van Scheltinga}, {Marcandalli}, {McClure}, {Hogerheijde}, \& {Linnartz}}]{TvS2021methylformate}
{Terwisscha van Scheltinga}, J., {Marcandalli}, G., {McClure}, M.~K., {Hogerheijde}, M.~R., \& {Linnartz}, H. 2021, \aap, 651, A95, \dodoi{10.1051/0004-6361/202140723}

\bibitem[{{Tielens} \& {Charnley}(1997)}]{Tielens1997}
{Tielens}, A.~G.~G.~M., \& {Charnley}, S.~B. 1997, Origins of Life and Evolution of the Biosphere, 27, 23, \dodoi{10.1023/A:1006513928588}

\bibitem[{{Viti} {et~al.}(2004){Viti}, {Collings}, {Dever}, {McCoustra}, \& {Williams}}]{Viti2004}
{Viti}, S., {Collings}, M.~P., {Dever}, J.~W., {McCoustra}, M. R.~S., \& {Williams}, D.~A. 2004, \mnras, 354, 1141, \dodoi{10.1111/j.1365-2966.2004.08273.x}

\bibitem[{{Watanabe} \& {Kouchi}(2002)}]{Watanabe2002}
{Watanabe}, N., \& {Kouchi}, A. 2002, \apjl, 571, L173, \dodoi{10.1086/341412}

\bibitem[{{Yang} {et~al.}(2022){Yang}, {Green}, {Pontoppidan}, {Bergner}, {Cleeves}, {Evans}, {Garrod}, {Jin}, {Kim}, {Kim}, {Lee}, {Sakai}, {Shingledecker}, {Shope}, {Tobin}, \& {van Dishoeck}}]{Yang2022}
{Yang}, Y.-L., {Green}, J.~D., {Pontoppidan}, K.~M., {et~al.} 2022, \apjl, 941, L13, \dodoi{10.3847/2041-8213/aca289}

\bibitem[{Öberg {et~al.}(2009)Öberg, Bottinelli, \& van Dishoeck}]{Oberg2009}
Öberg, K.~I., Bottinelli, S., \& van Dishoeck, E.~F. 2009, Astronomy \& Astrophysics, 494, L13, \dodoi{10.1051/0004-6361:200811471}

\end{thebibliography}

\appendix
\twocolumngrid

\section{FTIR Characterization of Pure Ices}
\label{app:ftir}

FTIR transmission spectra were used to verify the structural phase of the water and methanol ice substrates used in this study. The phase of \water\ ice was determined from the shape of the OH stretching band and the presence or absence of dangling–OH absorption features.

Porous ASW exhibits distinct dangling–OH bands near 3696 and 3720~cm$^{-1}$ arising from uncoordinated OH groups located on internal pore surfaces \citep{Rowland1991,Jenniskens1994}. The intensity of these bands decreases as the ice compacts or the pore structure collapses during heating \citep{Raut2007,Stevenson1999}.

\begin{figure*}[t]
\centering
\includegraphics[width=0.32\textwidth]{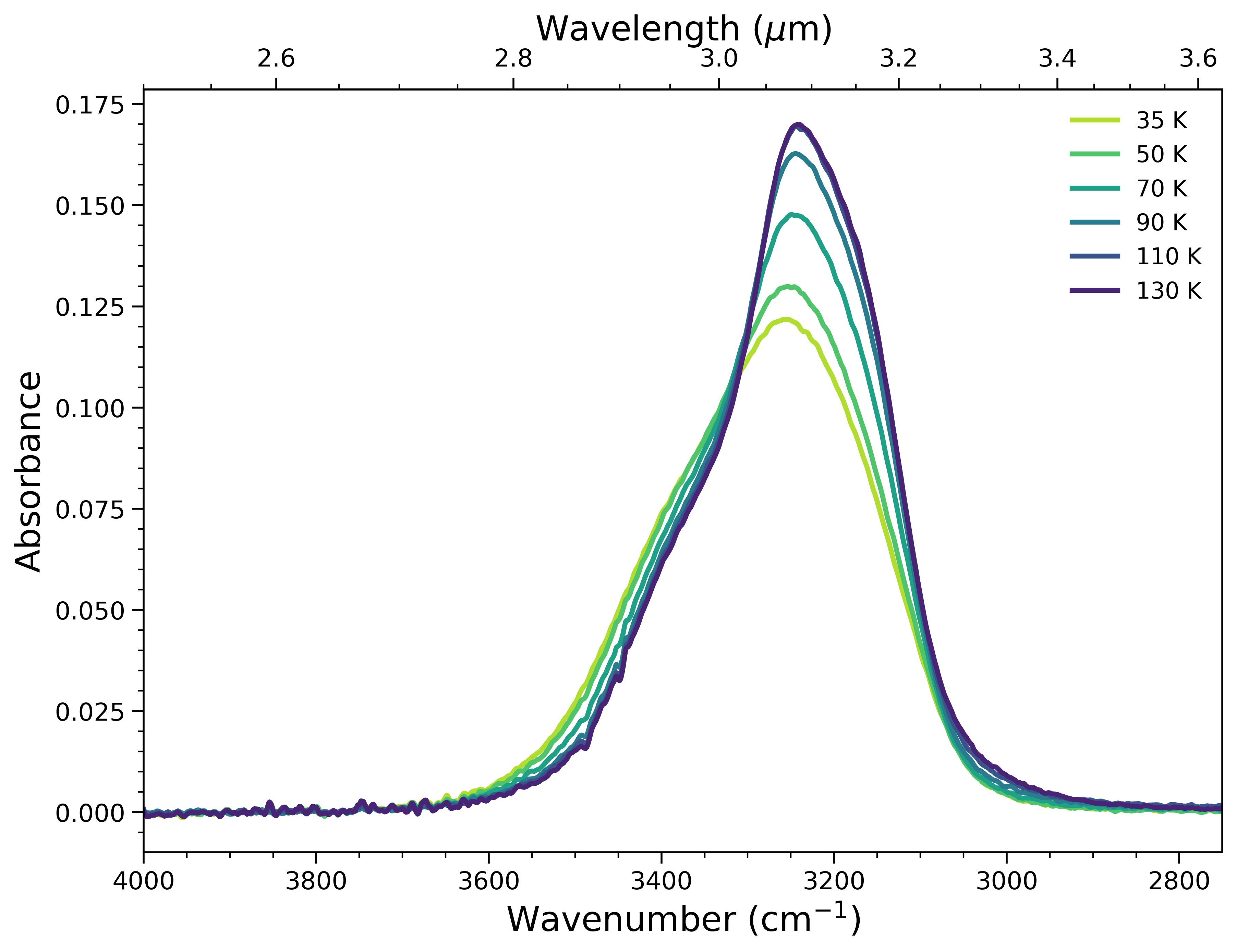}
\hfill
\includegraphics[width=0.32\textwidth]{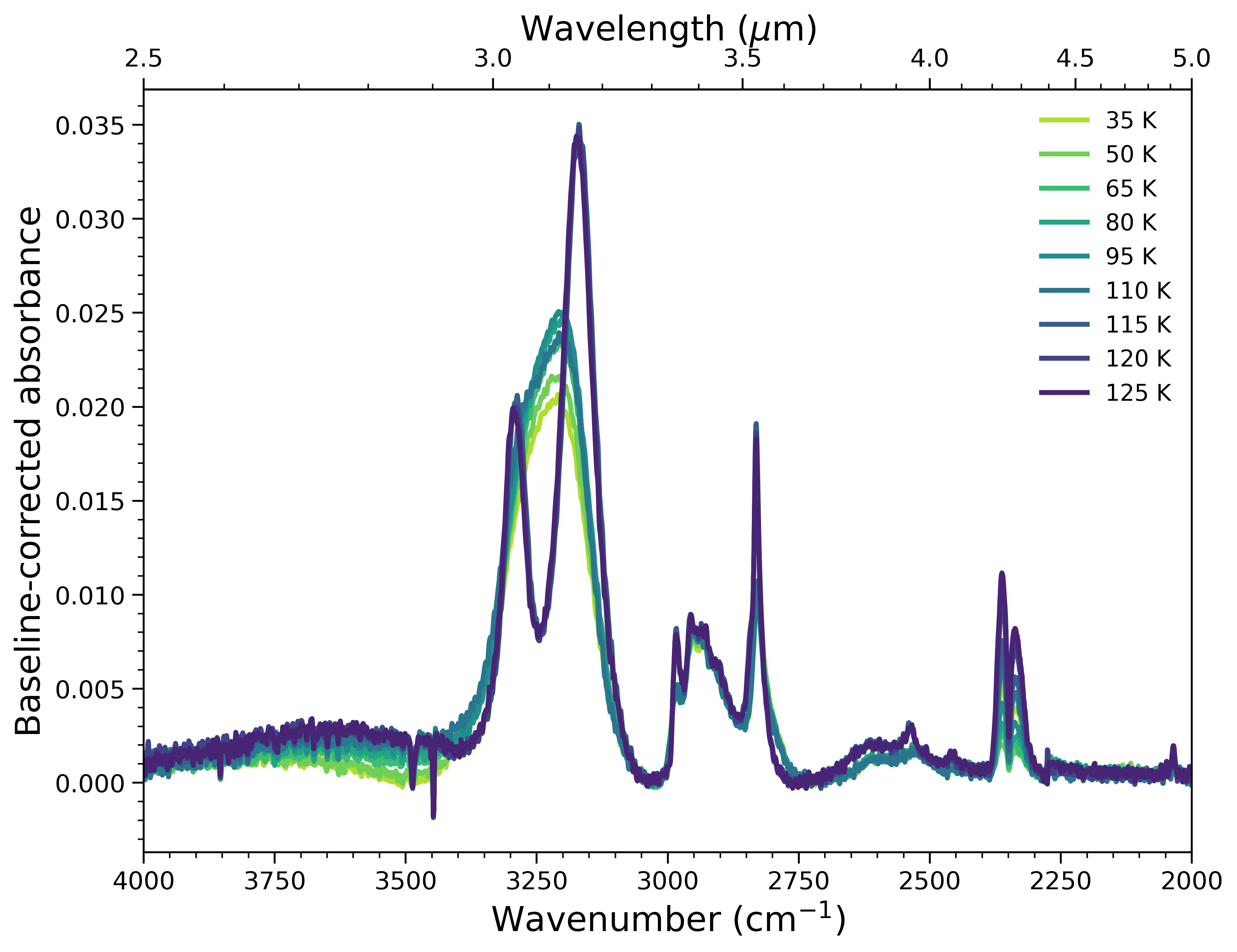}
\hfill
\includegraphics[width=0.32\textwidth]{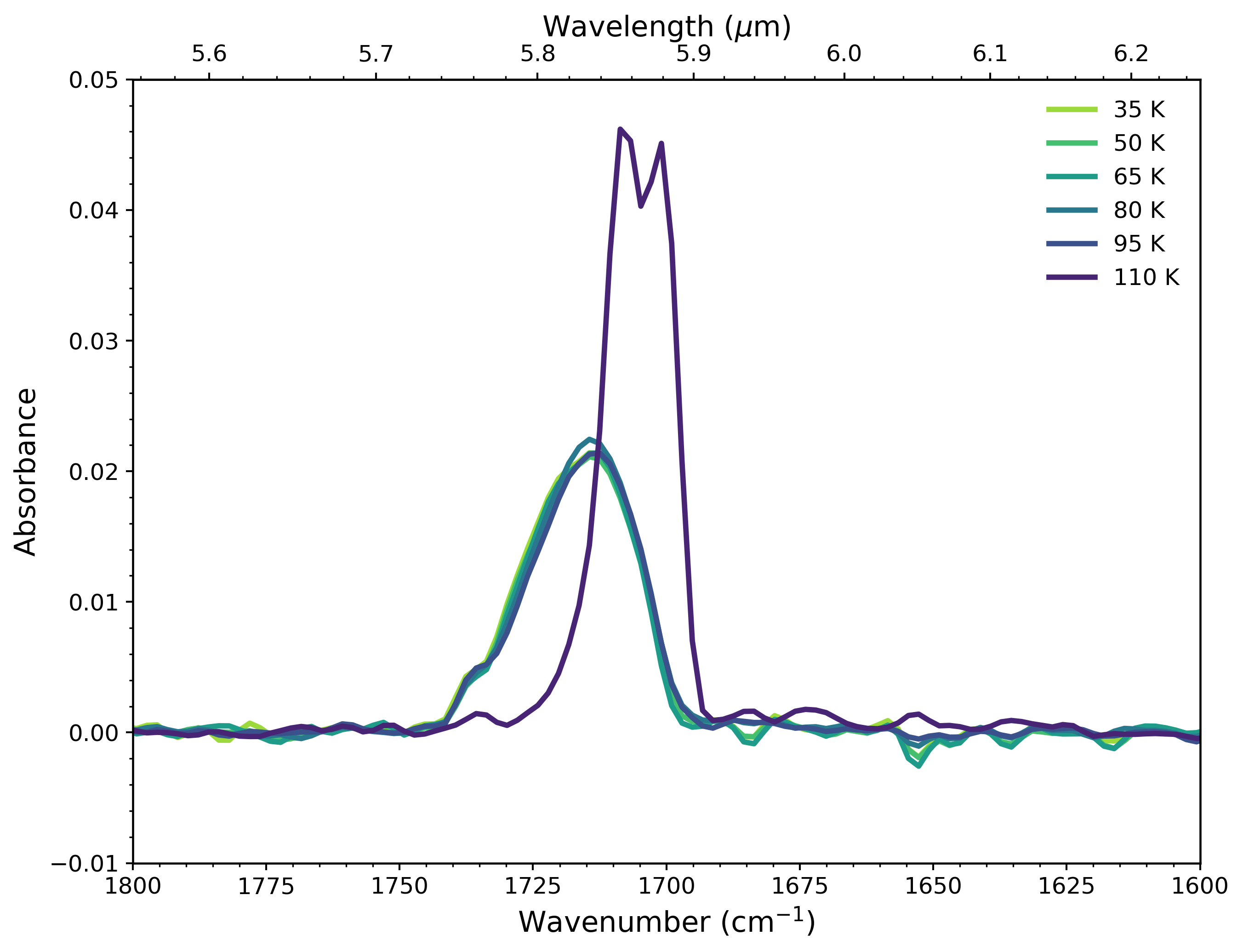}
\caption{
FTIR spectra used to verify the phase and composition of the ice substrates prior to TPD experiments. 
Left: \water\ ice deposited at 35~K. 
Center: \methanol\ ice during warm-up. 
Right: pure methyl formate (MF) ice spectra.
}
\label{fig:ftir_all}
\end{figure*}

In spectra of compact ASW deposited at 35~K in this work (Figure \ref{fig:ftir_all}), no distinct dangling–OH features are observed in the 3650–3700~cm$^{-1}$ region. This indicates that the deposited films form compact amorphous ice under our experimental conditions. The absence of dangling–OH features does not necessarily imply the complete absence of internal pores, as compact amorphous ices may still contain buried micro pore structure capable of trapping molecules \citep{Kouchi1994,MartinDomenech2014}.

\methanol\ shows the characteristic crystallization around 125~K where we see a sharp double peak in the OH stretching around 3250 wavenumbers. MF shows that the carbonyl C=O stretch at around 1700 wavenumbers becomes sharply double peaked by 100~K indicating a crystalline structure.

\section{Instrument Cracking Patterns for MF and \methanol}
\label{sec:cracking}

\begin{figure*}[t]
\centering

\includegraphics[width=0.9\linewidth]{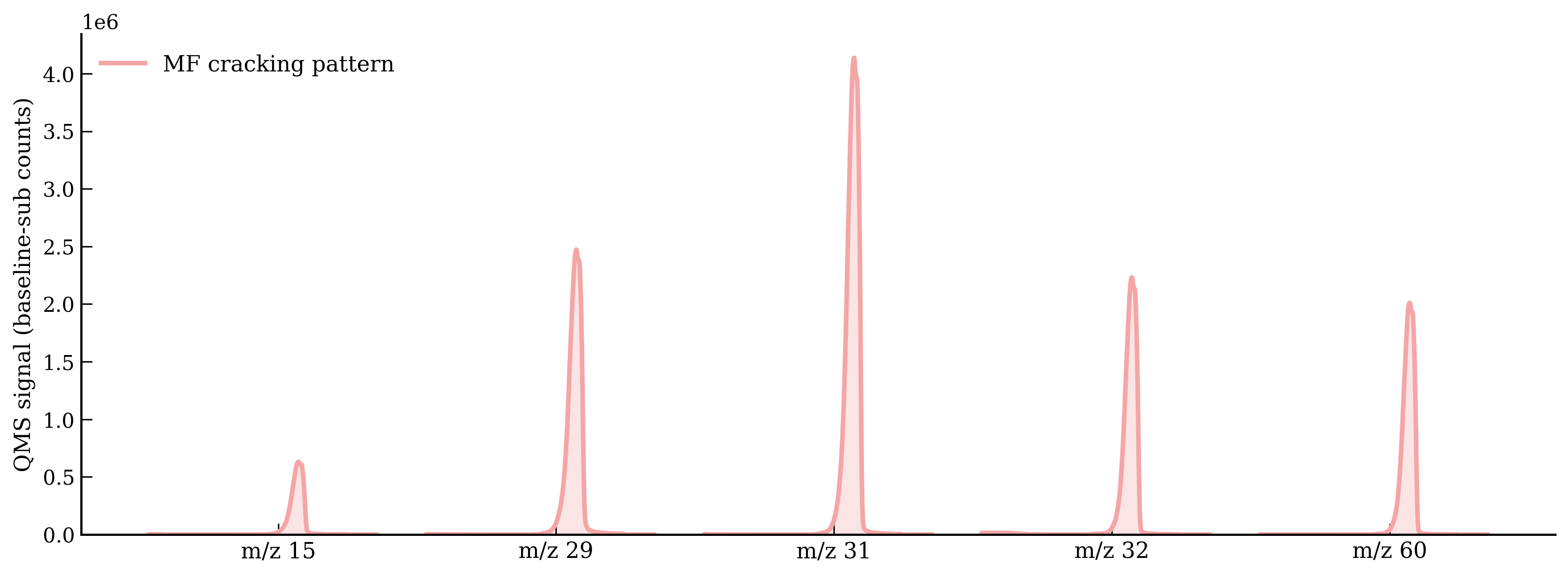}

\vspace{0.15cm}

{\small (a) MF cracking pattern}

\vspace{0.5cm}

\includegraphics[width=0.9\linewidth]{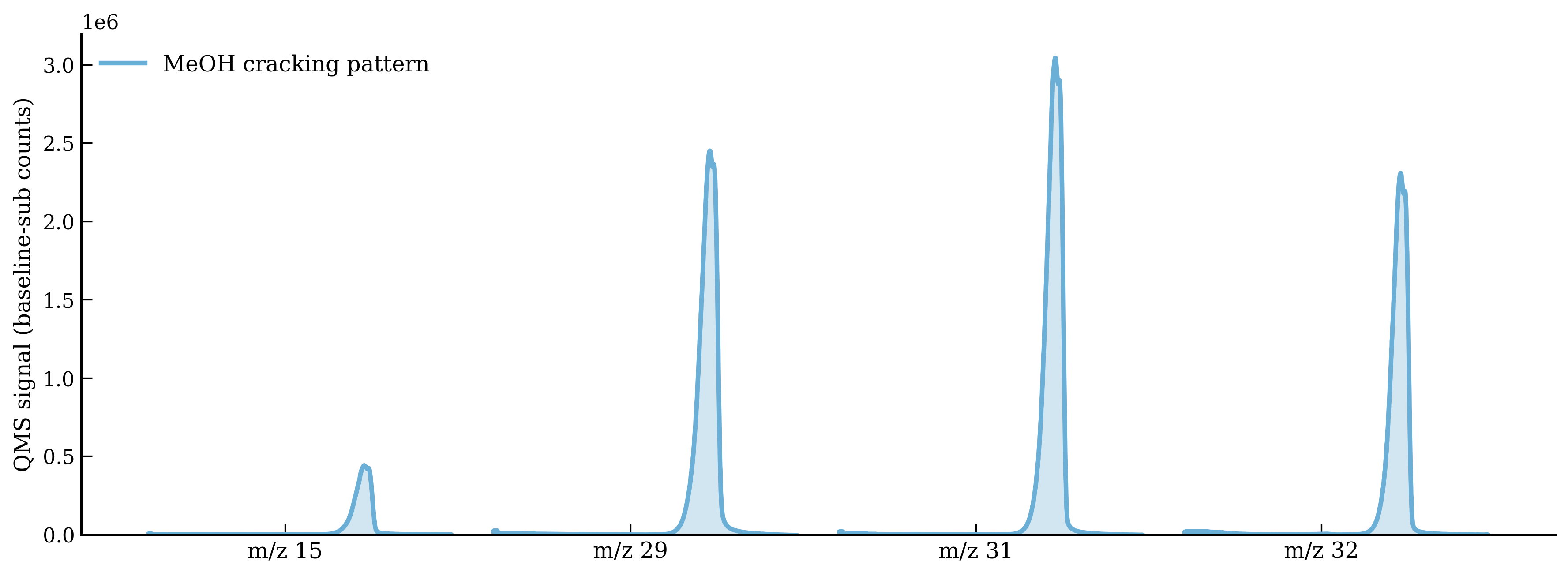}

\vspace{0.15cm}

{\small (b) MeOH cracking pattern}

\caption{
Instrument-specific QMS cracking patterns measured from pure MF and pure \methanol\ TPD experiments under identical conditions. Signals are baseline-subtracted and shown as a function of fragment mass channel.
}

\label{fig:cracking_patterns}

\end{figure*}

To account for overlapping fragments between MF and \methanol, we measured pure MF and pure \methanol\ TPDs under identical QMS tuning and acquisition settings, shown in Fig.\ref{fig:cracking_patterns}. From the pure MF run, we derived instrument-specific cracking ratios, defined as the relative fragmentation pattern of a molecule in the mass spectrometer, expressed as the ratio of the integrated signal in a given mass channel to that of a reference channel. Specifically, we compute $A(m/z)/A(60)$ for shared channels (15, 29, 31, 32), where $A(m/z)$ is the integrated signal over the MF desorption window defined by the diagnostic $m/z = 60$ channel. These ratios quantify the expected contribution of MF to each shared fragment channel in layered experiments, where MF is traced using $m/z = 60$.

\begin{table}[t]
\centering
\caption{Instrument-specific QMS cracking ratios derived from baseline-subtracted integrated TPD signals. MF ratios are normalized to the MF tracer m/z 60, and \methanol\ ratios are normalized to the \methanol\ tracer m/z 31.}
\label{tab:cracking_ratios}

\begin{tabular}{ccc}
\hline
m/z & MF $A(m/z)/A(60)$ & \methanol\ $A(m/z)/A(31)$ \\
\hline
15 & 0.330 & 0.150 \\
29 & 1.331 & 0.856 \\
31 & 2.149 & 1.000 \\
32 & 1.129 & 0.743 \\
60 & 1.000 & --- \\
\hline
\end{tabular}
\end{table}

\section{TST Prefactor and the Adsorbed-State Partition Function}
\label{app:qads}

In the main analysis, we adopt the localized-adsorbate approximation used by \citet{Ligterink2023}, in which the adsorbed molecule is assumed to be confined to a binding site and $q_{\rm ads}\approx 1$. In the full TST expression,
\begin{equation}
\nu(T) = \frac{k_B T}{h} \frac{q^{\ddagger}}{q_{\rm ads}},
\end{equation}
$q_{\rm ads}$ represents the partition function of the adsorbed reactant state. This term depends on the surface-bound degrees of freedom available to the molecule, including hindered translations, hindered rotations, and low-frequency vibrational modes.

The adsorbed-state partition function has not been experimentally constrained for MF on H$_2$O- or CH$_3$OH-rich ice surfaces. In reality, $q_{\rm ads}$ may vary with substrate composition, morphology, and local binding geometry. Statistical-thermodynamic treatments of desorption have shown that the assumed adsorbate model and adsorbed-state entropy can change desorption prefactors by orders of magnitude \citep{KnopfAmmann2021,Tait2005_II}. The binding energies reported in this work should therefore be interpreted within the context of the localized-adsorbate approximation adopted for the TST prefactor.

Because $q_{\rm ads}$ appears in the denominator of the TST expression, a larger adsorbed-state partition function lowers the inferred prefactor. The resulting effect on $E_{\rm des}$ is not linear, because the Redhead relation depends logarithmically on the prefactor. Therefore, order-of-magnitude uncertainty in $q_{\rm ads}$ can translate into systematic shifts of several hundred Kelvin in the derived binding energies for the desorption temperatures measured here.

If $q_{\rm ads}$ is comparable across the investigated substrates, this uncertainty primarily affects the absolute binding-energy scale. However, if $q_{\rm ads}$ differs strongly between H$_2$O-rich, CH$_3$OH-rich, and MF-rich binding environments, it could also modify the inferred magnitude of the substrate-dependent binding-energy differences.

A more rigorous determination of $q_{\rm ads}$ would require explicit modeling of MF bound to representative H$_2$O- and CH$_3$OH-rich ice surfaces, including the translational and rotational modes that contribute to the adsorbed-state partition function. Such calculations would provide useful surface-dependent desorption parameters for astrochemical models, but are beyond the scope of the present work.

\section{NNLS Inversion Results for Individual Runs}
\label{app:NNLS_validation_table}

Table \ref{tab:nnls_runs} lists the binding-energy distribution metrics derived from the NNLS inversion for each individual experiment. For each run we report the median binding energy of the recovered distribution together with the uncertainty components derived from Monte Carlo noise propagation and the temperature systematic. The combined per-run uncertainty is obtained by adding these two terms in quadrature.

For substrates where multiple low-coverage experiments were analyzed, the representative binding energy reported in Table \ref{tab:binding_energy_summary} corresponds to the mean of the median energies from the selected runs. The final quoted uncertainty includes both the mean per-run uncertainty and the run-to-run scatter added in quadrature.

\begin{table*}[t]
\centering
\caption{Binding-energy distribution metrics derived from NNLS inversion for each individual experiment.}
\label{tab:nnls_runs}

\begin{tabular}{ccccccc}
\hline
Run ID & Substrate & Coverage (ML) & Median $E$ (K) & $\sigma_{\rm MC}$ (K) & $\sigma_T$ (K) & $\sigma_{\rm run}$ (K) \\
\hline

y13 & Amorphous H$_2$O & 0.20 & 6265 & 3.2 & 37.4 & 37.6 \\
y15 & Amorphous H$_2$O & 0.30 & 6251 & 4.9 & 37.9 & 38.2 \\
y16 & Amorphous H$_2$O & 0.55 & 6225 & 1.1 & 38.1 & 38.1 \\

\hline
\multicolumn{7}{c}{\textit{Amorphous CH$_3$OH (sub-monolayer runs)}} \\

y25 & Amorphous CH$_3$OH & 0.25 & 5458 & 2.0 & 38.0 & 38.1 \\
y26 & Amorphous CH$_3$OH & 0.45 & 5465 & 1.2 & 38.2 & 38.2 \\
y27 & Amorphous CH$_3$OH & 0.6 & 5472 & 0.8 & 38.1 & 38.1 \\
y28 & Amorphous CH$_3$OH & 1.00 & 5480 & 0.6 & 38.0 & 38.0 \\

\hline
\multicolumn{7}{c}{\textit{Crystalline H$_2$O}} \\

y12 & Crystalline H$_2$O & 0.20 & 6213 & 1.6 & 38.2 & 38.3 \\
y14 & Crystalline H$_2$O & 0.30 & 6181 & 0.7 & 37.8 & 37.9 \\
y11 & Crystalline H$_2$O & 0.40 & 6205 & 0.5 & 38.1 & 38.1 \\
y8  & Crystalline H$_2$O & 1.00 & 6184 & 0.3 & 37.9 & 37.9 \\
y4  & Crystalline H$_2$O & 1.50 & 6418 & 1.5 & 38.3 & 38.4 \\

\hline
\end{tabular}

\vspace{0.2cm}

\parbox{0.9\textwidth}{\footnotesize
\textbf{Note.} $\sigma_{\rm MC}$ represents the Monte Carlo uncertainty derived from 1000 individual noise propagations. $\sigma_T$ represents the temperature-scale systematic obtained from inversions performed at $T \pm \Delta T$. The combined per-run uncertainty is $\sigma_{\rm run} = \sqrt{\sigma_{\rm MC}^2 + \sigma_T^2}$. Representative binding energies reported in Table \ref{tab:binding_energy_summary} are derived from the mean of the selected runs for each substrate: amorphous H$_2$O (0.20, 0.30, and 0.55 ML), crystalline H$_2$O (0.20 ML), and all selected runs for amorphous CH$_3$OH.. A coverage-to-coverage scatter term is added in quadrature with the mean per-run uncertainty to obtain the final quoted uncertainties.}
\end{table*}

\section{Leading-Edge Arrhenius and Rate Panels}
\label{appendix:arrhenius}

\begin{figure*}[t]
\centering
\includegraphics[width=0.9\textwidth]{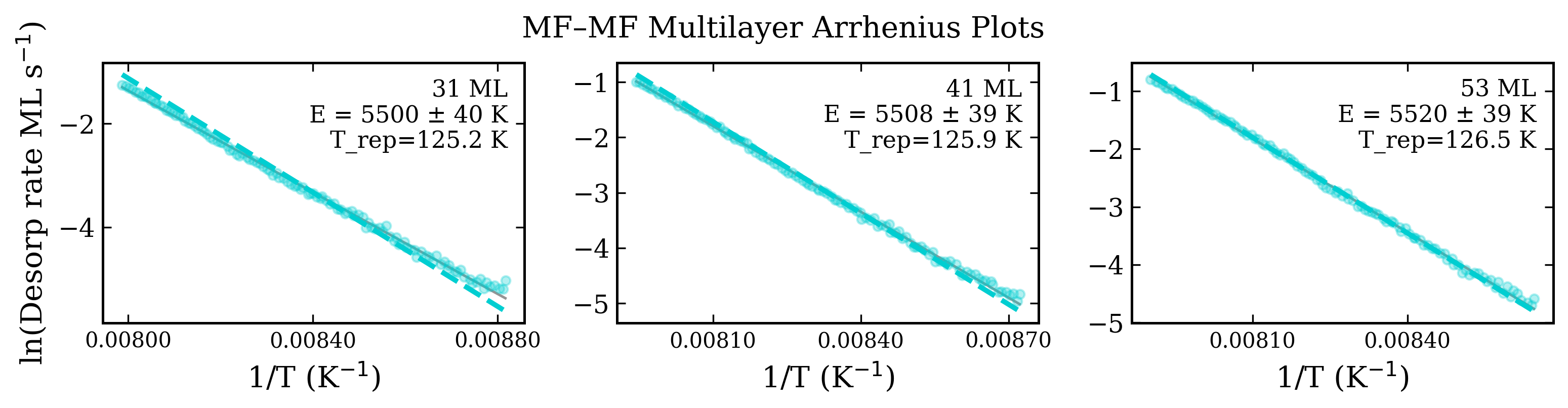}
\caption{
Arrhenius plots for pure multilayer MF (31-53 ML).
Points show $\ln r_{\rm des}$ (ML\,s$^{-1}$) plotted against $1/T$ for the objectively selected leading-edge interval.
Dashed lines show the fixed-$\nu$ fits, where $\nu=\nu_{\rm TST}(T_{\rm rep})$ is held constant and $E_{\rm des}$ is determined from the slope.
Thin solid lines show the corresponding free-intercept linear fits for comparison.
The linear behavior confirms coverage-independent multilayer desorption kinetics.
}
\label{fig:arrhenius_pureMF_appendix}
\end{figure*}

\begin{figure*}[t]
\centering
\includegraphics[width=0.9\textwidth]{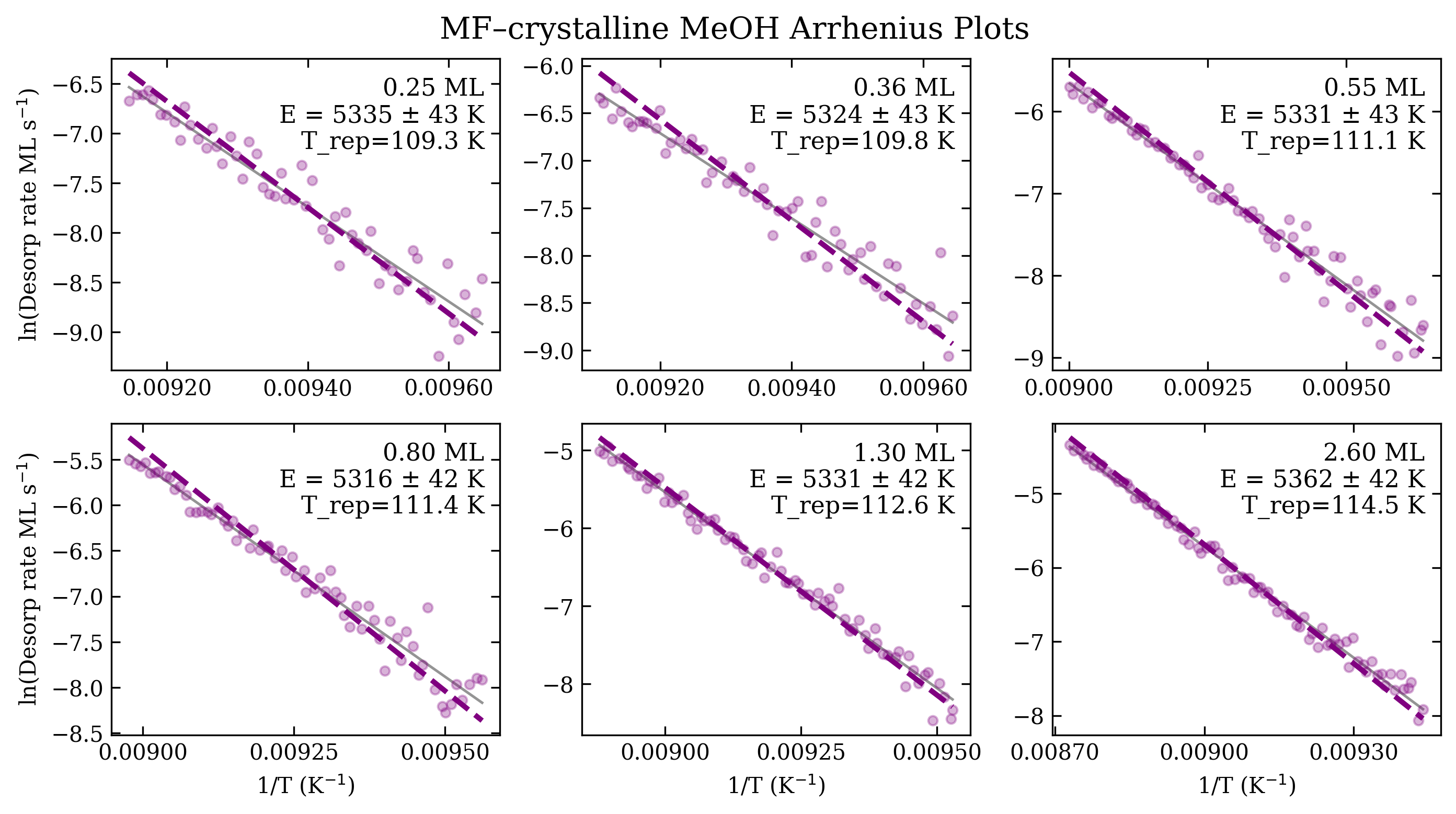}
\caption{
Arrhenius plots for MF deposited on crystalline methanol. Points show $\ln r_{\rm des}$ (ML\,s$^{-1}$) plotted against $1/T$ for the selected leading-edge intervals. Dashed lines show the corresponding fixed-$\nu$ fits, where $\nu=\nu_{\rm TST}(T_{\rm rep})$.
}
\label{fig:arrhenius_cMeOH_appendix}
\end{figure*}

Figures \ref{fig:arrhenius_pureMF_appendix} and \ref{fig:arrhenius_cMeOH_appendix} show the Arrhenius plots used in the leading-edge analysis for pure multilayer MF and for MF deposited on crystalline \methanol. In each panel, the points show the linear $\ln r_{\rm des}$ (ML\,s$^{-1}$) plotted against $1/T$ for the leading-edge interval identified from the corresponding rate panels (Section \ref{sec:leading_edge}).

For pure multilayer MF (Figure \ref{fig:arrhenius_pureMF_appendix}), the three thick-ice experiments (31--53 ML) define closely similar linear trends in Arrhenius space. The dashed lines show the fixed-$\nu$ fits, in which the prefactor is fixed to $\nu_{\rm TST}(T_{\rm rep})$ and the binding energy is determined from the slope. Thin solid lines show the corresponding free-intercept fits for comparison. The consistent slopes across all three coverages support coverage-independent multilayer desorption kinetics. The weighted mean binding energy derived from these experiments is $5509 \pm 10$ K.

Figure \ref{fig:arrhenius_cMeOH_appendix} shows the corresponding Arrhenius analysis for MF deposited on crystalline methanol. Despite the sub-monolayer nominal coverages, the selected leading-edge intervals define approximately linear trends in Arrhenius space. The dashed lines show the fixed-$\nu$ fits using $\nu=\nu_{\rm TST}(T_{\rm rep})$. The resulting binding energies are consistent with multilayer-like desorption from MF islands on the crystalline methanol surface, as discussed in Section \ref{sec:crysMeOH}.

\section{Forward-Model Validation of the NNLS Binding-Energy Inversion}
\label{app:distribution_inversion}

To assess the robustness of the NNLS binding-energy inversion, we forward-modeled the desorption profiles using the recovered binding-energy distributions for each selected coverage and substrate. In this validation step, the distribution $f(E)$ recovered from the inversion was combined with the same fixed-prefactor Polanyi-Wigner basis functions used in the inversion procedure to reconstruct a model desorption profile, which was then compared directly to the normalized experimental TPD curve.

\begin{figure*}[t]
\centering
\includegraphics[width=0.85\textwidth]{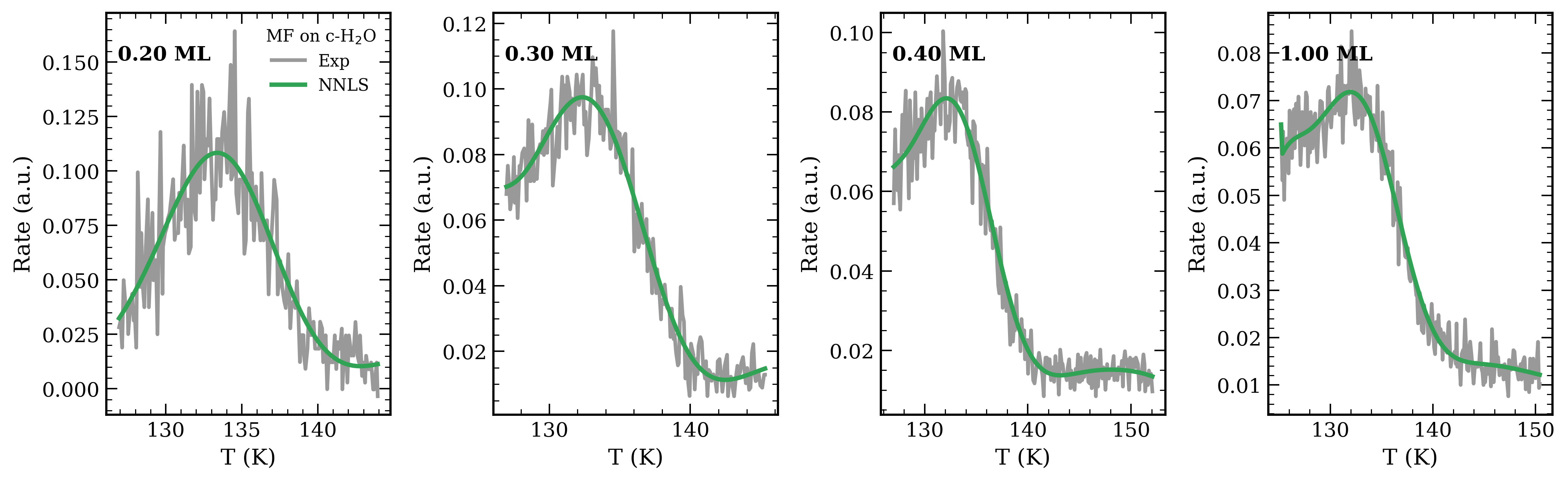}
\includegraphics[width=0.85\textwidth]{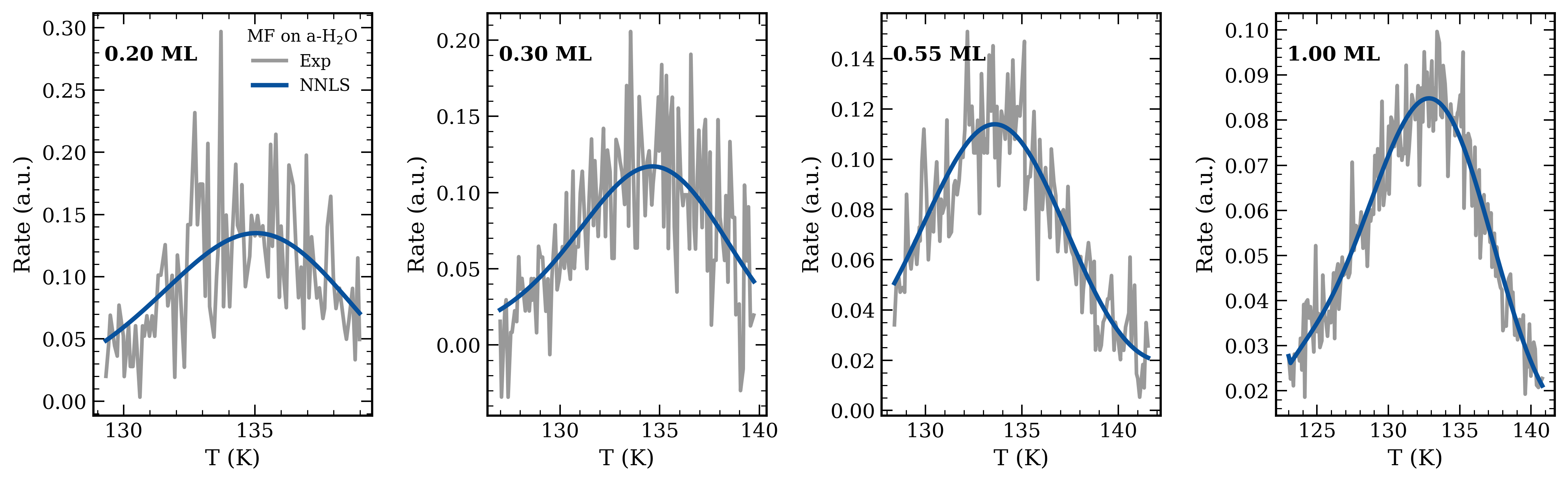}
\includegraphics[width=0.85\textwidth]{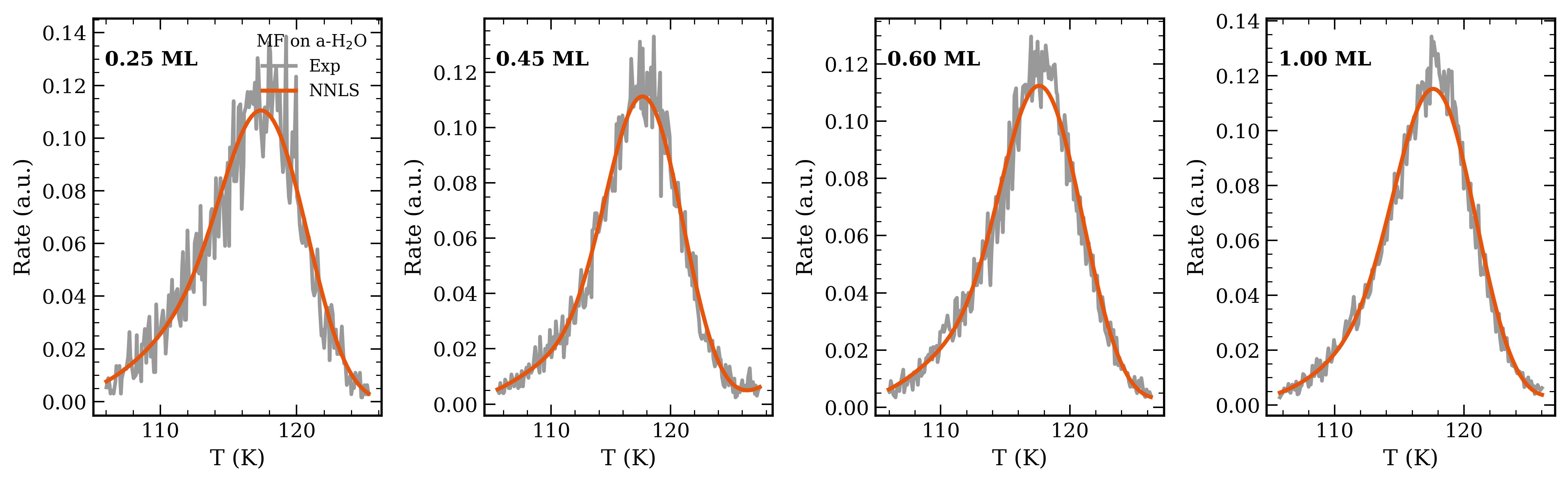}
\caption{Forward-model validation of the NNLS binding-energy inversion for MF desorption from different substrates. Gray curves show the normalized experimental desorption profiles, while the colored curves show the forward-model reconstructions generated from the recovered binding-energy distributions. Top: crystalline H$_2$O. Middle: amorphous H$_2$O. Bottom: amorphous CH$_3$OH.}
\label{fig:nnls_validation_all}
\end{figure*}

Figure~\ref{fig:nnls_validation_all} shows the resulting forward-model reconstructions for crystalline \water\ (top), amorphous \water\ (middle), and amorphous \methanol\ (bottom). In each panel, the gray curve is the normalized experimental desorption profile and the colored curve is the forward-model reconstruction obtained from the recovered distribution. The agreement between the reconstructed and experimental profiles demonstrates that the NNLS inversion captures the overall peak positions, widths, and asymmetries of the selected desorption components.

For amorphous \water, the selected low-coverage spectra are reproduced well, supporting the interpretation that these profiles are dominated by MF-H$_2$O interactions and are suitable for determining a representative MF-H$_2$O binding energy. At higher coverages, the reconstructed profiles become broader, consistent with the emergence of mixed MF-MF and MF-H$_2$O contributions discussed in Section \ref{binding_energy_compare}. For crystalline \water, only the 0.20 ML profile yields a compact and well-defined reconstruction, whereas higher coverages broaden rapidly, consistent with the early onset of intermolecular MF-MF interactions. For amorphous \methanol, the recovered distributions and forward-model reconstructions remain similar across the investigated coverages, supporting the use of an effective binding energy for MF desorption from the amorphous \methanol\ surface.

These forward-model comparisons provide an internal consistency check on the inversion results and confirm that the representative binding energies adopted in Table \ref{tab:binding_energy_summary} are derived from distributions that reproduce the corresponding experimental desorption profiles.

\section{Reference TPD} \label{app:reference_tpd}

\begin{figure}[t]
\centering
\includegraphics[width=0.45\textwidth]{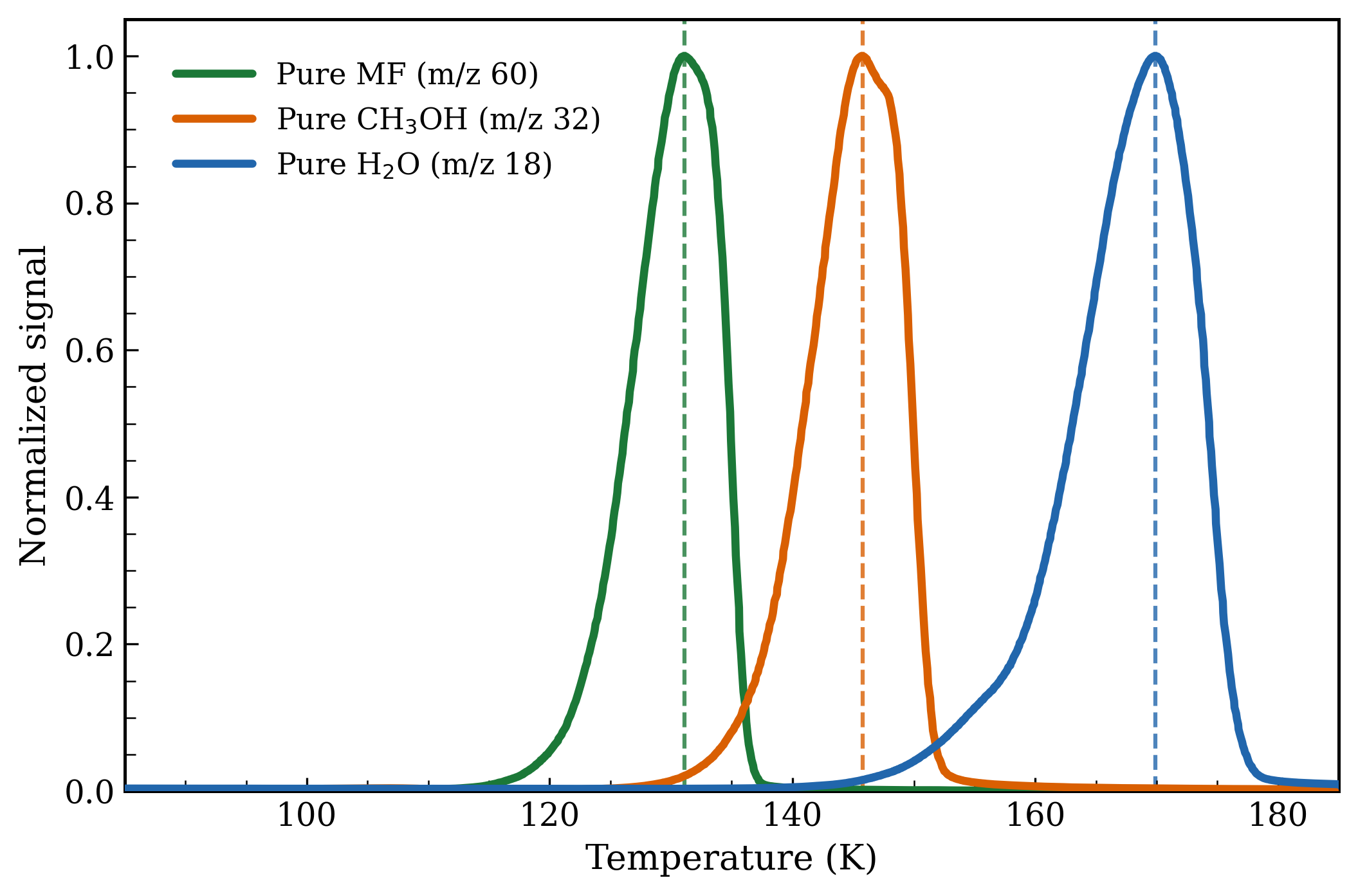}

\caption{Normalized temperature-programmed desorption (TPD) spectra of pure MF (m/z 60), \methanol\ (m/z 32), and H$_2$O (m/z 18). Each trace is normalized to its maximum intensity to highlight the relative ordering of desorption temperatures for the three pure ices. Dashed vertical lines indicate the peak desorption temperature for each species.}

\label{reference_tpd}
\end{figure}

The three spectra shown in Figure \ref{reference_tpd} are the baseline TPD curves obtained for pure MF, \methanol, and \water\ ices that were deposited directly on the QCM.  All experiments were carried out under identical instrumental conditions.

\end{document}